\documentclass
[prd,tightenlines,nofootinbib,secnumarabic,byrevtex,groupedaddress,11pt,showpacs]{revtex4}%
\usepackage{amsfonts}
\usepackage{amsmath}
\usepackage{amssymb}
\usepackage{graphicx}
\usepackage{fancyheadings}%
\begin{document}
\title[C\&Q Nambu Mech]{Classical and Quantum Nambu Mechanics}
\author{Thomas Curtright}
\email{curtright@physics.miami.edu}
\affiliation{Department of Physics, University of Miami, Coral Gables, FL 33124-8046, USA}
\author{Cosmas Zachos}
\email{zachos@hep.anl.gov}
\affiliation{High Energy Physics Division, Argonne National Laboratory, Argonne, IL
60439-4815, USA}
\keywords{one two three}
\pacs{02.30.Ik,11.30.Rd,11.25.Yb}

\begin{abstract}
The classical and quantum features of Nambu mechanics are analyzed and
fundamental issues are resolved. \ The classical theory is reviewed and
developed utilizing varied examples. The quantum theory is discussed in a
parallel presentation, and illustrated with detailed specific cases.
\ Quantization is carried out with standard Hilbert space methods. \ With the
proper physical interpretation, obtained by allowing for different time scales
on different invariant sectors of a theory, the resulting non-Abelian approach
to quantum Nambu mechanics is shown to be fully consistent.

\end{abstract}
\volumeyear{year}
\volumenumber{number}
\issuenumber{number}
\eid{identifier}
\received{26 December 2002}

\accepted{28 April 2003}

\startpage{1}
\endpage{ }
\maketitle
\tableofcontents

\newpage

\section{Introduction}

\paragraph{\underline{A brief historical overview}}

Nambu \cite{Nambu} introduced an elegant generalization of the classical
Hamiltonian formalism by suggesting to supplant the Poisson Bracket (PB) with
a $3$- or $n$-linear, fully antisymmetric bracket, the Classical Nambu Bracket
(CNB), a volume-element Jacobian determinant in a higher dimensional space.
\ This bracket, involving a dynamical quantity and two or more
\textquotedblleft Hamiltonians\textquotedblright, provides the time-evolution
of that quantity in a generalization of Hamilton's equations of motion for
selected physical systems. \ It was gradually realized
\cite{Bayen,Mukunda,Chatterjee} that Nambu Brackets in phase space describe
the generic classical evolution of all systems with sufficiently many
independent integrals of motion beyond those required for complete
integrability of the systems. \ That is to say, all such \textquotedblleft
superintegrable systems\textquotedblright\ \cite{Tempesta}\ are automatically
described by Nambu's mechanics \cite{Gonera,CurtrightZachos}, whether one
chooses to take cognizance of this alternate expression of their time
development or not. \ This approach to time evolution for superintegrable
systems is supplementary to the standard Hamiltonian dynamics evolution and
provides additional tools for analyzing such systems. \ The power of Nambu's
method is evident in a manifest and simultaneous accounting of a maximal
number of the symmetries of these systems, and in an efficient application of
algebraic methods to yield results even without detailed knowledge of their
specific dynamics.

As a bonus, the classical volume-preserving features of Nambu brackets suggest
that they are useful for membrane theory \cite{Minic}. \ There are in the
literature several persuasive but inconclusive arguments that Nambu brackets
are a natural language for describing extended objects, for example:
\cite{Awata,Azcarraga97,Baker,CurtrightFairlie,DitoFST,Hoppe,Kerner,Matsuo,MinicTze,Pioline,Vainerman}%
.

In his original paper \cite{Nambu}, Nambu also introduced operator versions of
his brackets as tools to implement the quantization of his approach to
mechanics. \ He enumerated various logical possibilities involving them,
arguing that some structures were either inconsistent or uninteresting, but he
did not advocate the position that the remaining possibilities were untenable:
\ Quantization was left as an open issue. \ 

Unfortunately, subsequent unwarranted insistence on algebraic structures
ill-suited to the solution of the relevant physics problems resulted in a
widely held belief that quantization of Nambu mechanics was
problematic\footnote{{\footnotesize A few representative statements from the
literature are: \ \textquotedblleft Associated statistical mechanics and
quantization are unlikely.\textquotedblright\ \cite{Estabrook};
\ \textquotedblleft A quantum generalization of these algebras is shown to be
impossible.\textquotedblright\ and \textquotedblleft\ \ldots\ the quantum
analog of Nambu mechanics does not exist.\textquotedblright\ \cite{Sahoo};
\ \textquotedblleft Usual approaches to quantization have failed to give an
appropriate solution...\textquotedblright\ \cite{DitoFlato};
\ \textquotedblleft...direct application of deformation quantization to
Nambu-Poisson structures is not possible.\textquotedblright\ \cite{DitoFST};
\ \textquotedblleft The quantization of Nambu bracket turns out to be a quite
non-trivial problem.\textquotedblright\ \cite{Xiong}; \ \textquotedblleft This
problem is still outstanding.\textquotedblright\ \cite{Takhtajan}.}},
especially when that quantization was formulated as a one-parameter
deformation of classical structures. \ In marked contrast to this prevailing
pessimism, several illustrative superintegrable systems were quantized in
\cite{CurtrightZachos} in a phase-space framework, both without and
\emph{with} the construction of Quantum Nambu Brackets (QNBs). \ However, the
phase-space quantization utilized there, while most appropriate for comparing
quantum expressions with their classical limits, is still unfamiliar to many
readers and will not be used in this paper. \ Here, the quantization of all
systems will be carried out in a conventional Hilbert space operator formalism.

It turns out \cite{CurtrightZachos} that all perceived difficulties in
quantizing Nambu mechanics may be traced mathematically to the algebraic
inconsistencies inherent in selecting constraints in a top-down approach, with
little regard to the correct phase-space structure which already provides full
and consistent answers, and with insufficient attention towards obtaining
specific answers compatible with those produced in the quantized Hamiltonian
description of these systems. \ Moreover, the physics underlying these
perceived difficulties is simple, and involves only basic principles in
quantum mechanics.

\paragraph{\underline{Evolution scales in quantum physics}}

Some physicists might hold, without realizing it, the prejudice that
continuous time evolution in quantum mechanics must always be formulated
infinitesimally as a derivation. \ Accordingly, they implicitly \emph{assume}
the instantaneous temporal change in all dynamical variables is always given
by nothing but a simple derivative, so that for all products of linear
operators%
\begin{equation}
\frac{d}{dt}\left(  AB\right)  =\left(  \frac{d}{dt}A\right)  B+A\left(
\frac{d}{dt}B\right)  \;.
\end{equation}
This assumption allows time development on physical Hilbert spaces to be
expressed algebraically in terms of commutators with a Hamiltonian, since
commutators are also derivations.\footnote{For simplicity we will assume,
unless otherwise stated, that the operators have no \emph{explicit} time
dependence, although it is an elementary exercise to relax this assumption.}%
\begin{equation}
\left[  H,AB\right]  =\left[  H,A\right]  B+A\left[  H,B\right]  \;.
\end{equation}
Evidently this approach leads to the simplest possible formalism. \ But is it
really necessary to make this assumption and follow this approach? \ 

It is not. \ Time evolution can also be expressed algebraically using quantum
Nambu brackets. \ These quantum brackets are defined as totally
antisymmetrized multi-linear products of any number of linear operators acting
on Hilbert space. \ When QNBs are used to implement time evolution in quantum
mechanics, the result is usually \emph{not} a derivation, but contains
derivations entwined within more elaborate structures (although there are some
interesting special exceptions that are described in the following).

This more general point of view towards time development can be arrived at
just by realizing a physical idea. When a system has a number of conserved
quantities, it is possible to partition the system's Hilbert space into
invariant sectors. \ Time evolution on those various sectors may then be
formulated using different time scales for the different sectors\footnote{In
fact, the choice of time variables in the different invariant sectors of a
quantum theory is very broad. \ They need not be just multiples of one
another, but could have complicated functional dependencies, as discussed in
\cite{Curtright} and \cite{MinicTze}. \ The closest classical counter-part of
this is found in the general method of \emph{analytic} \emph{time}, recently
exploited so effectively in \cite{Calogero,Francoise}.} \ The resulting
expression of instantaneous changes in time is then not a derivation, in
general, when acting on the full Hilbert space, and therefore is not given by
a simple commutator. \ Remarkably, however, it often turns out to be given
compactly in terms of QNBs. \ Conversely, if QNBs are used to describe time
development, they usually impose different time scales on different invariant
sectors of a system \cite{CurtrightZachos}.

Nevertheless, so long as the different time scales are implemented in such a
way as to produce evolving phase differences between nondegenerate energy
eigenstates, there is no loss of information in this more general approach to
time evolution. \ In the classical limit, this method is not really different
from the usual Hamiltonian approach. \ A given classical trajectory has fixed
values for all invariants, and hence would have a fixed time scale in Nambu
mechanics. \ Time development of any dynamical quantity along a single
classical trajectory would therefore always be just a derivation, with no
possibility of mixing time scales. \ Quantum mechanics, on the other hand, is
more subtle, since the preparation of a state may yield a superposition of
components from different invariant sectors. \ Such superpositions will, in
general, involve multiple time scales in Nambu mechanics.

Technically, the various time scales arise in quantum Nambu mechanics as the
entwined eigenvalues of generalized Jordan spectral problems, where selected
invariants of the model in question appear as operators in the spectral
equation. \ The resulting structure represents a new class of eigenvalue
problems for mathematical physics. \ Fortunately, solutions of this new class
can be found using traditional methods. \ (All this is explained explicitly in
the context of the first example of \S 3.2.)

\paragraph{\underline{Related studies in mathematics}}

Algebras which involve multi-linear products have also been considered at
various times in the mathematical literature, partly as efforts to understand
or generalize Jordan algebras \cite{Jacobson,Jordan,Lister} (cf. especially
the ``associator''), but more generally following Higgins' study in the
mid-1950s \cite{Baranovich,Filippov,Higgins,Kurosh}. \ This eventually
culminated in the investigations of certain cohomology questions, by
Schlesinger and Stasheff \cite{Schlesinger}, by Hanlon and Wachs
\cite{Hanlon}, and by Azc\'{a}rraga, Izquierdo, Perelomov, and P\'{e}rez Bueno
\cite{Azcarraga96a,Azcarraga96b,Azcarraga97}, that led to results most
relevant to Nambu's work.

\paragraph{\underline{Summary of material to follow}}

After a few motivational remarks on the geometry of Hamiltonian flows in
phase-space, \S 2.1, we describe the most important features of classical
Nambu brackets, \S 2.2, with emphasis on practical, algebraic, evaluation
methods. \ We delve into several examples, \S 2.3, to gain physical insight
for the classical theory.

We then give a parallel discussion of the quantum theory, \S 3.1, so far as
algebraic features and methods of evaluation are concerned. \ We define QNBs,
as well as Generalized Jordan Products (GJPs) that naturally arise in
conjunction with QNBs, when the latter are resolved into products of
commutators. We define derivators as measures of the failure of the Leibniz
rule for QNBs, and discuss Jacobi and Fundamental identities in a quantum
setting. \ Then we again turn to various examples, \S 3.2, to illustrate both
the elegance and peculiarities of quantization. \ We deal with essentially the
same examples in both classical and quantum frameworks, as a means of
emphasizing the similarities and, more importantly, delineating the
differences between CNBs and QNBs. \ The examples chosen are all models based
on Lie symmetry algebras: \ $so\left(  3\right)  =su\left(  2\right)  $,
$so\left(  4\right)  =su\left(  2\right)  \times su\left(  2\right)  $,
$so\left(  n\right)  $, $u\left(  n\right)  $, $u\left(  n\right)  \times
u\left(  m\right)  $, and $g\times g$. \ 

We conclude by summarizing our results, and by suggesting some topics for
further study. \ An Appendix discusses the formal solution of linear equations
in Lie and Jordan algebras, with suggestions for techniques to bypass the
effects of divisors of zero.

A hurried reader may wish to consider only \S 2.1, \S 2.2 through
Eqn(\ref{2n=2+(2n-2)}), \S 2.3 through Eqn(\ref{AnyLie4CNB}), \S 3.1 through
Eqn(\ref{MapsTo}), and \S 3.2 through Eqn(\ref{dA/dtAnyLie}). \ This abridged
material contains our main points.

\section{Classical Theory}

We begin with a brief geometrical discussion of phase-space dynamics, to
motivate the definition of classical Nambu brackets. \ We then describe
properties of CNBs, with emphasis on practical evaluation methods, including
various recursion relations among the brackets and simplifications that result
from classical Lie symmetries being imposed on the entries in the bracket.
\ We summarize the theory of the fundamental identity and explain its
subsidiary role. \ We then go through several examples to gain physical
insight for the classical formalism. \ All the examples are based on systems
with Lie symmetries: \ $so(3)=su\left(  2\right)  $, $u(n)$,
$so(4)=su(2)\times su(2)$, and $g\times g$.

\subsection{Phase-space geometry}

A Hamiltonian system with $N$ degrees of freedom is \emph{integrable} in the
Liouville sense if it has $N$ invariants in involution (globally defined and
functionally independent), and \emph{superintegrable} \cite{Tempesta} if it
has additional independent conservation laws up to a maximum total number of
$2N-1$ invariants. \ For a maximally superintegrable system, the total
multi-linear cross product of the $2N-1$ local phase-space gradients of the
invariants (each such gradient being perpendicular to its corresponding
invariant isocline) is always locally tangent to the classical trajectory.

\begin{center}%
\begin{center}
\includegraphics[
height=3.9539in,
width=5.5158in
]%
{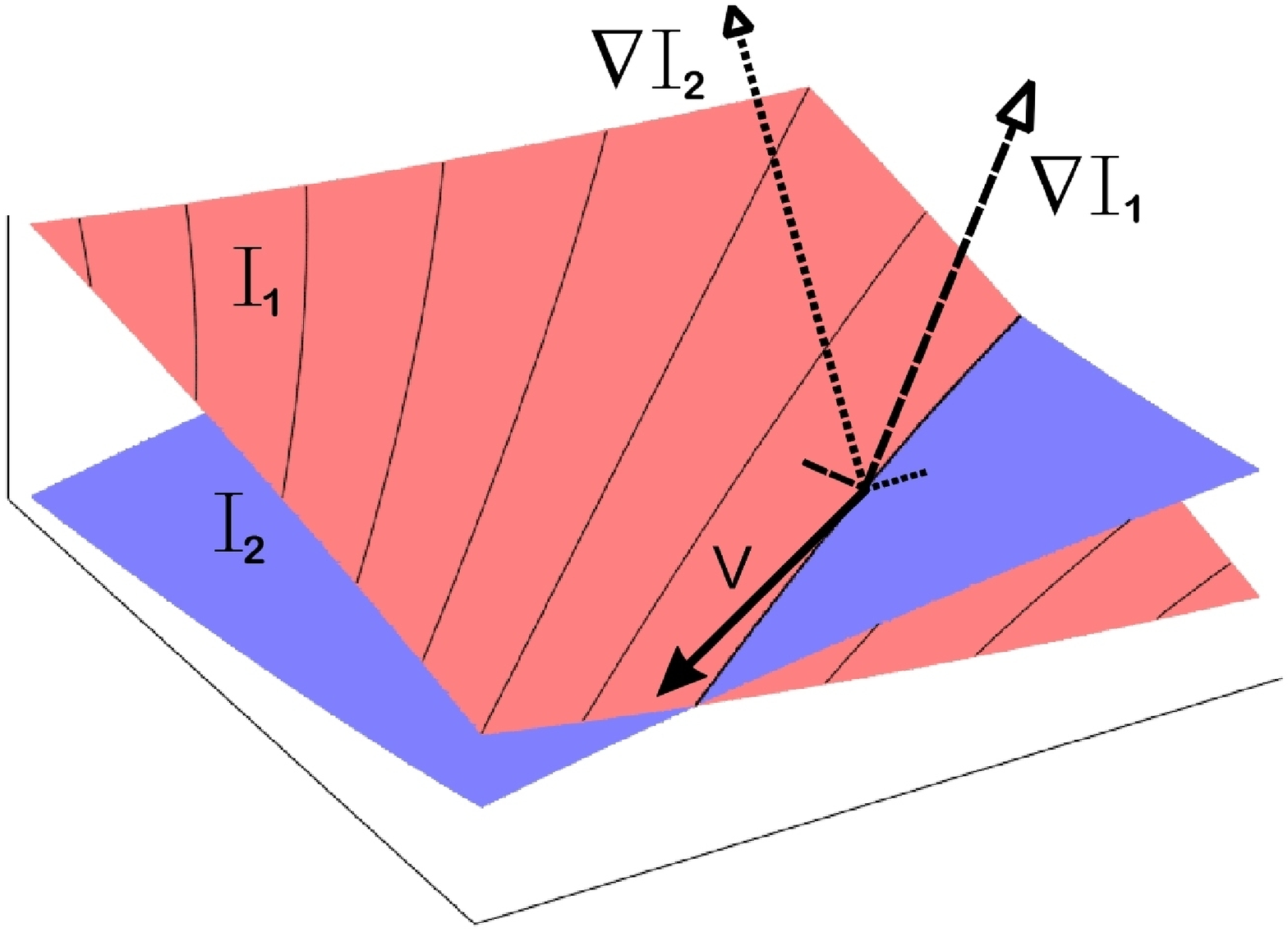}%
\\
Some classical phase-space geometry
\end{center}

\end{center}

\begin{quotation}
The illustrated surfaces are isoclines for two different invariants,
respectively $I_{1}$ and $I_{2}$. \ A particular trajectory lies along the
intersection of these two surfaces. \ The local phase-space tangent
$\mathsf{v}$ to this trajectory at the point depicted is given by the
cross-product of the local phase-space gradients of the invariants. \ (Other
possible trajectories along the $I_{1}$ surface are also shown as contours
representing other values for $I_{2}$, but the corresponding intersecting
$I_{2}$ surfaces are not shown for those other trajectories.)\bigskip
\end{quotation}

Thus, in $2N$-dimensional phase space, for any phase-space function
$A(\mathbf{x},\mathbf{p})$ with no explicit time dependence, the convective
motion is fully specified by a phase-space Jacobian which amounts to the
classical Nambu bracket,%
\begin{equation}
\frac{dA}{dt}\equiv\mathbf{v}\cdot\nabla A\;\;\propto\;\;\partial_{i_{1}%
}A\;\epsilon^{i_{1}i_{2}\cdots i_{2N}}\;\partial_{i_{2}}I_{1}\cdots
\partial_{i_{2N}}I_{2N-1}=\frac{\partial(A,I_{1},\cdots,I_{2N-1})}%
{\partial(x_{1},p_{1},x_{2},p_{2},\cdots,x_{N},p_{N})}\;, \label{dA/dt}%
\end{equation}
where $\mathbf{v}=(\mathbf{\dot{x}},\mathbf{\dot{p}})$ is the phase-space
velocity, and the phase-space gradients are $\nabla=(\partial_{\mathbf{x}%
},\partial_{\mathbf{p}})$. \ Evidently, the flow is divergenceless,
$\nabla\cdot\mathbf{v}=0$ (Liouville's theorem \cite{Nambu}). \ In short, a
superintegrable system in phase space \emph{can} \emph{hardly} \emph{avoid}
having its classical evolution described by CNBs \cite{CurtrightZachos}.

\subsection{Properties of the classical brackets}

\paragraph{\underline{Definitions}}

For a system with $N$ degrees of freedom, and hence a $2N$-dimensional
phase-space, we define the maximal classical Nambu bracket (CNB) of rank $2N$
to be the determinant%
\begin{equation}
\{A_{1},A_{2},\cdots,A_{2N}\}_{\text{NB}}=\frac{\partial(A_{1},A_{2}%
,\cdots,A_{2N})}{\partial(x_{1},p_{1},x_{2},p_{2},\cdots,x_{N},p_{N}%
)}=\epsilon^{i_{1}i_{2}\cdots i_{2N}}\;\partial_{i_{1}}A_{1}\cdots
\partial_{i_{2N}}A_{2N}\;. \label{CNBDefn}%
\end{equation}
This bracket is linear in its arguments, and completely antisymmetric in them.
\ It may be thought of as the Jacobian induced by transforming to new phase
space variables $A_{i}$, the \textquotedblleft elements\textquotedblright\ in
the bracket. \ As expected for such a Jacobian, two functionally dependent
elements cause the bracket to collapse to zero. \ So, in particular, adding to
any element an arbitrary linear combination of the other elements will not
change the value of the bracket.

Odd dimensional brackets are also defined identically \cite{Nambu}, in an
odd-dimensional space.

\paragraph{\underline{Recursion relations}}

The simplest of these are immediate consequences of the properties of the
totally antisymmetric Levi-Civita symbols.%
\begin{equation}
\frac{\partial\left(  A_{1},A_{2},\cdots,A_{k}\right)  }{\partial\left(
z_{1},z_{2}\cdots,z_{k}\right)  }=\frac{\epsilon^{i_{1}\cdots i_{k}}}{\left(
k-1\right)  !}\left(  \frac{\partial A_{1}}{\partial z_{i_{1}}}\right)
\frac{\partial\left(  A_{2},\cdots,A_{k}\right)  }{\partial\left(  z_{i_{2}%
},\cdots,z_{i_{k}}\right)  }=\frac{\epsilon^{j_{1}\cdots j_{k}}}{\left(
k-1\right)  !}\left(  \frac{\partial A_{j_{1}}}{\partial z_{1}}\right)
\frac{\partial\left(  A_{j_{2}},\cdots,A_{j_{k}}\right)  }{\partial\left(
z_{2},\cdots,z_{k}\right)  }\;.
\end{equation}
However, these $k=1+\left(  k-1\right)  $ resolutions are not especially
germane to a phase-space discussion, since they reduce even brackets into
products of odd brackets.

More usefully, any maximal even rank CNB can also be resolved into products of
Poisson brackets. \ For example, for systems with two degrees of freedom,
$\left\{  A,B\right\}  _{\text{PB}}=\frac{\partial\left(  A,B\right)
}{\partial\left(  x_{1},p_{1}\right)  }+\frac{\partial\left(  A,B\right)
}{\partial\left(  x_{2},p_{2}\right)  }$, and the 4-bracket $\left\{
A,B,C,D\right\}  _{\text{NB}}\equiv\frac{\partial\left(  A,B,C,D\right)
}{\partial\left(  x_{1},p_{1},x_{2},p_{2}\right)  }$ resolves
as\footnote{These PB resolutions are somewhat simpler than their quantum
counterparts, to be given below in \S 3.1, since ordering of products is not
an issue here. \ }%
\begin{equation}
\left\{  A,B,C,D\right\}  _{\text{NB}}=\left\{  A,B\right\}  _{\text{PB}%
}\left\{  C,D\right\}  _{\text{PB}}-\left\{  A,C\right\}  _{\text{PB}}\left\{
B,D\right\}  _{\text{PB}}-\left\{  A,D\right\}  _{\text{PB}}\left\{
C,B\right\}  _{\text{PB}}\;, \label{4BracketAsPBs}%
\end{equation}
in comportance with full antisymmetry under permutations of $A,B,C,$ and $D$.
\ The general result for maximal rank $2N$ brackets for systems with a
$2N$-dimensional phase-space is\footnote{This is essentially a special case of
Laplace's theorem on the general minor expansions of determinants (cf. Ch. 4
in \cite{Aitken}), although it must be said that we have never seen it
written, let alone used, in exactly this form, either in treatises on
determinants or in textbooks on classical mechanics.}%
\begin{equation}
\left\{  A_{1},A_{2},\cdots,A_{2N-1},A_{2N}\right\}  _{\text{NB}}%
=\sum_{\substack{\text{all }\left(  2N\right)  !\text{\ perms }\\\left\{
\sigma_{1},\sigma_{2},\cdots,\sigma_{2N}\right\}  \\\text{of the indices
}\\\left\{  1,2,\cdots,2N\right\}  }}\frac{\operatorname{sgn}\left(
\sigma\right)  }{2^{N}N!}\,\left\{  A_{\sigma_{1}},A_{\sigma_{2}}\right\}
_{\text{PB}}\left\{  A_{\sigma_{3}},A_{\sigma_{4}}\right\}  _{\text{PB}}%
\cdots\left\{  A_{\sigma_{2N-1}},A_{\sigma_{2N}}\right\}  _{\text{PB}}\;,
\label{2NBracketAsPBs}%
\end{equation}
where $\operatorname{sgn}\left(  \sigma\right)  =\left(  -1\right)
^{\pi\left(  \sigma\right)  }$ with $\pi\left(  \sigma\right)  $\ the parity
of the permutation $\left\{  \sigma_{1},\sigma_{2},\cdots,\sigma_{2N}\right\}
$. \ The sum only gives $\left(  2N-1\right)  !!=\left(  2N\right)  !/\left(
2^{N}N!\right)  $ distinct products of PBs on the RHS, not $\left(  2N\right)
!$ \ Each such distinct product appears with net coefficient $\pm1$. \ 

The proof of the relation (\ref{2NBracketAsPBs}) is elementary. \ Both left-
and right-hand sides of the expression are sums of $2N$-th degree monomials
linear in the $2N$ first-order partial derivatives of each of the $A$s. Both
sides are totally antisymmetric under permutations of the $A$s. \ Hence, both
sides are also totally antisymmetric under interchanges of partial
derivatives. \ Thus, the two sides must be proportional. \ The only issue left
is the constant of proportionality. \ This is easily determined to be $1,$ by
comparing the coefficients of any given term appearing on both sides of the
equation, e.g. $\partial_{x_{1}}A_{1}\partial_{p_{1}}A_{2}\cdots
\partial_{x_{N}}A_{2N-1}\partial_{p_{N}}A_{2N}$.

For similar relations to hold for sub-maximal even rank Nambu brackets, these
must first be defined. \ It is easiest to just \emph{define} sub-maximal even
rank CNBs by their Poisson bracket resolutions as in (\ref{2NBracketAsPBs}%
)\footnote{This definition is consistent with the classical limits of quantum
$\star$-brackets presented and discussed in \cite{CurtrightZachos}, from which
the same Poisson bracket resolutions follow as a consequence of taking the
classical limit of $\star$-commutator resolutions of even $\star$-brackets.
\ It is also consistent with taking symplectic traces of maximal CNBs, again
as presented in \cite{CurtrightZachos} (see (\ref{SymplecticTrace}) to
follow).}:%
\begin{equation}
\left\{  A_{1},A_{2},\cdots,A_{2n-1},A_{2n}\right\}  _{\text{NB}}%
=\sum_{\substack{\left(  2n\right)  !\text{\ perms }\sigma}}\frac
{\operatorname{sgn}\left(  \sigma\right)  }{2^{n}n!}\,\left\{  A_{\sigma_{1}%
},A_{\sigma_{2}}\right\}  _{\text{PB}}\left\{  A_{\sigma_{3}},A_{\sigma_{4}%
}\right\}  _{\text{PB}}\cdots\left\{  A_{\sigma_{2n-1}},A_{\sigma n}\right\}
_{\text{PB}}\;, \label{2nBracketAsPBs}%
\end{equation}
only here we allow $n<N$. \ So defined, these sub-maximal CNBs enter in
further recursive expressions. \ For example, for systems with three or more
degrees of freedom, $\left\{  A,B\right\}  _{\text{PB}}=\frac{\partial\left(
A,B\right)  }{\partial\left(  x_{1},p_{1}\right)  }+\frac{\partial\left(
A,B\right)  }{\partial\left(  x_{2},p_{2}\right)  }+\frac{\partial\left(
A,B\right)  }{\partial\left(  x_{3},p_{3}\right)  }+\cdots$, and a general
6-bracket expression resolves as
\begin{align}
\left\{  A_{1},A_{2},A_{3},A_{4},A_{5},A_{6}\right\}  _{\text{NB}}  &
=\left\{  A_{1},A_{2}\right\}  _{\text{PB}}\left\{  A_{3},A_{4},A_{5}%
,A_{6}\right\}  _{\text{NB}}-\left\{  A_{1},A_{3}\right\}  _{\text{PB}%
}\left\{  A_{2},A_{4},A_{5},A_{6}\right\}  _{\text{NB}}\nonumber\\
&  +\left\{  A_{1},A_{4}\right\}  _{\text{PB}}\left\{  A_{2},A_{3},A_{5}%
,A_{6}\right\}  _{\text{NB}}-\left\{  A_{1},A_{5}\right\}  _{\text{PB}%
}\left\{  A_{2},A_{3},A_{4},A_{6}\right\}  _{\text{NB}}\nonumber\\
&  +\left\{  A_{1},A_{6}\right\}  _{\text{PB}}\left\{  A_{2},A_{3},A_{4}%
,A_{5}\right\}  _{\text{NB}}\;,
\end{align}
with the 4-brackets resolvable into PBs as in (\ref{4BracketAsPBs}). \ This
permits the building-up of higher even rank brackets proceeding from initial
PBs involving all degrees of freedom. \ The general recursion relation with
this $2n=2+\left(  2n-2\right)  $ form is%
\begin{align}
\left\{  A_{1},A_{2},\cdots,A_{2n-1},A_{2n}\right\}  _{\text{NB}}  &
=\left\{  A_{1},A_{2}\right\}  _{\text{PB}}\left\{  A_{3},\cdots
,A_{2n}\right\}  _{\text{NB}}\nonumber\\
&  +\sum_{j=3}^{2n-1}\left(  -1\right)  ^{j}\left\{  A_{1},A_{j}\right\}
_{\text{PB}}\left\{  A_{2},\cdots,A_{j-1},A_{j+1},\cdots,A_{2n}\right\}
_{\text{NB}}\nonumber\\
&  +\left\{  A_{1},A_{2n}\right\}  _{\text{PB}}\left\{  A_{2},\cdots
,A_{2n-1}\right\}  _{\text{NB}}\;, \label{2n=2+(2n-2)}%
\end{align}
and features $2n-1$ terms on the RHS. \ This recursive result is equivalent to
taking (\ref{2nBracketAsPBs}) as a definition for $2n<2N$ elements, as can be
seen by substituting the PB resolutions of the $\left(  2n-2\right)
$-brackets on the RHS of (\ref{2n=2+(2n-2)}). \ Similar relations obtain\ when
the $2n$ elements in the CNB are partitioned into sets of $\left(
2n-2k\right)  $ and $2k$ elements, with suitable antisymmetrization with
respect to exchanges between the two sets.

These results may be extended beyond maximal CNBs to super-maximal brackets,
in a useful way. \ All such super-maximal classical brackets vanish, for the
simple reason that there are not enough independent partial derivatives to
avoid repeating columns of the implicit matrix whose determinant is under
consideration. \ Another way to say this is as the impossibility to
antisymmetrize more than $2N$ coordinate and momentum indices in
$2N$-dimensional phase space, so for any phase-space function $V$, we have
$\ \epsilon^{\lbrack j_{1}j_{2}\cdots j_{2N}}\partial^{i]}V\equiv0\;,$ with
$\partial^{i}=\partial/\partial x^{i},\;\partial^{1+i}=\partial/\partial
p^{i},\;1\leq i$ (odd) $\leq2N-1.$ \ Consequently, $\partial^{j_{1}}%
A_{1}\cdots\partial^{j_{2N}}A_{2N}\,\epsilon^{\lbrack j_{1}j_{2}\cdots j_{2N}%
}\partial^{i]}V=0$, for any $2N$ phase-space functions $A_{j}$, $j=1,\cdots
,2N,$ and any $V$, a result that may be thought of as the vanishing of the
$\left(  2N+1\right)  $-th super-maximal CNB. \ As a further consequence, we
have on $2N$-dimensional phase spaces other super-maximal identities of the
form
\begin{gather}
\{B_{1},\cdots,B_{k},V\}_{\text{NB}}\{A_{1},A_{2},\cdots,A_{2N}\}_{\text{NB}%
}=\{B_{1},\cdots,B_{k},A_{1}\}_{\text{NB}}\{V,A_{2},\cdots,A_{2N}%
\}_{\text{NB}}\label{preModFI}\\
+\{B_{1},\cdots,B_{k},A_{2}\}_{\text{NB}}\{A_{1},V,A_{3},\cdots,A_{2N}%
\}_{\text{NB}}+\cdots+\{B_{1},\cdots,B_{k},A_{2N}\}_{\text{NB}}\{A_{1}%
,A_{2},\cdots,A_{2N-1},V\}_{\text{NB}}\;,\nonumber
\end{gather}
for any choice of $V$, $k$, $A$s, and $B$s. \ We have distinguished here a
$\left(  2N+1\right)  $-th phase-space function as $V$ in anticipation of
using the result later (cf. the discussion of the modified fundamental
identity, (\ref{ModFI}) et seq.). \ The expansions in (\ref{2nBracketAsPBs})
and (\ref{2n=2+(2n-2)}) also apply to the super-maximal case as well, where
they provide vanishing theorems for the sums on the RHSs of those relations. \ 

\paragraph{\underline{Reductions for classical Lie symmetries}}

When the phase-space functions involved in a classical bracket obey a Poisson
bracket algebra (possibly even an infinite one), the NB reduces to become a
sum of products, each product involving half as many phase-space functions
(\emph{reductio ad dimidium}). \ It follows as an elementary consequence of
the PB resolution of even CNBs. For any PB Lie algebra given by%
\begin{equation}
\left\{  B_{i},B_{j}\right\}  _{\text{PB}}=\sum_{m}c_{ij}^{\;\;m}B_{m}\;,
\end{equation}
the PB resolution then gives (sum over all repeated $m$s\ is to be understood)%
\begin{gather}
\left\{  B_{1},\cdots,B_{2k+1},A\right\}  _{\text{NB}}=\sum_{\left(
2k+1\right)  !\;\text{perms}\;\sigma}\frac{\operatorname{sgn}\left(
\sigma\right)  }{2^{k}k!}\,\left\{  B_{\sigma_{1}},B_{\sigma_{2}}\right\}
_{\text{PB}}\left\{  B_{\sigma_{3}},B_{\sigma_{4}}\right\}  _{\text{PB}}%
\cdots\left\{  B_{\sigma_{2k-1}},B_{\sigma_{2k}}\right\}  _{\text{PB}}\left\{
B_{\sigma_{2k+1}},A\right\}  _{\text{PB}}\nonumber\\
=\sum_{\left(  2k+1\right)  !\;\text{perms}\;\sigma}\frac{\operatorname{sgn}%
\left(  \sigma\right)  }{2^{k}k!}\,c_{\sigma_{1}\sigma_{2}}^{\;\;\;\;m_{1}%
}c_{\sigma_{3}\sigma_{4}}^{\;\;\;\;m_{2}}\cdots c_{\sigma_{2k-1}\sigma_{2k}%
}^{\;\;\;\;\;\;\;\;m_{k}}\;B_{m_{1}}B_{m_{2}}\cdots B_{m_{k}}\left\{
B_{\sigma_{2k+1}},A\right\}  _{\text{PB}}\;,
\end{gather}
where $A$ is arbitrary. \ Of course, if $A$ is also an element of the Lie
algebra, then the last PB also reduces.

\paragraph{\underline{Traces}}

Define the \emph{symplectic trace} of a classical bracket as
\begin{equation}
\sum_{i}\left\{  x_{i},p_{i},A_{1},\cdots,A_{2k}\right\}  _{\text{NB}}=\left(
N-k\right)  \left\{  A_{1},\cdots,A_{2k}\right\}  _{\text{NB}}\;.
\label{SymplecticTrace}%
\end{equation}
A complete reduction of maximal CNBs to PBs follows by inserting $N-1$
conjugate pairs of phase-space coordinates and summing over them.
\begin{equation}
\left\{  A,B\right\}  _{\text{PB}}=\frac{1}{\left(  N-1\right)  !}\left\{
A,B,x_{i_{1}},p_{i_{1}},\cdots,x_{i_{N-1}},p_{i_{N-1}}\right\}  _{\text{NB}%
}\;,
\end{equation}
where summation over all pairs of repeated indices is understood. \ Fewer
traces lead to relations between CNBs of maximal rank, $2N$, and those of
lesser rank, $2k$.
\begin{equation}
\left\{  A_{1},\cdots,A_{2k}\right\}  _{\text{NB}}=\frac{1}{\left(
N-k\right)  !}\left\{  A_{1},\cdots,A_{2k},x_{i_{1}},p_{i_{1}},\cdots
,x_{i_{N-k}},p_{i_{N-k}}\right\}  _{\text{NB}}\;. \label{MultiTrace}%
\end{equation}
This is consistent with the PB resolutions (\ref{2nBracketAsPBs}) used to
define the lower rank CNBs previously, and provides another practical
evaluation tool for these CNBs.

Through the use of such symplectic traces, Hamilton's equations for a general
system---not necessarily superintegrable---admit an NB expression different
from Nambu's original one, namely
\begin{equation}
\frac{dA}{dt}=\{A,H\}_{\text{PB}}={\frac{1}{(N-1)!}}\{A,H,x_{i_{1}},p_{i_{1}%
},...,x_{i_{N-1}},p_{i_{N-1}}\}_{\text{NB}}\;,
\end{equation}
where $H$ is the system Hamiltonian.

\paragraph{\underline{Derivations and the classical \textquotedblleft
Fundamental Identity\textquotedblright}}

CNBs are all derivations with respect to each of their arguments \cite{Nambu}.
\ For even brackets, this follows from (\ref{CNBDefn}) for maximal CNBs, and
from (\ref{2nBracketAsPBs}) (or (\ref{MultiTrace})) for sub-maximal brackets.
\begin{equation}
\delta_{\mathbf{B}}A=\left\{  A,B_{1},B_{2},\cdots,B_{2n-1}\right\}
_{\text{NB}}\;,
\end{equation}
where $\mathbf{B}$ is a short-hand for the string $B_{1},B_{2},\cdots
,B_{2n-1}$. \ By derivation, we mean that Leibniz's elementary rule is
satisfied,
\begin{equation}
\delta_{\mathbf{B}}(A\mathcal{A})=\left(  \delta_{\mathbf{B}}A\right)
\mathcal{A}+A\left(  \delta_{\mathbf{B}}\mathcal{A}\right)  =\left\{
A,B_{1},\cdots,B_{2n-1}\right\}  _{\text{NB}}\mathcal{A}+A\left\{
\mathcal{A},B_{1},\cdots,B_{2n-1}\right\}  _{\text{NB}}\;.
\end{equation}
Moreover, when these derivations act on other \emph{maximal} CNBs, they yield
simple bracket identities \cite{Nambu,Filippov,Sahoo},
\begin{equation}
\delta_{\mathbf{B}}\{C_{1},\cdots,C_{2N}\}_{\text{NB}}=\{\delta_{\mathbf{B}%
}C_{1},\cdots,C_{2N}\}_{\text{NB}}+\cdots+\{C_{1},\cdots,\delta_{\mathbf{B}%
}C_{2N}\}_{\text{NB}}\;,
\end{equation}
alternatively
\begin{align}
\{\{C_{1},\cdots,C_{2N}\}_{\text{NB}},B_{1},\cdots,B_{2n-1}\}_{\text{NB}}  &
=\{\{C_{1},B_{1},\cdots,B_{2n-1}\}_{\text{NB}},\cdots,C_{2N}\}_{\text{NB}%
}\nonumber\\
&  +\cdots+\{C_{1},\cdots,\{C_{2N},B_{1},\cdots,B_{2n-1}\}_{\text{NB}%
}\}_{\text{NB}}\;.
\end{align}
In particular, any maximal CNB acting on any other maximal CNB always obeys
the $(4N-1)$-element, $(2N+1)$-term identity \cite{Filippov,Sahoo}.%
\begin{equation}
0=\left\{  \left\{  A_{1},A_{2},\cdots,A_{2N}\right\}  _{\text{NB}}%
,B_{1},\cdots,B_{2N-1}\right\}  _{\text{NB}}-\sum_{j=1}^{2N}\left\{
A_{1},\cdots,\left\{  A_{j},B_{1},\cdots,B_{2N-1}\right\}  _{\text{NB}}%
,\cdots,A_{2N}\right\}  _{\text{NB}}\;. \label{simpleFI}%
\end{equation}
This has been designated \textquotedblleft the Fundamental
Identity\textquotedblright\ (FI) \cite{Takhtajan}, although its essentially
subsidiary role should be apparent in this classical context.

\paragraph{\underline{Invariant coefficients}}

The fact that all CNBs are derivations, and that all super-maximal classical
brackets vanish, leads to a slightly modified form of $4N$-element, $\left(
2N+1\right)  $-term fundamental identities, for a system in a $2N$-dimensional
phase-space \cite{CurtrightZachos}. \
\begin{gather}
\left\{  B_{1},\cdots,B_{2N-1},V\left\{  A_{1},A_{2},\cdots,A_{2N}\right\}
_{\text{NB}}\right\}  _{\text{NB}}=\left\{  V\left\{  B_{1},\cdots
,B_{2N-1},A_{1}\right\}  _{\text{NB}},A_{2},\cdots,A_{2N}\right\}
_{\text{NB}}\label{ModFI}\\
+\left\{  A_{1},V\left\{  B_{1},\cdots,B_{2N-1},A_{2}\right\}  _{\text{NB}%
},A_{3},\cdots,A_{4N-1}\right\}  _{\text{NB}}+\cdots+\left\{  A_{1}%
,\cdots,A_{2N-1},V\left\{  B_{1},\cdots,B_{2N-1},A_{2N}\right\}  _{\text{NB}%
}\right\}  _{\text{NB}}\;,\nonumber
\end{gather}
for any choice of $V$, $A$s, and $B$s. \ This identity is just the sum of the
super-maximal identity (\ref{preModFI}), for $k=2N-1$, plus $V$ times the FI
(\ref{simpleFI}) for the derivation $\left\{  B_{1},\cdots,B_{2N-1},\left\{
A_{1},A_{2},\cdots,A_{2N}\right\}  _{\text{NB}}\right\}  _{\text{NB}}$. \ 

As a consequence of this modified FI, any proportionality constant $V$
appearing in (\ref{dA/dt}), i.e.%
\begin{equation}
\frac{dA}{dt}=V\{A,I_{1},\cdots,I_{2N-1}\}_{\text{NB}}\;,
\label{ClassicalScaledTime}%
\end{equation}
has to be a time-invariant if it has no \emph{explicit} time dependence
\cite{Gonera}. \ As proof \cite{CurtrightZachos}, since the time derivation
satisfies the conditions for the above $\delta$, we have
\begin{equation}
\frac{d}{dt}(V\{A_{1},\cdots,A_{2N}\}_{\text{NB}})=\dot{V}\{A_{1}%
,\cdots,A_{2N}\}_{\text{NB}}+V\{\dot{A}_{1},\cdots,A_{2N}\}_{\text{NB}}%
+\cdots+V\{A_{1},\cdots,\dot{A}_{2N}\}_{\text{NB}}\;.
\end{equation}
Consistency with (\ref{ClassicalScaledTime}) requires this to be the same as
\begin{align}
V\{V\{A_{1},\cdots,A_{2N}\}_{\text{NB}},I_{1},\cdots,I_{2N-1}\}_{\text{NB}}
&  =\dot{V}\{A_{1},\cdots,A_{2N}\}_{\text{NB}}\\
+V\{V\{A_{1},I_{1},\cdots,I_{2N-1}\}_{\text{NB}},\cdots,A_{2N}\}_{\text{NB}}
&  +\cdots+V\{A_{1},\cdots,V\{A_{2N},I_{1},\cdots,I_{2N-1}\}_{\text{NB}%
}\}_{\text{NB}}\;.\nonumber
\end{align}
By substitution of (\ref{ModFI}) with $B_{j}\equiv I_{j}$, $\dot{V}=0$ follows.

\subsection{Illustrative classical examples}

It is useful to consider explicit examples of classical dynamical systems
described by Nambu brackets, to gain insight and develop intuition concerning
CNBs. \ Previous classical examples were given by Nambu \cite{Nambu},\ and
more recently, by Chatterjee \cite{Chatterjee}, and by Gonera and Nutku
\cite{Gonera,Nutku}. \ We offer an eclectic selection based on those in
\cite{CurtrightZachos}.

\paragraph{\underline{SO(3) as a special case}}

For example, consider a particle constrained to the surface of a unit radius
two-sphere $S^{2}$, but otherwise moving freely. \ Three independent
invariants of this maximally superintegrable system are the angular momenta
about the center of the sphere: \ $L_{x},\;L_{y},\;L_{z}$. \ Actually, no two
of these are in involution, but this is quickly remedied, and moreover it is
not a hindrance since in the Nambu approach to mechanics all invariants are on
a more equal footing.

To be more explicit, we may coordinate the upper and lower ($\pm$) hemispheres
by projecting the particle's location onto the equatorial disk, $\left\{
\left(  x,y\right)  \;|\;x^{2}+y^{2}\leq1\right\}  $. \ The invariants are
then%
\begin{equation}
L_{z}=xp_{y}-yp_{x}\;,\;\;\;L_{y}=\pm\sqrt{1-x^{2}-y^{2}}\;p_{x}%
\;,\;\;\;L_{x}=\mp\sqrt{1-x^{2}-y^{2}}\;p_{y}\;.
\end{equation}
The last two are the de Sitter momenta, or nonlinearly realized axial charges
corresponding to the ``pions'' $x,y$ of this truncated $\sigma$-model.

The Poisson brackets of these expressions close into the expected $SO(3)$
algebra,
\begin{equation}
\{L_{x},L_{y}\}_{\text{PB}}=L_{z}\;,\;\;\;\;\;\{L_{y},L_{z}\}_{\text{PB}%
}=L_{x}\;,\;\;\;\;\;\{L_{z},L_{x}\}_{\text{PB}}=L_{y}\;.
\end{equation}
The usual Hamiltonian of the free particle system is the Casimir invariant
\cite{CurtrightZachos}%
\begin{equation}
H=\frac{1}{2}\left(  L_{x}L_{x}+L_{y}L_{y}+L_{z}L_{z}\right)  =\frac{1}%
{2}\left(  1-x^{2}\right)  p_{x}^{2}+\frac{1}{2}\left(  1-y^{2}\right)
p_{y}^{2}-xyp_{x}p_{y}\;.
\end{equation}
Thus, it immediately follows algebraically that PBs of $H$ with the
$\mathbf{L}$ vanish, and their time-invariance holds.
\begin{equation}
\frac{d}{dt}\mathbf{L}=\{\mathbf{L},H\}_{\text{PB}}=0\;.
\end{equation}
So any one of the $L$s and this Casimir constitute a pair of invariants in involution.

The corresponding $so\left(  3\right)  $ CNB dynamical evolution, found in
\cite{CurtrightZachos}, is untypically concise.
\begin{equation}
\frac{dA}{dt}=\{A,H\}_{\text{PB}}=\left\{  A,L_{x},L_{y},L_{z}\right\}
_{\text{NB}}=\frac{\partial(A,L_{x},L_{y},L_{z})}{\partial(x,p_{x},y,p_{y}%
)}\;. \label{so(3)ClassicalBeauty}%
\end{equation}
The simplicity of this result actually extends to more general contexts, upon
use of suitable linear combinations. Special sums of such 4-brackets can be
used to express time evolution for any classical system with a continuous
symmetry algebra underlying the dynamics and whose Hamiltonian is just the
quadratic Casimir of that symmetry algebra. \ The system need not be
superintegrable, or even integrable, in general.

Any simple Lie algebra allows a PB with a quadratic Casimir to be rewritten as
a \emph{sum} of 4-brackets. \ Suppose
\begin{equation}
\left\{  Q_{a},Q_{b}\right\}  _{\text{PB}}=f_{abc}Q_{c}\;,
\end{equation}
in a basis where $f_{abc}$ is totally antisymmetric. \ Then, for the following
linear combination of Nambu 4-brackets weighted by the structure constants,
use the PB resolution of the 4-bracket (\ref{4BracketAsPBs}) to obtain (sum
over repeated indices)%
\begin{equation}
f_{abc}\left\{  A,Q_{a},Q_{b},Q_{c}\right\}  _{\text{NB}}=3f_{abc}\left\{
A,Q_{a}\right\}  _{\text{PB}}\left\{  Q_{b},Q_{c}\right\}  _{\text{PB}%
}=3f_{abc}f_{bcd}\left\{  A,Q_{a}\right\}  _{\text{PB}}Q_{d}\;.
\end{equation}
Now, for simple Lie algebras (with appropriately normalized charges) one has
\begin{equation}
f_{abc}f_{bcd}=c_{\text{adjoint}}\,\delta_{ad}\;, \label{CAdjoint}%
\end{equation}
where $c_{\text{adjoint}}$ is a \emph{number} (For example, $c_{\text{adjoint}%
}=N$ for $su\left(  N\right)  $). Thus the classical 4-bracket reduces to a PB
with the Casimir $Q_{a}Q_{a}$,
\begin{equation}
f_{abc}\left\{  A,Q_{a},Q_{b},Q_{c}\right\}  _{\text{NB}}=3\,c_{\text{adjoint}%
}\,\left\{  A,Q_{a}\right\}  _{\text{PB}}Q_{a}=\frac{3}{2}\,c_{\text{adjoint}%
}\,\left\{  A,Q_{a}Q_{a}\right\}  _{\text{PB}}\;, \label{AnyLie4CNB}%
\end{equation}
For $su\left(  2\right)  =so\left(  3\right)  $, $c_{\text{adjoint}}=2$,
$f_{abc}\left\{  A,Q_{a},Q_{b},Q_{c}\right\}  _{\text{NB}}=6\left\{
A,L_{x},L_{y},L_{z}\right\}  _{\text{NB}}$, and we establish
(\ref{so(3)ClassicalBeauty}) above.

\paragraph{\underline{U(n) and isotropic oscillators}}

If we realize the $U\left(  n\right)  $ algebra in the oscillator basis, where
the phase-space \textquotedblleft charges\textquotedblright\ $N_{jk}=\left(
x_{j}-ip_{j}\right)  \left(  x_{k}+ip_{k}\right)  /2$ obey the PB relations
\begin{equation}
\left\{  N_{jk},N_{lm}\right\}  _{\text{PB}}=-i\left(  N_{jm}\delta
_{kl}-N_{lk}\delta_{jm}\right)  \;,\;\;\;j,k,l,m=1,\cdots,n\;,
\label{ClassicalU(n)Alg}%
\end{equation}
then the isotropic Hamiltonian is
\begin{equation}
H=\omega\sum_{i=1}^{n}N_{i}\;,\;\;\;N_{i}\equiv N_{ii}\;.
\end{equation}
This gives the $n^{2}$ conservation laws
\begin{equation}
\left\{  H,N_{ij}\right\}  _{\text{PB}}=0\;.
\end{equation}
However, only $2n-1$ of the $N_{ij}$ are functionally independent for a
classical system with a $2n$-dimensional phase-space. \ This follows because
all full phase-space Jacobians (i.e., maximal CNBs) involving $2n$ of the
$N_{ij}$ vanish. \ (For details, see the upcoming discussion surrounding
(\ref{ClassicalU(n)Perm}).)

Following the logic that led to the previous reductio ad dimidium for general
Lie symmetries, we obtain the main result for classical isotropic oscillator
$2n$-brackets.

\paragraph{\underline{Classical isotropic oscillator brackets}}

(the $U\left(  n\right)  $ reductio ad dimidium): \ Let $N=N_{1}+N_{2}%
+\cdots+N_{n}$, and intercalate the $n-1$ non-diagonal charges $N_{i\;i+1}$,
for $i=1,\cdots,n-1$, into a classical Nambu $2n$-bracket with the $n$
mutually involutive $N_{j}$, for $j=1,\cdots,n$, to find\footnote{The
non-diagonal charges are not real,\ but neither does this present a real
problem. \ The proof leading to (\ref{classicalTLCresult}) also goes through
if non-diagonal charges have their subscripts transposed. \ This allows
replacing $N_{i\;i+1}$ with real or purely imaginary combinations
$N_{i\;i+1}\pm N_{i+1\;i}$ in the LHS $2n$-bracket, to obtain the alternative
linear combinations $N_{i\;i+1}\mp N_{i+1\;i}$ in the product on the RHS.}
\begin{align}
\left\{  A,N_{1},N_{12},N_{2},N_{23},\cdots,N_{n-1},N_{n-1\;n},N_{n}\right\}
_{\text{NB}}  &  =\left(  -i\right)  ^{n-1}\left\{  A,N\right\}  _{\text{PB}%
}N_{12}N_{23}\cdots N_{n-1\;n}\nonumber\\
&  =\left(  -i\right)  ^{n-1}\left\{  AN_{12}N_{23}\cdots N_{n-1\;n}%
,N\right\}  _{\text{PB}}\;. \label{classicalTLCresult}%
\end{align}
This result follows from the $U\left(  n\right)  $ PB algebra of the charges,
(\ref{ClassicalU(n)Alg}). \ When the algebra is realized specifically by
harmonic oscillators, the RHS factor may also be written as $N_{12}%
N_{23}\cdots N_{n-1\;n}=\left(  N_{2}N_{3}\cdots N_{n-1}\right)  N_{1\;n}$.

\noindent\textbf{Proof:} \ Linearity in each argument and total antisymmetry
of the CNB allow us to replace any one of the $N_{i}$ by the sum $N$. Replace
$N_{n}\rightarrow N$, to obtain
\begin{equation}
\left\{  A,N_{1},N_{12},N_{2},\cdots,N_{n-1},N_{n-1\;n},N_{n}\right\}
_{\text{NB}}=\left\{  A,N_{1},N_{12},N_{2},\cdots,N_{n-1},N_{n-1\;n}%
,N\right\}  _{\text{NB}}\;.
\end{equation}
Now since $\left\{  N,N_{ij}\right\}  _{\text{PB}}=0$, the PB resolution of
the $2n$-bracket implies that $N$ must appear \textquotedblleft
locked\textquotedblright\ in a PB with $A$, and therefore $A$ cannot appear in
any other PB. \ But then $N_{1}$ is in involution with all the remaining free
$N_{ij}$ except $N_{12}$. \ So $N_{1}$ must be locked in $\left\{
N_{1},N_{12}\right\}  _{\text{PB}}$. \ Continuing in this way, $N_{2}$ must be
locked in $\left\{  N_{2},N_{23}\right\}  _{\text{PB}}$, etc., until finally
$N_{n-1}$ is locked in $\left\{  N_{n-1},N_{n-1\;n}\right\}  _{\text{PB}}$.
\ Thus, all $2n$ entries have been paired and locked in the indicated $n$ PBs,
i.e. they are all \textquotedblleft zipped-up\textquotedblright.
Consequently,
\begin{equation}
\left\{  A,N_{1},N_{12},N_{2},\cdots,N_{n-1},N_{n-1\;n},N_{n}\right\}
_{\text{NB}}=\left\{  A,N\right\}  _{\text{PB}}\left\{  N_{1},N_{12}\right\}
_{\text{PB}}\cdots\left\{  N_{n-1},N_{n-1\;n}\right\}  _{\text{PB}}\;.
\end{equation}
All the paired $N_{jk}$ Poisson brackets evaluate as $\left\{  N_{j-1}%
,N_{j-1\;j}\right\}  _{\text{PB}}=-iN_{j-1\;j}$, so
\begin{equation}
\left\{  A,N_{1},N_{12},N_{2},\cdots,N_{n-1},N_{n-1\;n},N_{n}\right\}
_{\text{NB}}=\left(  -i\right)  ^{n-1}\left\{  A,N\right\}  _{\text{PB}}%
N_{12}\cdots N_{n-1\;n}\;.
\end{equation}
Finally, the PB with $N$ may be performed either before or after the product
of $A$ with all the $N_{j-1\;j}$, since again $\left\{  N,N_{ij}\right\}
_{\text{PB}}=0$, and the PB is a derivation. \ Hence
\begin{equation}
\left\{  A,N\right\}  _{\text{PB}}N_{12}\cdots N_{n-1\;n}=\left\{
AN_{12}\cdots N_{n-1\;n},N\right\}  _{\text{PB}}\;.\;\;\;\text{{\small QED}}%
\end{equation}

Remarkably, in (\ref{classicalTLCresult}), the invariants which are in
involution (i.e. the Cartan subalgebra of $u\left(  n\right)  $) are separated
out of the CNB into a single PB involving their sum (the Hamiltonian,
$H=\omega N$), while the invariants which are not in involution ($n-1$ of
them, corresponding in number to the rank of $SU\left(  n\right)  $) are
effectively swept into a simple product. \ Time evolution for the isotropic
oscillator is then given by \cite{CurtrightZachos}%
\begin{equation}
\left(  -i\right)  ^{n-1}N_{12}\cdots N_{n-1\;n}\frac{dA}{dt}=\omega\left\{
A,N_{1},N_{12},N_{2},\cdots,N_{n-1},N_{n-1\;n},N_{n}\right\}  _{\text{NB}}\;.
\label{u(n)CNBdA/dt}%
\end{equation}
This result reveals a possible degenerate situation for the Nambu approach. \ 

When any two or more of the phase-space gradients entering into the bracket
are parallel, or when one or more of them vanish, the corresponding bracket
also vanishes, even if $\frac{dA}{dt}\neq0$. \ Under these conditions, the
bracket does not give any temporal change of $A$: \ Such changes are ``lost''
by the bracket. \ This can occur for the $u\left(  n\right)  $ bracket under
consideration whenever $0=N_{12}\cdots N_{n-1\;n}$, i.e. whenever any
$N_{i-1\;i}=0$ for some $i$. \ Initial classical configurations for which this
is the case are not evolved by this particular bracket. \ This is not really a
serious problem, since on the one hand, the configurations for which it
happens are so easily catalogued, and on the other hand, there are other
choices for the bracket entries which can be used to recover the lost temporal
changes. \ It is just necessary to be aware of any such ``kernel'' when using
any given bracket.

With that caveat in mind, there is another way to write (\ref{u(n)CNBdA/dt})
since the classical bracket is a derivation of each of its entries. \ Namely,%
\begin{equation}
\frac{dA}{dt}=i^{n-1}\omega\left\{  A,N_{1},\ln\left(  N_{12}\right)
,N_{2},\ln\left(  N_{23}\right)  ,N_{3},\cdots,N_{n-1},\ln\left(
N_{n-1\;n}\right)  ,N_{n}\right\}  _{\text{NB}}\;. \label{TakingLogs}%
\end{equation}
The logarithms intercalated between the diagonal $N_{j}$s on the RHS now have
branch points corresponding to the classical bracket's kernel.

The selection of $2n-1$ invariants to be used in the maximal $U\left(
n\right)  $ bracket is not unique, of course. \ In the list that we have
selected, the indices, $1,2,\cdots,n,$ can be replaced by any permutation,
$\sigma_{1},\sigma_{2},\cdots,\sigma_{n}$, so long as the correlations between
indices for elements in the list are maintained. \ That is, we may replace the
elements $N_{1},N_{12},N_{2},N_{23},\cdots,N_{n-1},N_{n-1\;n},N_{n}$ by
$N_{\sigma_{1}},N_{\sigma_{1}\sigma_{2}},N_{\sigma_{2}},N_{\sigma_{2}%
\sigma_{3}},\cdots,N_{\sigma_{n-1}},N_{\sigma_{n-1}\;\sigma_{n}},N_{\sigma
_{n}}$, and the reductio ad dimidium still holds.%
\begin{equation}
\left\{  A,N_{\sigma_{1}},N_{\sigma_{1}\sigma_{2}},N_{\sigma_{2}}%
,N_{\sigma_{2}\sigma_{3}},\cdots,N_{\sigma_{n-1}},N_{\sigma_{n-1}\;\sigma_{n}%
},N_{\sigma_{n}}\right\}  _{\text{NB}}=\left(  -i\right)  ^{n-1}\left\{
A,N\right\}  _{\text{PB}}N_{\sigma_{1}\sigma_{2}}N_{\sigma_{2}\sigma_{3}%
}\cdots N_{\sigma_{n-1}\;\sigma_{n}}\;. \label{ClassicalU(n)Perm}%
\end{equation}
Whatever list is selected, any invariant in that list is manifestly conserved
by the $2n$-bracket. \ All other $U\left(  n\right)  $ charges are also
conserved by the bracket, even though they are not among the selected list of
invariants. \ This last statement follows immediately from the $\left\{
A,H\right\}  _{\text{PB}}$ factor on the RHS of (\ref{ClassicalU(n)Perm}).

\paragraph{\underline{SO(n+1) and free particles on n-spheres}}

For a particle moving freely on the surface of an $n$-sphere, $S^{n}$, one now
has a choice of $2n-1$ of the $n(n+1)/2$ invariant charges of $so(n+1)$, whose
PB Lie algebra is conveniently written in terms of the $n(n-1)/2$ rotation
generators, $L_{ab}=x^{a}p_{b}-x^{b}p_{a}$, for $a,b=1,\cdots,n$, and in terms
of the de Sitter momenta, $P_{a}=\sqrt{1-q^{2}}~p_{a}$, for $a=1,\cdots,n$,
where $q^{2}=\sum_{a=1}^{n}\left(  x^{a}\right)  ^{2}$. That PB algebra is%
\begin{equation}
\left\{  P_{a},P_{b}\right\}  _{\text{PB}}=L_{ab}\;,\;\;\;\left\{
L_{ab},P_{c}\right\}  _{\text{PB}}=\delta_{ac}P_{b}-\delta_{bc}P_{a}%
\;,\;\;\;\left\{  L_{ab},L_{cd}\right\}  _{\text{PB}}=L_{ac}\delta_{bd}%
-L_{ad}\delta_{bc}-L_{bc}\delta_{ad}+L_{bd}\delta_{ac}\;.
\end{equation}
By direct calculation, one of several possible expressions for time-evolution
as a $2n$-bracket is \cite{CurtrightZachos}
\begin{equation}
\left(  -1\right)  ^{n-1}P_{2}P_{3}\cdots P_{n-1}\frac{dA}{dt}=\frac
{\partial\left(  A,P_{1},L_{12},P_{2},L_{23},P_{3},\cdots,P_{n-1}%
,L_{n-1\;n},P_{n}\right)  }{\partial\left(  x_{1},p_{1},x_{2},p_{2}%
,\cdots,x_{n},p_{n}\right)  }\;, \label{ClassicalSphere}%
\end{equation}
where $\frac{dA}{dt}=\left\{  A,H\right\}  _{\text{PB}}$ and%
\begin{equation}
H=\frac{1}{2}\sum_{a=1}^{n}P_{a}P_{a}+\frac{1}{4}\sum_{a,b=1}^{n}L_{ab}%
L_{ab}\;. \label{HforNsphere}%
\end{equation}
The CNB expressing classical time-evolution may also be written, more
compactly, as a derivation.%
\begin{equation}
\frac{dA}{dt}=\left(  -1\right)  ^{n-1}\left\{  A,P_{1},L_{12},\ln\left(
P_{2}\right)  ,L_{23},\ln\left(  P_{3}\right)  ,\cdots,\ln\left(
P_{n-1}\right)  ,L_{n-1\;n},P_{n}\right\}  _{\text{NB}}\;.
\label{ClassicalSphereLogs}%
\end{equation}
Once again, the branch points in the intercalated logarithms are indicators of
this particular bracket's kernel.

\paragraph{\underline{SO(4)=SU(2)$\times$SU(2) as another special case}}

The treatment of the 3-sphere $S^{3}$ also accords to the standard chiral
model technology using left- and right-invariant Vielbeine
\cite{CurtrightZachos}. \ Specifically, the two choices for such Dreibeine for
the 3-sphere are \cite{CurtrightUematsuZachos}: $\ q^{2}=x^{2}+y^{2}+z^{2}$%
\begin{equation}
^{(\pm)}V_{a}^{i}=\epsilon^{iab}x^{b}\pm\sqrt{1-q^{2}}~g_{ai}~,\qquad
\qquad^{(\pm)}V^{ai}=\epsilon^{iab}x^{b}\pm\sqrt{1-q^{2}}~\delta^{ai}\;.
\end{equation}
The corresponding right and left conserved charges (left- and right-invariant,
respectively) then are
\begin{equation}
\mathcal{R}^{i}=~^{(+)}V_{a}^{i}~\frac{d}{dt}x^{a}=~^{(+)}V^{ai}p_{a}%
~,\qquad\qquad\mathcal{L}^{i}=~^{(-)}V_{a}^{i}~\frac{d}{dt}x^{a}=~^{(-)}%
V^{ai}p_{a}~.
\end{equation}
Perhaps more intuitive are the linear combinations into Axial and Isospin
charges (again linear in the momenta),
\begin{equation}
\tfrac{1}{2}\left(  \mathcal{R}\mathbf{-}\mathcal{L}\right)  =\sqrt{1-q^{2}%
}~\mathbf{p}\equiv\mathbf{A\;},\qquad\tfrac{1}{2}\left(  \mathcal{R}%
\mathbf{+}\mathcal{L}\right)  =\mathbf{x}\times\mathbf{p}\equiv\mathbf{I\;}.
\label{Axial&Isospin}%
\end{equation}
It can easily be seen that the $\mathcal{L}$s and the $\mathcal{R}$s have PBs
closing into the standard $SU(2)\times SU(2)$ algebra, i.e.%
\begin{equation}
\left\{  \mathcal{L}_{i},\mathcal{L}_{j}\right\}  _{\text{NB}}=-2\varepsilon
_{ijk}\mathcal{L}_{k}\;,\;\;\;\left\{  \mathcal{R}_{i},\mathcal{R}%
_{j}\right\}  _{\text{NB}}=-2\varepsilon_{ijk}\mathcal{R}_{k}\;,\;\;\;\left\{
\mathcal{L}_{i},\mathcal{R}_{j}\right\}  _{\text{NB}}=0\;.
\label{Axial&IsospinPBs}%
\end{equation}
Thus they are seen to be constant, since the Hamiltonian (and also the
Lagrangian) can, in fact, be written in terms of either quadratic Casimir
invariant,
\begin{equation}
H=\tfrac{1}{2}\mathcal{L}\cdot\mathcal{L}=\tfrac{1}{2}\mathcal{R}%
\cdot\mathcal{R}\;.
\end{equation}

The classical dynamics of this algebraic system is, like the single $SU\left(
2\right)  $ invariant dynamics that composes it, elegantly expressed on the
six dimensional phase-space with maximal CNBs. \ We find various 6-bracket
relations such as
\begin{equation}
\frac{\partial\left(  A,H,\mathcal{R}_{1},\mathcal{R}_{2},\mathcal{L}%
_{1},\mathcal{L}_{2}\right)  }{\partial\left(  x_{1},p_{1},x_{2},p_{2}%
,x_{3},p_{3}\right)  }\equiv\left\{  A,H,\mathcal{R}_{1},\mathcal{R}%
_{2},\mathcal{L}_{1},\mathcal{L}_{2}\right\}  _{\text{NB}}=-4\mathcal{L}%
_{3}\mathcal{R}_{3}\,\frac{dA}{dt}\;, \label{HinBracket}%
\end{equation}
where $2H=\mathcal{R}_{1}^{2}+\mathcal{R}_{2}^{2}+\mathcal{R}_{3}%
^{2}=\mathcal{L}_{1}^{2}+\mathcal{L}_{2}^{2}+\mathcal{L}_{3}^{2}$, and $A$ is
an arbitrary function of the phase-space dynamical variables. \ Also,
\begin{equation}
\left\{  A,\mathcal{R}_{1},\mathcal{R}_{2},\mathcal{L}_{3},\mathcal{L}%
_{1},\mathcal{L}_{2}\right\}  _{\text{NB}}=-4\mathcal{R}_{3}\,\frac{dA}{dt}\;,
\label{3SphereR3Kernel}%
\end{equation}
and similarly ($\mathcal{R}\longleftrightarrow\mathcal{L}$)%
\begin{equation}
\left\{  A,\mathcal{R}_{1},\mathcal{R}_{2},\mathcal{R}_{3},\mathcal{L}%
_{1},\mathcal{L}_{2}\right\}  _{\text{NB}}=-4\mathcal{L}_{3}\,\frac{dA}{dt}\;.
\label{3SphereL3Kernel}%
\end{equation}
The kernels of these various brackets are evident from the factors multiplying
$\frac{dA}{dt}$. \ None of these particular 6-bracket relations directly
permits the $\mathcal{L}_{3}$ or $\mathcal{R}_{3}$ factors on their RHSs to be
absorbed into logarithms, through use of the Leibniz rule. \ But, by
subtracting the last two to obtain%
\begin{equation}
\left\{  A,\mathcal{R}_{1},\mathcal{R}_{2},\mathcal{L}_{3}-\mathcal{R}%
_{3},\mathcal{L}_{1},\mathcal{L}_{2}\right\}  _{\text{NB}}=4\left(
\mathcal{L}_{3}-\mathcal{R}_{3}\right)  \,\frac{dA}{dt}\;,
\label{3SphereDifference}%
\end{equation}
we can now introduce a logarithm to produce just a numerical factor
multiplying the time derivative,
\begin{equation}
\left\{  A,\mathcal{R}_{1},\mathcal{R}_{2},\ln\left(  \mathcal{L}%
_{3}-\mathcal{R}_{3}\right)  ^{2},\mathcal{L}_{1},\mathcal{L}_{2}\right\}
_{\text{NB}}=8\,\frac{dA}{dt}\;.
\end{equation}
Similarly, by adding (\ref{3SphereR3Kernel}) and (\ref{3SphereL3Kernel}), we
find%
\begin{equation}
\left\{  A,\mathcal{R}_{1},\mathcal{R}_{2},\ln\left(  \mathcal{L}%
_{3}+\mathcal{R}_{3}\right)  ^{2},\mathcal{L}_{1},\mathcal{L}_{2}\right\}
_{\text{NB}}=-8\,\frac{dA}{dt}\;.
\end{equation}

\paragraph{\underline{G$\times$G chiral particles}}

In general, the preceding discussion also applies to all chiral models, with
the algebra $g$ for a chiral group $G$ replacing $su(2)$. \ The
Vielbein-momenta combinations $V^{aj}p_{a}$ represent algebra generator
invariants, whose quadratic Casimir group invariants yield the respective
Hamiltonians. \ 

That is to say, for \cite{BraatenCurtrightZachos} group matrices $U$ generated
by exponentiated constant group algebra matrices $T$ weighted by functions of
the particle coordinates $x$, with $U^{-1}=U^{\dagger}$, we have
\begin{equation}
iU^{-1}\frac{d}{dt}U=~^{(+)}V_{a}^{j}T_{j}\frac{d}{dt}x^{a}=~^{(+)}V^{aj}%
p_{a}T_{j}\;,\;\;\;\;\;iU\frac{d}{dt}U^{-1}=~^{(-)}V^{aj}p_{a}T_{j}\;,
\end{equation}
It follows that PBs of left- and right-invariant charges (designated by
$\mathcal{R}$s and $\mathcal{L}$s, respectively), as defined by the traces,%
\begin{equation}
\mathcal{R}_{j}\equiv\frac{i}{2}\operatorname{tr}\left(  T_{j}U^{-1}\frac
{d}{dt}U\right)  =~^{(+)}V^{aj}p_{a}\;,\;\;\;\mathcal{L}_{j}\equiv\frac{i}%
{2}\operatorname{tr}\left(  T_{j}U\frac{d}{dt}U^{-1}\right)  =~^{(-)}%
V^{aj}p_{a}\;,
\end{equation}
close to the identical PB Lie algebras,
\begin{equation}
\left\{  \mathcal{R}_{i},\mathcal{R}_{j}\right\}  _{\text{PB}}=-2f_{ijk}%
\mathcal{R}_{k}\;,\;\;\;\left\{  \mathcal{L}_{i},\mathcal{L}_{j}\right\}
_{\text{PB}}=-2f_{ijk}\mathcal{L}_{k} \label{GxGPBs}%
\end{equation}
and PB commute with each other,
\begin{equation}
\left\{  \mathcal{R}_{i},\mathcal{L}_{j}\right\}  _{\text{PB}}=0\;.
\end{equation}
These two statements are implicit in \cite{BraatenCurtrightZachos} and
throughout the literature, and are explicitly proven in \cite{CurtrightZachos}.

The Hamiltonian for a particle moving freely on the $G\times G$ group manifold
is the simple form
\begin{equation}
H=\frac{1}{2}(p_{a}V^{ai})(V^{bi}p_{b})\;,
\end{equation}
with either choice, $V^{aj}=~^{(\pm)}V^{aj}$. \ That is,%
\begin{equation}
H=\tfrac{1}{2}\mathcal{L}_{j}\mathcal{L}_{j}=\tfrac{1}{2}\mathcal{R}%
_{j}\mathcal{R}_{j}\;,
\end{equation}
just as in the previous $SO\left(  4\right)  =SU\left(  2\right)  \times
SU\left(  2\right)  $ case. \ There are now several ways to present time
evolution as CNBs for these models. \ 

One way is as sums of 6-brackets. \ Making use of (\ref{CAdjoint}) and summing
repeated indices:%
\begin{gather}
f_{ijk}f_{imn}\left\{  A,H,\mathcal{R}_{j},\mathcal{R}_{k},\mathcal{L}%
_{m},\mathcal{L}_{n}\right\}  _{\text{NB}}=f_{ijk}f_{imn}\left\{
\mathcal{R}_{j},\mathcal{R}_{k}\right\}  _{\text{PB}}\left\{  \mathcal{L}%
_{m},\mathcal{L}_{n}\right\}  _{\text{PB}}\left\{  A,H\right\}  _{\text{PB}%
}\nonumber\\
=4f_{ijk}f_{imn}f_{jkl}f_{mno}\,\mathcal{R}_{l}\mathcal{L}_{o}\left\{
A,H\right\}  _{\text{PB}}=4c_{\text{adjoint}}^{2}\mathcal{R}_{l}%
\mathcal{L}_{l}\left\{  A,H\right\}  _{\text{PB}}\;.
\end{gather}
Thus we have%
\begin{equation}
\frac{dA}{dt}=\frac{1}{4c_{\text{adjoint}}^{2}\mathcal{R}_{l}\mathcal{L}_{l}%
}\;f_{ijk}f_{imn}\left\{  A,H,\mathcal{R}_{j},\mathcal{R}_{k},\mathcal{L}%
_{m},\mathcal{L}_{n}\right\}  _{\text{NB}}\;. \label{GxG6CNB}%
\end{equation}
The bracket kernel here is given by zeroes of $\left(  \mathcal{R}_{l}%
\pm\mathcal{L}_{l}\right)  ^{2}-4H=\pm2\mathcal{R}_{l}\mathcal{L}_{l}$.

Another way to specify the time development for these chiral models is to use
a maximal set of invariants in the CNB, selected from both left and right
charges. \ Take $n$ to be the dimension of the group $G$, then all charge
indices range from $1$ to $n$. \ For a point particle moving on the group
manifold $G\times G$, the maximal bracket involves $2n$ elements. \ So, for
example, we have (note the ranges of all the sums here are truncated to $n-1$,
as are the indices on the Levi-Civita symbols)%
\begin{align}
&  \left\{  A,H,\mathcal{L}_{1},\cdots,\mathcal{L}_{n-1},\mathcal{R}%
_{1},\cdots,\mathcal{R}_{n-1}\right\}  _{\text{NB}}\nonumber\\
&  =\frac{1}{\left[  \left(  n-1\right)  !\right]  ^{2}}\,\sum_{\text{all
}i,j=1}^{n-1}\varepsilon_{i_{1}\cdots i_{n-1}}\varepsilon_{j_{1}\cdots
j_{n-1}}\left\{  A,H,\mathcal{L}_{i_{1}},\cdots,\mathcal{L}_{i_{n-1}%
},\mathcal{R}_{j_{1}},\cdots,\mathcal{R}_{j_{n-1}}\right\}  _{\text{NB}}\;.
\label{AnotherWay}%
\end{align}
The RHS here vanishes for even $n$, so we take odd $n$, say $n=1+2s$. \ (To
obtain a nontrivial result for even $n$, we may replace $H$ by either
$\mathcal{L}_{n}$ or $\mathcal{R}_{n}$. \ We leave this as an exercise in the
classical case. \ The relevant combinatorics are discussed later, in the
context of the quantized model.) \ So, since $\left\{  H,\mathcal{L}%
_{i}\right\}  _{\text{PB}}=0=\left\{  H,\mathcal{R}_{i}\right\}  _{\text{PB}}%
$, by the PB resolution we can write
\begin{align}
&  \left\{  A,H,\mathcal{L}_{1},\cdots,\mathcal{L}_{n-1},\mathcal{R}%
_{1},\cdots,\mathcal{R}_{n-1}\right\}  _{\text{NB}}\label{GxGMaximalCNB}\\
&  =K_{n}\,\sum_{\text{all }i,j=1}^{n-1}\varepsilon_{i_{1}\cdots i_{n-1}%
}\varepsilon_{j_{1}\cdots j_{n-1}}\left\{  A,H\right\}  _{\text{PB}}\left\{
\mathcal{L}_{i_{1}},\mathcal{L}_{i_{2}}\right\}  _{\text{PB}}\cdots\left\{
\mathcal{L}_{i_{n-2}},\mathcal{L}_{i_{n-1}}\right\}  _{\text{PB}}\left\{
\mathcal{R}_{j_{1}},\mathcal{R}_{j_{2}}\right\}  _{\text{PB}}\cdots\left\{
\mathcal{R}_{j_{n-2}},\mathcal{R}_{j_{n-1}}\right\}  _{\text{PB}}\;,\nonumber
\end{align}
where\footnote{The number of ways of picking the $n$ PBs in the formula
(\ref{GxGMaximalCNB}), taking into account both $\varepsilon$'s, is $\left(
n-2\right)  \left(  n-4\right)  \cdots\left(  1\right)  \times\left(
n-2\right)  \left(  n-4\right)  \cdots\left(  1\right)  $, so $K_{n}=\left(
\frac{\left(  n-2\right)  \left(  n-4\right)  \cdots\left(  1\right)
}{\left(  n-1\right)  !}\right)  ^{2}$.}
\begin{equation}
K_{n=1+2s}=\frac{1}{4^{s}\left(  s!\right)  ^{2}} \label{Kn}%
\end{equation}
is a numerical combinatoric factor incorporating the number of equivalent ways
to obtain the list of PBs in the product as written in (\ref{GxGMaximalCNB}). \ 

Introducing a completely symmetric tensor, $\sigma_{\left\{  k_{1}\cdots
k_{s}\right\}  }$, defined by%
\begin{equation}
\sigma_{\left\{  k_{1}\cdots k_{s}\right\}  }=\sum_{\text{all }i=1}%
^{n-1}\varepsilon_{i_{1}\cdots i_{n-1}}\;f_{i_{1}i_{2}k_{1}}\cdots
f_{i_{n-2}i_{n-1}k_{s}}\;, \label{SigmaTensor}%
\end{equation}
and using (\ref{GxGPBs}), we may rewrite (\ref{GxGMaximalCNB}) as (note the
sums\ over $k$s and $m$s here are not truncated)%
\begin{align}
&  \left\{  A,H,\mathcal{L}_{1},\cdots,\mathcal{L}_{n-1},\mathcal{R}%
_{1},\cdots,\mathcal{R}_{n-1}\right\}  _{\text{NB}}\nonumber\\
&  =\left(  -2\right)  ^{n-1}K_{n}\,\sum_{\text{all }k,m=1}^{n}\sigma
_{\left\{  k_{1}\cdots k_{s}\right\}  }\sigma_{\left\{  m_{1}\cdots
m_{s}\right\}  }\left\{  A,H\right\}  _{\text{PB}}\mathcal{L}_{k_{1}}%
\cdots\mathcal{L}_{k_{s}}\mathcal{R}_{m_{1}}\cdots\mathcal{R}_{m_{s}}\;.
\end{align}
Thus we arrive at a maximal CNB expression of time evolution, for
odd-dimensional $G$:%
\begin{equation}
\frac{dA}{dt}=V\left\{  A,H,\mathcal{L}_{1},\cdots,\mathcal{L}_{n-1}%
,\mathcal{R}_{1},\cdots,\mathcal{R}_{n-1}\right\}  _{\text{NB}}\;,
\label{GxGClassicalTimeEvolution}%
\end{equation}
where the invariant factor $V$ on the RHS is given by%
\begin{equation}
\frac{1}{V}=\frac{1}{\left(  s!\right)  ^{2}}\,\sum_{\text{all }k=1}^{n}%
\sigma_{\left\{  k_{1}\cdots k_{s}\right\}  }\mathcal{L}_{k_{1}}%
\cdots\mathcal{L}_{k_{s}}\sum_{\text{all }m=1}^{n}\sigma_{\left\{  m_{1}\cdots
m_{s}\right\}  }\mathcal{R}_{m_{1}}\cdots\mathcal{R}_{m_{s}}\;,\;\;\;s\equiv
\frac{n-1}{2}\;.
\end{equation}
This factor determines the kernel of the bracket in question.

All this extends in a straightforward way to even-dimensional groups $G$, and
to the algebras of symmetry groups involving arbitrary numbers of factors,
$G_{1}\times G_{2}\times\cdots$. \ 

\section{Quantum Theory}

We now consider the quantization of Nambu mechanics. \ Despite contrary claims
in the literature, it turns out that the quantization is straightforward using
the Hilbert space operator methods as originally proposed by Nambu. \ All that
is needed is a properly consistent physical interpretation of the results, by
allowing for dynamical time scales, as summarized in the Introduction. \ We
provide a very detailed description of that interpretation in the following,
but first we develop the techniques and machinery that are used to reach and
implement it. \ Our presentation parallels the previous classical discussion
as much as possible.

\subsection{Properties of the quantum brackets}

\paragraph{\underline{Definition of QNBs}}

Define the quantum Nambu bracket, or QNB \cite{Nambu}, as a fully
antisymmetrized multilinear sum of operator products in an associative
enveloping algebra,
\begin{equation}
\left[  A_{1},A_{2},\cdots,A_{k}\right]  \equiv\sum_{\substack{\text{all
}k!\text{\ perms }\left\{  \sigma_{1},\sigma_{2},\cdots,\sigma_{k}\right\}
\\\text{of the indices }\left\{  1,2,\cdots,k\right\}  }}\operatorname{sgn}%
\left(  \sigma\right)  \,A_{\sigma_{1}}A_{\sigma_{2}}\cdots A_{\sigma_{k}}\;,
\label{QNBDefinition}%
\end{equation}
where $\operatorname{sgn}\left(  \sigma\right)  =\left(  -1\right)
^{\pi\left(  \sigma\right)  }$ with $\pi\left(  \sigma\right)  $\ the parity
of the permutation $\left\{  \sigma_{1},\sigma_{2},\cdots,\sigma_{k}\right\}
$. \ The bracket is unchanged by adding to any one element a linear
combination of the others, in analogy with the usual row or column
manipulations on determinants.

\paragraph{\underline{Recursion relations}}

There are various ways to obtain QNBs recursively, from products involving
fewer operators. \ For example, a QNB involving $k$ operators has both left-
and right-sided resolutions of single operators multiplying QNBs of $k-1$
operators.%
\begin{align}
\left[  A_{1},A_{2},\cdots,A_{k}\right]   &  =\sum_{k!\;\text{perms}\;\sigma
}\frac{\operatorname{sgn}\left(  \sigma\right)  }{\left(  k-1\right)
!}\,A_{\sigma_{1}}\left[  A_{\sigma_{2}},\cdots,A_{\sigma_{k}}\right]
\nonumber\\
&  =\sum_{k!\;\text{perms}\;\sigma}\frac{\operatorname{sgn}\left(
\sigma\right)  }{\left(  k-1\right)  !}\,\left[  A_{\sigma_{1}},\cdots
,A_{\sigma_{k-1}}\right]  A_{\sigma_{k}}\;.
\end{align}
On the RHS there are actually only $k$ distinct products of single elements
with $\left(  k-1\right)  $-brackets, each such product having a net
coefficient $\pm1$. \ The denominator compensates for replication of these
products in the sum over permutations. \ (We leave it as an elementary
exercise for the reader to prove this result.)

For example, the 2-bracket is obviously just the commutator $\left[
A,B\right]  =AB-BA$,\ while the 3-bracket may be written in either of two
\cite{Nambu} or three convenient ways.%
\begin{align}
\left[  A,B,C\right]   &  =A\left[  B,C\right]  +B\left[  C,A\right]
+C\left[  A,B\right] \nonumber\\
&  =\left[  A,B\right]  C+\left[  B,C\right]  A+\left[  C,A\right]
B\nonumber\\
&  =\tfrac{3}{2}\left\{  \left[  A,B\right]  ,C\right\}  +\tfrac{1}{2}\left[
\left\{  A,B\right\}  ,C\right]  -\left[  A,\left\{  B,C\right\}  \right]  \;.
\label{QNB3}%
\end{align}
Summing the first two lines gives anticommutators containing commutators on
the RHS.%
\begin{equation}
2\,\left[  A,B,C\right]  =\left\{  A,\left[  B,C\right]  \right\}  +\left\{
B,\left[  C,A\right]  \right\}  +\left\{  C,\left[  A,B\right]  \right\}  \;.
\end{equation}
The last expression is to be contrasted to the Jacobi identity obtained by
taking the difference of the first two RHS lines in (\ref{QNB3}).
\begin{equation}
0=\left[  A,\left[  B,C\right]  \right]  +\left[  B,\left[  C,A\right]
\right]  +\left[  C,\left[  A,B\right]  \right]  \;. \label{2Jacobi}%
\end{equation}
Similarly for the 4-bracket,%
\begin{align}
\left[  A,B,C,D\right]   &  =A\left[  B,C,D\right]  -B\left[  C,D,A\right]
+C\left[  D,A,B\right]  -D\left[  A,B,C\right] \nonumber\\
&  =-\left[  B,C,D\right]  A+\left[  C,D,A\right]  B-\left[  D,A,B\right]
C+\left[  A,B,C\right]  D\;. \label{QNB4}%
\end{align}
Summing these two lines gives%
\begin{equation}
2\,\left[  A,B,C,D\right]  =\left[  A,\left[  B,C,D\right]  \right]  -\left[
B,\left[  C,D,A\right]  \right]  +\left[  C,\left[  D,A,B\right]  \right]
-\left[  D,\left[  A,B,C\right]  \right]  \;,
\end{equation}
while taking the difference gives%
\begin{equation}
0=\left\{  A,\left[  B,C,D\right]  \right\}  -\left\{  B,\left[  C,D,A\right]
\right\}  +\left\{  C,\left[  D,A,B\right]  \right\}  -\left\{  D,\left[
A,B,C\right]  \right\}  \;. \label{Baby3Jacobi}%
\end{equation}
There may be some temptation to think of the last of these as something like a
generalization of the Jacobi identity, and, in principle, it is, but in a a
crucially limited way, so that temptation should be checked. The more
appropriate and complete generalization of the Jacobi identity is given
systematically below (cf. (\ref{QJI})).

\paragraph{\underline{Jordan products}}

Define a fully symmetrized, generalized Jordan operator product (GJP).%
\begin{equation}
\left\{  A_{1},A_{2},\cdots,A_{k}\right\}  \equiv\sum_{\substack{\text{all
}k!\text{\ perms }\left\{  \sigma_{1},\sigma_{2},\cdots,\sigma_{k}\right\}
\\\text{of the indices }\left\{  1,2,\cdots,k\right\}  }}A_{\sigma_{1}%
}A_{\sigma_{2}}\cdots A_{\sigma_{k}}\;,
\end{equation}
as introduced, in the bilinear form at least, by Jordan \cite{Jordan} to
render non-Abelian algebras into Abelian algebras at the expense of
non-associativity. \ The generalization to multi-linears was suggested by
Kurosh \cite{Kurosh}, but the idea was not used in any previous physical
application, as far as we know. \ A GJP also has left- and right-sided
recursions,
\begin{align}
\left\{  A_{1},A_{2},\cdots,A_{k}\right\}   &  =\sum_{k!\;\text{perms}%
\;\sigma}\frac{1}{\left(  k-1\right)  !}\,A_{\sigma_{1}}\left\{  A_{\sigma
_{2}},A_{\sigma_{3}},\cdots,A_{\sigma_{k}}\right\} \nonumber\\
&  =\sum_{k!\;\text{perms}\;\sigma}\frac{1}{\left(  k-1\right)  !}\,\left\{
A_{\sigma_{2}},A_{\sigma_{3}},\cdots,A_{\sigma_{k-1}}\right\}  A_{\sigma_{k}%
}\;.
\end{align}
On the RHS there are again only $k$ distinct products of single elements with
$\left(  k-1\right)  $-GJPs, each such product having a net coefficient $+1$.
\ The denominator again compensates for replication of these products in the
sum over permutations. \ (We leave it as another elementary exercise for the
reader to prove this result.)

For example, a Jordan 2-product is obviously just an anticommutator $\left\{
A,B\right\}  =AB+BA$, while a 3-product is given by%
\begin{align}
\left\{  A,B,C\right\}   &  =\left\{  A,B\right\}  C+\left\{  A,C\right\}
B+\left\{  B,C\right\}  A\nonumber\\
&  =A\left\{  B,C\right\}  +B\left\{  A,C\right\}  +C\left\{  A,B\right\}
\nonumber\\
&  =\tfrac{3}{2}\left\{  \left\{  A,B\right\}  ,C\right\}  +\tfrac{1}%
{2}\left[  \left[  A,B\right]  ,C\right]  -\left[  A,\left[  B,C\right]
\right]  \;. \label{GJP3}%
\end{align}
Equivalently, taking sums and differences, we obtain%
\begin{equation}
2\,\left\{  A,B,C\right\}  =\left\{  A,\left\{  B,C\right\}  \right\}
+\left\{  B,\left\{  A,C\right\}  \right\}  +\left\{  C,\left\{  A,B\right\}
\right\}  \;,
\end{equation}
as well as the companion of the Jacobi identity often encountered in
super-algebras,%
\begin{equation}
0=\left[  A,\left\{  B,C\right\}  \right]  +\left[  B,\left\{  A,C\right\}
\right]  +\left[  C,\left\{  A,B\right\}  \right]  \;. \label{Super2Jacobi}%
\end{equation}
Similarly for the 4-product,%
\begin{align}
\left\{  A,B,C,D\right\}   &  =A\left\{  B,C,D\right\}  +B\left\{
C,D,A\right\}  +C\left\{  D,A,B\right\}  +D\left\{  A,B,C\right\} \nonumber\\
&  =\left\{  A,B,C\right\}  D+\left\{  B,C,D\right\}  A+\left\{
C,D,A\right\}  B+\left\{  D,A,B\right\}  C\;.
\end{align}
Summing gives%
\begin{equation}
2\,\left\{  A,B,C,D\right\}  =\left\{  A,\left\{  B,C,D\right\}  \right\}
+\left\{  B,\left\{  C,D,A\right\}  \right\}  +\left\{  C,\left\{
D,A,B\right\}  \right\}  +\left\{  D,\left\{  A,B,C\right\}  \right\}  \;,
\end{equation}
while subtracting gives%
\begin{equation}
0=\left[  A,\left\{  B,C,D\right\}  \right]  +\left[  B,\left\{
C,D,A\right\}  \right]  +\left[  C,\left\{  D,A,B\right\}  \right]  +\left[
D,\left\{  A,B,C\right\}  \right]  \;. \label{BabySuper3Jacobi}%
\end{equation}
Again the reader is warned off the temptation to think of the last of these as
a bona fide generalization of the super-Jacobi identity. \ While it is a valid
identity, of course, following from nothing but associativity, there is a
superior and complete set of identities to be given later (cf. (\ref{QJI}) to follow).

\paragraph{\underline{(Anti)Commutator resolutions}}

As in the classical case, Section 2.2, it is always possible to resolve even
rank brackets into sums of commutator products, very usefully. \ For example,
\begin{align}
\left[  A,B,C,D\right]   &  =\left[  A,B\right]  \left[  C,D\right]  -\left[
A,C\right]  \left[  B,D\right]  -\left[  A,D\right]  \left[  C,B\right]
\nonumber\\
&  +\left[  C,D\right]  \left[  A,B\right]  -\left[  B,D\right]  \left[
A,C\right]  -\left[  C,B\right]  \left[  A,D\right]  \;.
\label{4QNBasCommutators}%
\end{align}
An arbitrary even bracket of rank $2n$ breaks up into $\left(  2n\right)
!/\left(  2^{n}\right)  =n!\left(  2n-1\right)  !!$ such products. \ Another
way to say this is that even QNBs can be written in terms of GJPs of
commutators. \ The general result is%
\begin{equation}
\left[  A_{1},A_{2},\cdots,A_{2n-1},A_{2n}\right]  =\sum_{\left(  2n\right)
!\;\text{perms}\;\sigma}\frac{\operatorname{sgn}\left(  \sigma\right)  }%
{2^{n}n!}\,\left\{  \left[  A_{\sigma_{1}},A_{\sigma_{2}}\right]  ,\left[
A_{\sigma_{3}},A_{\sigma_{4}}\right]  ,\cdots,\left[  A_{\sigma_{2n-1}%
},A_{\sigma_{2n}}\right]  \right\}  \;. \label{2nQNBasCommutators}%
\end{equation}
An even GJP also resolves into symmetrized products of anticommutators.%
\begin{equation}
\left\{  A_{1},A_{2},\cdots,A_{2n-1},A_{2n}\right\}  =\sum_{\left(  2n\right)
!\;\text{perms}\;\sigma}\frac{1}{2^{n}n!}\,\left\{  \left\{  A_{\sigma_{1}%
},A_{\sigma_{2}}\right\}  ,\left\{  A_{\sigma_{3}},A_{\sigma_{4}}\right\}
,\cdots,\left\{  A_{\sigma_{2n-1}},A_{\sigma_{2n}}\right\}  \right\}  \;.
\label{2nGJPasAnticommutators}%
\end{equation}
This resolution makes it transparent that all such even brackets will vanish
if one or more of the $A_{i}$ are central (i.e. commute with all the other
elements in the bracket). \ For instance, if any one $A_{i}$ is a multiple of
the unit operator, the $2n$-bracket will vanish. \ (This same statement does
not apply to odd brackets, as Nambu realized originally for 3-brackets
\cite{Nambu}, and consequently there are additional hurdles to be overcome
when using odd QNBs.)

As in the classical bracket formalism, the proofs of the (anti)commutator
resolution relations are elementary. \ Both left- and right-hand sides of the
expressions are sums of $2n$-th degree monomials linear in each of the $A$s.
\ Both sides are either totally antisymmetric, in the case of
(\ref{2nQNBasCommutators}), or totally symmetric, in the case of
(\ref{2nGJPasAnticommutators}), under permutations of the $A$s. \ Thus the two
sides must be proportional. \ The only open issue is the constant of
proportionality. \ This is easily determined to be $1,$ just by comparing the
coefficients of any given term appearing on both sides of the equation, e.g.
$A_{1}A_{2}\cdots A_{2N-1}A_{2N}$. \ 

It is clear from the commutator resolution of even QNBs that totally
symmetrized GJPs and totally antisymmetrized QNBs are not unrelated. \ In
fact, the relationship is most pronounced in quantum mechanical applications
where the operators form a Lie algebra.

\paragraph{\underline{Reductions for Lie algebras}}

In full analogy to the classical case above, when the operators involved in a
QNB close into a Lie algebra, even if an infinite one, the Nambu bracket
reduces in rank to become a sum of GJPs involving about half as many operators
(\emph{quantum} \emph{reductio ad dimidium}). It follows as an elementary
consequence of the commutator resolution of the Nambu bracket. \ First,
consider even brackets, since the commutator reduction applies directly to
that case. From the commutator resolution, it follows that for any Lie algebra
given by%
\begin{equation}
\left[  B_{i},B_{j}\right]  =i\hbar\sum_{m}c_{ij}^{\;\;m}B_{m}\;,
\label{LieAlg}%
\end{equation}
we have for arbitrary $A$ (sum over repeated $m$s)\footnote{After obtaining
this result, and using it in \cite{CurtrightZachos}, we learned that similar
statements appeared previously in \cite{Azcarraga96b,Azcarraga97}.}%
\begin{align}
\left[  B_{1},\cdots,B_{2k+1},A\right]   &  =\sum_{\left(  2k+1\right)
!\;\text{perms}\;\sigma}\frac{\operatorname{sgn}\left(  \sigma\right)  }%
{2^{k}k!}\,\left\{  \left[  B_{\sigma_{1}},B_{\sigma_{2}}\right]  ,\left[
B_{\sigma_{3}},B_{\sigma_{4}}\right]  ,\cdots,\left[  B_{\sigma_{2k-1}%
},B_{\sigma_{2k}}\right]  ,\left[  B_{\sigma_{2k+1}},A\right]  \right\}
\nonumber\\
&  =\sum_{\left(  2k+1\right)  !\;\text{perms}\;\sigma}\frac
{\operatorname{sgn}\left(  \sigma\right)  }{2^{k}k!}\,\left(  i\hbar\right)
^{k}\,c_{\sigma_{1}\sigma_{2}}^{\;\;\;\;m_{1}}c_{\sigma_{3}\sigma_{4}%
}^{\;\;\;\;m_{2}}\cdots c_{\sigma_{2k-1}\sigma_{2k}}^{\;\;\;\;\;\;\;\;m_{k}%
}\left\{  B_{m_{1}},B_{m_{2}},\cdots,B_{m_{k}},\left[  B_{\sigma_{2k+1}%
},A\right]  \right\}  \;. \label{NambuJordanLie}%
\end{align}
For odd brackets, it is first necessary to resolve the QNB into products of
single operators with even brackets, and then resolve the various even
brackets into commutators. \ This gives a larger sum of terms for odd
brackets, but again each term involves about half as many Jordan products
compared to the number of commutators resolving the original Nambu bracket.
\ The mixture of algebraic structures in (\ref{NambuJordanLie}) suggests
referring to this as a Nambu-Jordan-Lie (NJL) algebra.

\paragraph{\underline{The classical limit}}

Since Poisson brackets are straightforward classical limits of commutators,
\[
\lim_{\hbar\rightarrow0}\left(  \frac{1}{i\hbar}\right)  \left[  A,B\right]
=\left\{  A,B\right\}  _{\text{PB}}\;,
\]
it follows that the commutator resolution of all even QNBs directly specifies
their classical limit. \ (For a detailed approach to the classical limit,
including sub-dominant terms of higher order in $\hbar$, see, e.g., the Moyal
Bracket discussion in \cite{CurtrightZachos}.)

For example, from%
\begin{equation}
\left[  A,B,C,D\right]  =\left\{  \left[  A,B\right]  ,\left[  C,D\right]
\right\}  -\left\{  \left[  A,C\right]  ,\left[  B,D\right]  \right\}
-\left\{  \left[  A,D\right]  ,\left[  C,B\right]  \right\}  \;,
\end{equation}
with due attention to a critical factor of $2$ (i.e. the anticommutators on
the RHS become just twice the ordinary products of their entries), the
classical limit emerges as%
\begin{align}
\frac{1}{2}\,\lim_{\hbar\rightarrow0}\left(  \frac{1}{i\hbar}\right)
^{2}\left[  A,B,C,D\right]   &  =\left\{  A,B\right\}  _{\text{PB}}\left\{
C,D\right\}  _{\text{PB}}-\left\{  A,C\right\}  _{\text{PB}}\left\{
B,D\right\}  _{\text{PB}}-\left\{  A,D\right\}  _{\text{PB}}\left\{
C,B\right\}  _{\text{PB}}\nonumber\\
&  =\left\{  A,B,C,D\right\}  _{\text{NB}}\;. \label{4QNBClassicalLimit}%
\end{align}
And so it goes with all other even rank Nambu brackets. \ For a $2n$-bracket,
one sees that
\begin{align}
\frac{1}{n!}\,\lim_{\hbar\rightarrow0}\left(  \frac{1}{i\hbar}\right)
^{n}\left[  A_{1},A_{2},\cdots,A_{2n}\right]   &  =\sum_{\left(  2n\right)
!\text{\ perms }\sigma}\frac{\operatorname{sgn}\left(  \sigma\right)  }%
{2^{n}n!}\,\left\{  A_{\sigma_{1}},A_{\sigma_{2}}\right\}  _{\text{PB}%
}\left\{  A_{\sigma_{3}},A_{\sigma_{4}}\right\}  _{\text{PB}}\cdots\left\{
A_{\sigma_{2n-1}},A_{\sigma_{2n}}\right\}  _{\text{PB}}\nonumber\\
&  =\left\{  A_{1},A_{2},\cdots,A_{2n}\right\}  _{\text{NB}}\;.
\label{2nQNBClassicalLimit}%
\end{align}
This is another way to establish that there are indeed $\left(  2n-1\right)
!!$ independent products of $n$ Poisson brackets summing up to give the PB
resolution of the classical Nambu $2n$-bracket. \ Once again due attention
must be given to a critical additional factor of $n!$ (as in the denominator
on the LHS of (\ref{2nQNBClassicalLimit})) since the GJPs on the RHS of
(\ref{2nQNBasCommutators}) will, in the classical limit, always replicate the
same classical product $n!$ times.

\paragraph{\underline{The Leibniz rule failure and derivators}}

Define the \emph{derivator} to measure the failure of the simplest Leibniz
rule for QNBs,
\begin{equation}
^{k+1}\mathbf{\Delta}_{\mathbf{B}}\left(  A,\mathcal{A}\right)  \equiv\left(
A,\mathcal{A}\,|\,B_{1},\cdots,B_{k}\right)  \equiv\left[  \,A\mathcal{A\,}%
,B_{1},\cdots,B_{k}\right]  -A\left[  \mathcal{A},B_{1},\cdots,B_{k}\right]
-\left[  A,B_{1},\cdots,B_{k}\right]  \mathcal{A}\;. \label{DerivatorDefn}%
\end{equation}
The first term on the RHS is a $\left(  k+1\right)  $-bracket acting on just
the product of $A$\ and $\mathcal{A}$, the order of the bracket being evident
in the pre-superscript of the $\mathbf{\Delta}_{\mathbf{B}}$ notation. \ This
reads in an obvious way. \ For instance, $^{4}\mathbf{\Delta}_{\mathbf{B}}$ is
a ``4-delta of $B$s''. \ That notation also emphasizes that the $B$s act
\emph{on} the pair of $A$s. \ The second notation in (\ref{DerivatorDefn})
makes explicit all the $B$s and is useful for computer code. \ 

Any $\mathbf{\Delta}_{\mathbf{B}}$ acts on all pairs of elements in the
enveloping algebra $\mathfrak{A}$ to produce another element in $\mathfrak{A}%
$.%
\begin{equation}
\mathbf{\Delta}_{\mathbf{B}}:\mathfrak{A}\times\mathfrak{A}\longmapsto
\mathfrak{A}\;. \label{MapsTo}%
\end{equation}
When $\mathbf{\Delta}_{\mathbf{B}}$ does not vanish the corresponding bracket
with the $B$s does not define a derivation on $\mathfrak{A}$. \ The derivator
$\mathbf{\Delta}_{\mathbf{B}}\left(  A,\mathcal{A}\right)  $ is linear in both
$A$ and $\mathcal{A},$ as well as linear in each of the $B$s.

Less trivially, from explicit calculations, we find inhomogeneous recursion
relations for these derivators.%
\begin{align}
\left(  A,\mathcal{A}\,|\,B_{1},\cdots,B_{k}\right)   &  =\frac{1}{2}%
\sum_{k!\text{\ perms }\sigma}\frac{\operatorname{sgn}\left(  \sigma\right)
}{\left(  k-1\right)  !}\,\left(  \,\left(  A,\mathcal{A}\,|\,B_{\sigma_{1}%
},\cdots,B_{\sigma_{k-1}}\right)  B_{\sigma_{k}}+\left(  -1\right)
^{k}B_{\sigma_{k}}\left(  A,\mathcal{A}\,|\,B_{\sigma_{1}},\cdots
,B_{\sigma_{k-1}}\right)  \,\right) \nonumber\\
&  +\frac{1}{2}\sum_{k!\text{\ perms }\sigma}\frac{\operatorname{sgn}\left(
\sigma\right)  }{\left(  k-1\right)  !}\,\left(  \,\left[  A,B_{\sigma_{k}%
}\right]  \left[  B_{\sigma_{1}},\cdots,B_{\sigma_{k-1}},\mathcal{A}\right]
-\left[  A,B_{\sigma_{1}},\cdots,B_{\sigma_{k-1}}\right]  \left[
B_{\sigma_{k}},\mathcal{A}\right]  \,\right) \nonumber\\
&  +\frac{\left(  -1\right)  ^{k+1}-1}{2}\;A\left[  B_{1},\cdots,B_{k}\right]
\mathcal{A}.
\end{align}
Alternatively, we may write this so as to emphasize the number of distinct
terms on the RHS and distinguish between the even and odd bracket cases. \ The
first two terms under the sum on the RHS give a commutator/anticommutator for
$k$ odd/even, and the last term is absent for $k$ odd.

For even (2n+2)-brackets, this becomes
\begin{align}
2\,\left(  A,\mathcal{A}\,|\,B_{1},\cdots,B_{2n+1}\right)   &  =\left[
\left(  A,\mathcal{A}\,|\,B_{1},\cdots,B_{2n}\right)  ,B_{2n+1}\right]
\nonumber\\
&  +\left[  A,B_{2n+1}\right]  \left[  B_{1},\cdots,B_{2n},\mathcal{A}\right]
-\left[  A,B_{1},\cdots,B_{2n}\right]  \left[  B_{2n+1},\mathcal{A}\right]
\nonumber\\
&  +\text{ (2n signed permutations of the }B\text{s)\ ,} \label{2n+2derivator}%
\end{align}
where the first RHS line involves derivators of reduced rank, within
commutators. \ For odd (2n+1)-brackets, it becomes
\begin{align}
2\,\left(  A,\mathcal{A}\,|\,B_{1},\cdots,B_{2n}\right)   &  =\left\{  \left(
A,\mathcal{A}\,|\,B_{1},\cdots,B_{2n-1}\right)  ,B_{2n}\right\} \nonumber\\
&  +\left[  A,B_{2n}\right]  \left[  B_{1},\cdots,B_{2n-1},\mathcal{A}\right]
-\left[  A,B_{1},\cdots,B_{2n-1}\right]  \left[  B_{2n},\mathcal{A}\right]
\nonumber\\
&  +\text{ (2n-1 signed permutations of the }B\text{s)}\nonumber\\
&  -2\,A\left[  B_{1},\cdots,B_{2n}\right]  \mathcal{A}\;,
\label{2n+1derivator}%
\end{align}
where the first RHS line involves derivators of reduced rank, within
anticommutators. \ Note the additional inhomogeneity in the last RHS line of
these results. \ It may be viewed as a type of quantum obstruction in the
recursion relation for the odd (2n+1)-bracket.

The obstruction is clarified when we specialize to $n=1$, i.e. the 3-bracket
case. \ Since commutators are always derivations, one has $^{2}\mathbf{\Delta
}_{B}\left(  A,\mathcal{A}\right)  =0$, and the first RHS line vanishes in
(\ref{2n+1derivator})\ for the $^{3}\mathbf{\Delta}_{\mathbf{B}}\left(
A,\mathcal{A}\right)  $ case. \ So we have just
\begin{equation}
\left(  A,\mathcal{A}\,|\,B_{1},B_{2}\right)  =\left[  A,B_{2}\right]  \left[
B_{1},\mathcal{A}\right]  -\left[  A,B_{1}\right]  \left[  B_{2}%
,\mathcal{A}\right]  -A\left[  B_{1},B_{2}\right]  \mathcal{A}\;.
\end{equation}
The first two terms on the RHS are $O\left(  \hbar^{2}\right)  $\ while the
last is $O\left(  \hbar\right)  $. \ It is precisely this last term which was
responsible for some of Nambu's misgivings concerning his quantum 3-bracket.
\ In particular, even in the extreme case when both $A$ and $\mathcal{A}$
commute with the $B$s, $^{3}\mathbf{\Delta}_{\mathbf{B}}\left(  A,\mathcal{A}%
\right)  $ does not vanish:
\begin{equation}
\left.  \left(  A,\mathcal{A}\,|\,B_{1},B_{2}\right)  \right\vert _{\left[
A,B_{i}\right]  =0=\left[  \mathcal{A},B_{i}\right]  }=-A\mathcal{A}\,\left[
B_{1},B_{2}\right]  \;.
\end{equation}
By contrast, for the even (2n+2)-bracket, all terms on the RHS of
(\ref{2n+2derivator}) are generically of the same order, $O\left(  \hbar
^{n+1}\right)  $, and all terms vanish if $A$ and $\mathcal{A}$ commute with
all the $B$s. \ In terms of combinatorics, this seems to be the only feature
for the simple, possibly failed, Leibniz rule that distinguishes between even
and odd brackets. \ An even-odd QNB dichotomy has been previously noted
\cite{Hanlon} and stressed \cite{Azcarraga97}, for other reasons.

The size of the brackets involved in the derivators can be reduced when the
operators obey a Lie algebra as in (\ref{LieAlg}). \ The simplest situation
occurs when the bracket is even. \ For this situation we have%
\begin{gather}
\left(  A,\mathcal{A}\,|\,B_{1},\cdots,B_{2k+1}\right)  =\sum_{\left(
2k+1\right)  !\;\text{perms}\;\sigma}\frac{\operatorname{sgn}\left(
\sigma\right)  }{2^{k}k!}\,\left(  i\hbar\right)  ^{k}\,c_{\sigma_{1}%
\sigma_{2}}^{\;\;\;\;m_{1}}c_{\sigma_{3}\sigma_{4}}^{\;\;\;\;m_{2}}\cdots
c_{\sigma_{2k-1}\sigma_{2k}}^{\;\;\;\;m_{k}}\times\\
\times{\Large (}\left\{  B_{m_{1}},\cdots,B_{m_{k}},\left[  B_{\sigma_{2k+1}%
},A\mathcal{A}\right]  \right\}  -A\left\{  B_{m_{1}},\cdots,B_{m_{k}},\left[
B_{\sigma_{2k+1}},\mathcal{A}\right]  \right\}  -\left\{  B_{m_{1}}%
,\cdots,B_{m_{k}},\left[  B_{\sigma_{2k+1}},A\right]  \right\}  \mathcal{A}%
{\Large )\;.}\nonumber
\end{gather}

\paragraph{\underline{Generalized Jacobi identities and QFIs}}

We previously pointed out some elementary identities involving QNBs, e.g.
(\ref{Baby3Jacobi}) and (\ref{BabySuper3Jacobi}), which are suggestive of
generalizations of the Jacobi identity for commutators. \ Those particular
identities, while true, were not designated as ``generalized Jacobi
identities'' (GJIs), for the simple fact that they do \emph{not} involve the
case where QNBs of a given rank act on QNBs of the same rank. \ Here, we
explore QNB identities of the latter type. \ There are indeed acceptable
generalizations of the usual commutators-acting-on-commutators Jacobi identity
(i.e. quantum 2-brackets acting on quantum 2-brackets), and these
generalizations are indeed valid for \emph{all} higher rank QNBs (i.e. quantum
n-brackets acting on quantum n-brackets). \ However, there is an essential
distinction to be drawn between the even and odd quantum bracket cases
\cite{Hanlon,Azcarraga97}.

It is important to note that, historically, there have been some incorrect
guesses and false starts in this direction that originated from the so-called
fundamental identity obeyed by classical Nambu brackets, (\ref{simpleFI}).
\ This simple identity apparently misled several investigators \cite{Sahoo},
most recently \cite{Takhtajan} and \cite{DitoFlato,DitoFST}, to think of it as
a \textquotedblleft fundamental\textquotedblright\ generalization of the
Jacobi identity, without taking care to preserve the Jacobi Identity's
traditional role of encoding nothing but associativity. \ These same
investigators then insisted that a \textquotedblleft correct
quantization\textquotedblright\ of the classical Nambu bracket \emph{must}
satisfy an identity of the same form as (\ref{simpleFI}).

Unfortunately for them, QNBs do \emph{not} satisfy this particular identity,
in general, and thereby pose a formidable problem to proponents of that
identity's fundamental significance. This difficulty led \cite{Takhtajan} and
\cite{DitoFlato,DitoFST}, to seek alternative ways to quantize CNBs,
ultimately culminating in the so-called Abelian deformation method
\cite{DitoFlato,DitoFST}. This amounted to demanding that the quantized
brackets satisfy the mathematical postulates of an ``n-Lie algebra'' as
defined by Filippov \cite{Filippov} many years earlier. \ However, not only
are those postulates \emph{not} satisfied by generic QNBs, but more
importantly, those postulates are \emph{not} warranted by the physics of QNBs,
as will be clear in the examples to follow.

The correct generalizations of the Jacobi identities which \emph{do} encode
associativity were found independently by groups of mathematicians
\cite{Hanlon}\ and physicists \cite{Azcarraga96a,Azcarraga97}.
\ Interestingly, both groups were studying cohomology questions, so perhaps it
is not surprising that they arrived at the same result. \ (Fortunately for us
the result is sufficiently simple in its combinatorics that we do not need to
go through the cohomology issues.) \ The acceptable generalization of the
Jacobi identity that was found is satisfied by all QNBs, although for odd QNBs
there is a significant difference in the form of the final result: \ It
contains an ``inhomogeneity''. \ The correct generalization is obtained just
by totally antisymmetrizing the action of one n-bracket on the other.
\ Effectively, this amounts to antisymmetrizing the form of the RHS of
(\ref{simpleFI}) over all permutations of the $A$s and $B$s, including all
exchanges\emph{ }of $A$s with $B$s. \ 

We illustrate the correct quantum identity for the case of a 3-bracket acting
on a 3-bracket, where the classical result is%
\begin{equation}
0=\left\{  \left\{  A,B,C\right\}  _{NB},D,E\right\}  _{NB}-\left\{  \left\{
A,D,E\right\}  _{NB},B,C\right\}  _{NB}-\left\{  A,\left\{  B,D,E\right\}
_{NB},C\right\}  _{NB}-\left\{  A,B,\left\{  C,D,E\right\}  _{NB}\right\}
_{NB}\;, \label{simpleFI3}%
\end{equation}
i.e. (\ref{simpleFI}) for $n=3$. \ For ease in writing, we let $A_{1}\equiv
A$, $A_{2}\equiv B$, $A_{3}\equiv C$, $B_{1}\equiv D$, and $B_{2}\equiv E$.
\ Consider $\left[  \left[  A,B,C\right]  ,D,E\right]  $. \ This QNB
corresponds to the first term on the RHS of (\ref{simpleFI3}). \ If we
antisymmetrize $\left[  \left[  A,B,C\right]  ,D,E\right]  $ over all $5!$
permutations of $A,B,C,D,$ and $E$, we obtain, with a common overall
coefficient of $12=2!3!$, a total of $10=5!/\left(  2!3!\right)  $ distinct
terms as follows:%
\begin{align}
&  \left[  \left[  A,B,C\right]  ,D,E\right]  +\left[  \left[  A,D,E\right]
,B,C\right]  +\left[  \left[  D,B,E\right]  ,A,C\right]  +\left[  \left[
D,E,C\right]  ,A,B\right] \nonumber\\
&  -\left[  \left[  D,B,C\right]  ,A,E\right]  -\left[  \left[  E,B,C\right]
,D,A\right]  -\left[  \left[  A,D,C\right]  ,B,E\right] \nonumber\\
&  -\left[  \left[  A,E,C\right]  ,D,B\right]  -\left[  \left[  A,B,D\right]
,C,E\right]  -\left[  \left[  A,B,E\right]  ,D,C\right]  \;. \label{1stTerm}%
\end{align}
Now we determine the coefficient of any given monomial produced by this
sum\footnote{This line of argument is an adaptation of that in \cite{Hanlon}.
\ Equivalent methods are used in \cite{Azcarraga96a,Azcarraga97}.}. \ Since
the expression is totally antisymmetrized in all the five elements, the result
must be proportional to $\left[  A,B,C,D,E\right]  $. \ To determine the
constant of proportionality it suffices to consider the monomial $ABCDE$.
\ This particular monomial can be found in only 3\ terms out of the ten in
(\ref{1stTerm}), namely in
\begin{equation}
\left[  \left[  A,B,C\right]  ,D,E\right]  \;,\;\;\;\left[  A,\left[
B,C,D\right]  ,E\right]  =-\left[  \left[  D,B,C\right]  ,A,E\right]
\;,\;\text{and}\;\;\;\left[  A,B,\left[  C,D,E\right]  \right]  =\left[
\left[  D,E,C\right]  ,A,B\right]  \;. \label{TheOddCycle}%
\end{equation}
The various terms are obtained just by ``shifting'' the interior brackets from
left to right within the exterior brackets, while keeping all the bracket
entries in a fixed left-to-right order, and keeping track of the
$\operatorname{sgn}\left(  \sigma\right)  $ factors. \ (Call this the
``shifting bracket argument''.) \ The monomial $ABCDE$ appears in each of
these terms with coefficient $+1,$ for a total of $+3\times ABCDE$. \ Thus, we
conclude with a 5-element, 11-term identity,
\begin{align}
&  \left[  \left[  A,B,C\right]  ,D,E\right]  +\left[  \left[  A,D,E\right]
,B,C\right]  +\left[  \left[  D,B,E\right]  ,A,C\right]  +\left[  \left[
D,E,C\right]  ,A,B\right] \nonumber\\
&  -\left[  \left[  D,B,C\right]  ,A,E\right]  -\left[  \left[  E,B,C\right]
,D,A\right]  -\left[  \left[  A,D,C\right]  ,B,E\right] \nonumber\\
&  -\left[  \left[  A,E,C\right]  ,D,B\right]  -\left[  \left[  A,B,D\right]
,C,E\right]  -\left[  \left[  A,B,E\right]  ,D,C\right] \nonumber\\
&  =3\left[  A,B,C,D,E\right]  \;. \label{3Jacobi}%
\end{align}
This is the prototypical generalization of the Jacobi identity for odd QNBs,
and like the Jacobi identity, it is antisymmetric in all of its elements.
\ The RHS here is the previously designated inhomogeneity.

\begin{center}
\textit{The totally antisymmetrized action of odd }$n$\textit{ QNBs on other
odd }$n$\textit{ QNBs results in }$\left(  2n-1\right)  $\textit{-brackets.}
\end{center}

We recognize in the first line of (\ref{3Jacobi}) those QNB combinations which
correspond to all four of the individual terms on the RHS of (\ref{simpleFI3}%
). \ However, \emph{the signs are changed} for 3 of the 4 QNB terms relative
to those in (\ref{simpleFI3}). \ One might hope that changing these signs in
the QNB combinations will lead to some simplification, and indeed it does, but
it does not cause the resulting expression to vanish, as it did in
(\ref{simpleFI3}). \ To see this, consider in the same way the effects of
antisymmetrizing the QNBs corresponding to each of the other three terms on
the RHS of (\ref{simpleFI3}). $\ $The second RHS term would have as
correspondent $-\left[  \left[  A,D,E\right]  ,B,C\right]  $, which, when
totally antisymmetrized, gives an overall common coefficient of $2!3!$
multiplying$\ $ \
\begin{align}
&  -\left[  \left[  A,D,E\right]  ,B,C\right]  -\left[  \left[  A,B,C\right]
,D,E\right]  -\left[  \left[  B,D,C\right]  ,A,E\right]  -\left[  \left[
B,C,E\right]  ,A,D\right] \nonumber\\
&  +\left[  \left[  B,D,E\right]  ,A,C\right]  +\left[  \left[  C,D,E\right]
,B,A\right]  +\left[  \left[  A,B,E\right]  ,D,C\right] \nonumber\\
&  +\left[  \left[  A,C,E\right]  ,B,D\right]  +\left[  \left[  A,D,B\right]
,E,C\right]  +\left[  \left[  A,D,C\right]  ,B,E\right] \nonumber\\
&  =-3\left[  A,B,C,D,E\right]  \;. \label{2ndTerm}%
\end{align}
The third RHS term of (\ref{simpleFI3}) would have as correspondent $-\left[
A,\left[  B,D,E\right]  ,C\right]  =\left[  \left[  B,D,E\right]  ,A,C\right]
$, which, when totally antisymmetrized, gives an overall common coefficient of
$2!3!$ multiplying%
\begin{align}
&  \left[  \left[  B,D,E\right]  ,A,C\right]  +\left[  \left[  B,A,C\right]
,D,E\right]  +\left[  \left[  A,D,C\right]  ,B,E\right]  +\left[  \left[
A,C,E\right]  ,B,D\right] \nonumber\\
&  -\left[  \left[  A,D,E\right]  ,B,C\right]  -\left[  \left[  C,D,E\right]
,A,B\right]  -\left[  \left[  B,A,E\right]  ,D,C\right] \nonumber\\
&  -\left[  \left[  B,C,E\right]  ,A,D\right]  -\left[  \left[  B,D,A\right]
,E,C\right]  -\left[  \left[  B,D,C\right]  ,A,E\right] \nonumber\\
&  =-3\left[  A,B,C,D,E\right]  \;. \label{3rdTerm}%
\end{align}
The fourth and final RHS term of (\ref{simpleFI3}) would have as correspondent
$-\left[  A,B,\left[  C,D,E\right]  \right]  =\left[  \left[  C,D,E\right]
,B,A\right]  $, which, when totally antisymmetrized, gives an overall common
coefficient of $2!3!$ multiplying%
\begin{align}
&  \left[  \left[  C,D,E\right]  ,B,A\right]  +\left[  \left[  C,B,A\right]
,D,E\right]  +\left[  \left[  B,D,A\right]  ,C,E\right]  +\left[  \left[
B,A,E\right]  ,C,D\right] \nonumber\\
&  -\left[  \left[  B,D,E\right]  ,C,A\right]  -\left[  \left[  A,D,E\right]
,B,C\right]  -\left[  \left[  C,B,E\right]  ,D,A\right] \nonumber\\
&  -\left[  \left[  C,A,E\right]  ,B,D\right]  -\left[  \left[  C,D,B\right]
,E,A\right]  -\left[  \left[  C,D,A\right]  ,B,E\right] \nonumber\\
&  =-3\left[  A,B,C,D,E\right]  \;. \label{4thTerm}%
\end{align}
Adding (\ref{1stTerm}), (\ref{2ndTerm}), (\ref{3rdTerm}), and (\ref{4thTerm})
leads to the sum of QNB combinations that corresponds to the antisymmetrized
form of the RHS of (\ref{simpleFI3}), namely
\begin{gather}
{\Large (}\left[  \left[  A,B,C\right]  ,D,E\right]  -\left[  \left[
A,D,E\right]  ,B,C\right]  -\left[  A,\left[  B,D,E\right]  ,C\right]
-\left[  A,B,\left[  C,D,E\right]  \right]  {\Large )}\nonumber\\
\pm\text{(nine distinct permutations of all four terms)}=-6\left[
A,B,C,D,E\right]  \;.
\end{gather}
This result shows that the simple combination of QNB terms that corresponds to
(\ref{simpleFI3}) (without full antisymmetrization) cannot possibly vanish
unless the 5-bracket $\left[  A,B,C,D,E\right]  $ vanishes.

A similar consideration of the action of a 4-bracket on a 4-bracket
illustrates the general form of the GJI for even brackets and shows the
essential differences between the even and odd bracket cases. \ We proceed as
above by starting with the combination $\left[  \left[  A,B,C,D\right]
,E,F,G\right]  ,$\ and then totally antisymmetrizing with respect to
$A,B,C,D,E,F,$ and $G$. \ We find $35=7!/\left(  3!4!\right)  $ distinct terms
in the resulting sum. \ Now we determine the coefficient of any given monomial
that would appear in this sum. \ Since the expression is again totally
antisymmetrized in all the seven elements, the result must be proportional to
$\left[  A,B,C,D,E,F,G\right]  $. \ To determine the constant of
proportionality it suffices to consider the monomial $ABCDEFG$ and use the
shifting bracket argument, which shows that this particular monomial can be
found in only 4\ terms out of the 35 in the sum, namely in
\begin{equation}
\left[  \left[  A,B,C,D\right]  ,E,F,G\right]  \;,\;\;\;-\left[  A,\left[
B,C,D,E\right]  ,F,G\right]  \;,\;\;\;\left[  A,B,\left[  C,D,E,F\right]
,G\right]  \;,\;\text{and}\;\;\;-\left[  A,B,C,\left[  D,E,F,G\right]
\right]  \;. \label{TheEvenCycle}%
\end{equation}
The monomial $ABCDEFG$ appears in these 4 terms with coefficients $+1,-1,+1,$
and $-1,$ for a total of $0\times ABCDEFG$. \ Thus we conclude that
(\cite{Hanlon,Azcarraga96a}, and also \cite{Azcarraga97}, especially Eqn(32))%
\begin{equation}
\left[  \left[  A,B,C,D\right]  ,E,F,G\right]  \pm\text{(34 distinct
permutations)}=0\;.
\end{equation}
This is the prototypical generalization of the Jacobi identity for even QNBs,
and constitutes the full antisymmetrization of all arguments of the analogous
FI. \ There is no RHS inhomogeneity in this case.\ 

\begin{center}
\textit{The totally antisymmetrized action of even }$n$\textit{ QNBs on other
even }$n$\textit{ QNBs results in zero.}
\end{center}

The generalized Jacobi identity for arbitrary $n$-brackets follows from the
same simple analysis of coefficients of any given monomial, as in
(\ref{TheOddCycle}) and (\ref{TheEvenCycle}). \ The shifting bracket argument
actually leads to a larger set of results, where the action of any bracket on
any other is totally antisymmetrized. \ We present that larger generalization
here, calling it the \emph{quantum Jacobi identity}, or QJI. \ The GJI is the
QJI for $k=n-1$.

\paragraph{\underline{QJIs for QNBs}}%

\begin{align}
&  \sum_{\left(  n+k\right)  !\;\text{perms }\sigma}\operatorname{sgn}\left(
\sigma\right)  \left[  \left[  A_{\sigma_{1}},\cdots,A_{\sigma_{n}}\right]
,A_{\sigma_{n+1}},\cdots,A_{\sigma_{n+k}}\right] \nonumber\\
&  =\left[  A_{1},\cdots,A_{n+k}\right]  \times n!k!\times\left\{
\begin{array}
[c]{c}%
\left(  k+1\right)  \text{ \ \ if }n\text{\ is odd}\\
\frac{1}{2}\left(  1+\left(  -1\right)  ^{k}\right)  \text{ \ \ if
}n\text{\ is even}%
\end{array}
\right.  \;. \label{QJI}%
\end{align}
This result is proven just by computing the coefficient of the $A_{1}\cdots
A_{n+k}$\ monomial using the shifting bracket argument as given previously to
establish (\ref{TheOddCycle}) and (\ref{TheEvenCycle}). $\ $Other arguments
leading to the same result may be found in \cite{Azcarraga96a,Azcarraga97}.

This is the quantum identity that most closely corresponds to the general
classical result (see the second talk under \cite{CurtrightZachos}, Eqn (28))
for any even $n$ and any odd $k$ (only $n=2N,$ $k=2N-1$ is the FI),%
\begin{gather}
\left\{  \left\{  A_{1},A_{2},\cdots,A_{n}\right\}  _{NB},B_{1},\cdots
,B_{k}\right\}  _{NB}-\sum_{j=1}^{n}\left\{  A_{1},\cdots,\left\{  A_{j}%
,B_{1},\cdots,B_{k}\right\}  _{NB},\cdots,A_{n}\right\}  _{NB}\nonumber\\
=\left\{  B_{1},\left\{  B_{2},\cdots,B_{k}\right\}  _{NB},A_{1},\cdots
,A_{n}\right\}  _{NB}-\cdots+\left\{  B_{k},\left\{  B_{1},\cdots
,B_{k-1}\right\}  _{NB},A_{1},\cdots,A_{n}\right\}  _{NB}\ .
\end{gather}
While this classical identity holds without requiring full antisymmetrization
over all exchanges of $A$s and $B$s, in contrast the quantum identity
\emph{must} be totally antisymmetrized if it is to be a consequence of only
the associativity of the underlying algebra of Hilbert space operators. \ Note
that the $n!k!$ on the RHS of (\ref{QJI}) may be replaced by just $1$ if we
sum \emph{only} over permutations in which the $A_{i\leq n}$ are interchanged
with the $A_{i>n}$ in $\left[  \left[  A_{\sigma_{1}},\cdots,A_{\sigma_{n}%
}\right]  ,A_{\sigma_{n+1}},\cdots,A_{\sigma_{n+k}}\right]  $, and ignore all
permutations of the $A_{1},A_{2},\cdots,A_{n}$ among themselves, and of the
$A_{n+1},\cdots,A_{n+k}$ among themselves.

There is an important specialization of the QJI result
\cite{Hanlon,Azcarraga96a}: \ For any even $n$ and any odd $k$%
\begin{equation}
\sum_{\left(  n+k\right)  !\;\text{perms}\;\sigma}\operatorname{sgn}\left(
\sigma\right)  \left[  \left[  A_{\sigma_{1}},\cdots,A_{\sigma_{n}}\right]
,A_{\sigma_{n+1}},\cdots,A_{\sigma_{n+k}}\right]  =0\;.
\end{equation}
In particular, when $k=n-1,$ for $n$ even, the vanishing RHS obtains. \ All
other $n$-not-even and/or $k$-not-odd cases of the QJI have the $\left[
A_{1},\cdots,A_{n+k}\right]  $ inhomogeneity on the RHS.

The QJI also permits us to give the correct form of the so-called fundamental
identities valid for all QNBs. \ We accordingly call these \emph{quantum
fundamental identities} (QFIs), and present them in their general form.

\paragraph{\underline{QFIs for QNBs}}%

\begin{align}
&  \sum_{\left(  n+k\right)  !\;\text{perms}\;\sigma}\operatorname{sgn}\left(
\sigma\right)  {\Large (}\left[  \left[  A_{\sigma_{1}},\cdots,A_{\sigma_{n}%
}\right]  ,A_{\sigma_{n+1}},\cdots,A_{\sigma_{n+k}}\right]  -\sum_{j=1}%
^{n}\left[  A_{\sigma_{1}},\cdots,\left[  A_{\sigma_{j}},A_{\sigma_{n+1}%
},\cdots,A_{\sigma_{n+k}}\right]  ,\cdots,A_{\sigma_{n}}\right]
{\Large )}\nonumber\\
&  =\left[  A_{1},\cdots,A_{n+k}\right]  \times n!k!\times\left\{
\begin{array}
[c]{c}%
0\text{ \ \ if }k\text{\ is odd}\\
\left(  1-n\right)  \left(  k+1\right)  \text{ \ \ if }k\text{\ is even and
}n\text{\ is odd}\\
\left(  1-n\left(  k+1\right)  \right)  \text{ \ \ if }k\text{\ is even and
}n\text{\ is even}%
\end{array}
\right.  \;.
\end{align}
Aside from the trivial case of $n=1$, the only way the RHS vanishes without
conditions on the full $\left(  n+k\right)  $-bracket is when $k$ is odd.
\ All $n>1$, even $k$ result in the $\left[  A_{1},\cdots,A_{n+k}\right]  $
inhomogeneity on the RHS.

Partial antisymmetrizations of the individual terms in the general QFI may
also be entertained. \ The result is to find more complicated inhomogeneities,
and does not seem to be very informative. \ At best these partial
antisymmetrizations show in a supplemental way how the fully antisymmetrized
results are obtained.

In certain isolated, special cases (cf. the $su\left(  2\right)  $ example of
the next section, for which $k=3$), the bracket effects of select $B$s can act
as a derivation (essentially because the k-bracket is equivalent, in its
effects, to a commutator). \ If that is the case, then the quantum version of
the simple identity in (\ref{simpleFI}) holds trivially. \ It is also possible
in principle for that simple identity to hold, again in very special
situations, if the quantum bracket is not a derivation, through various
cancellations among terms. \ As an aid to finding such peculiar situations, it
is useful to resolve the quantum correspondents of the terms in the classical
FI into the derivators introduced previously (\ref{DerivatorDefn}). \ From the
definition of $\left[  A_{1},\cdots,A_{n}\right]  $ in (\ref{QNBDefinition}),
and some straightforward manipulations, we find
\begin{align}
&  \left[  \left[  A_{1},\cdots,A_{n}\right]  ,\mathbf{B}\right]  -\sum
_{j=1}^{n}\left[  A_{1},\cdots,\left[  A_{j},\mathbf{B}\right]  ,\cdots
,A_{n}\right] \label{FIasDerivators}\\
&  =\sum_{n!\text{\ perms }\sigma}\operatorname{sgn}\left(  \sigma\right)
{\Large (}\frac{1}{\left(  n-1\right)  !}\left(  A_{\sigma_{1}},\left[
A_{\sigma_{2}},\cdots,A_{\sigma_{n}}\right]  \,|\,\mathbf{B}\right)  +\frac
{1}{\left(  n-2\right)  !}A_{\sigma_{1}}\left(  A_{\sigma_{2}},\left[
A_{\sigma_{3}},\cdots,A_{\sigma_{n}}\right]  \,|\,\mathbf{B}\right)
\nonumber\\
&  +\frac{1}{2!\left(  n-3\right)  !}\left[  A_{\sigma_{1}},A_{\sigma_{2}%
}\right]  \left(  A_{\sigma_{3}},\left[  A_{\sigma_{4}},\cdots,A_{\sigma_{n}%
}\right]  \,|\,\mathbf{B}\right)  +\cdots+\frac{1}{\left(  n-1\right)
!}\left[  A_{\sigma_{1}},A_{\sigma_{2}},\cdots,A_{\sigma_{n-2}}\right]
\left(  A_{\sigma_{n-1}},A_{\sigma_{n}}\,|\,\mathbf{B}\right)  {\Large )\;,}%
\nonumber
\end{align}
with the abbreviation $\mathbf{B=}B_{1},\cdots,B_{k}$. \ The terms on the RHS
are a sum over $j=1,\cdots,n-1$ of derivators between solitary $A$s (i.e.
$1$-brackets) and various $\left(  n-j\right)  $-brackets, left-multiplied by
complementary rank $\left(  j-1\right)  $-brackets. \ (There is a similar
identity that involves right-multiplication by the complementary brackets.) \ 

For example, suppose $n=2$. \ Then we have for any number of $B$s%
\begin{equation}
\left[  \left[  A_{1},A_{2}\right]  ,\mathbf{B}\right]  -\left[  \left[
A_{1},\mathbf{B}\right]  ,A_{2}\right]  -\left[  A_{1},\left[  A_{2}%
,\mathbf{B}\right]  \right]  =\left(  A_{1},A_{2}\,|\,\mathbf{B}\right)
-\left(  A_{2},A_{1}\,|\,\mathbf{B}\right)  \;.
\end{equation}
In principle, this can vanish, even when the action of the $B$s is not a
derivation, if the k-derivator is symmetric in the first two arguments. \ That
is, if $\left(  A_{1},A_{2}\,|\,B_{1},\cdots,B_{k}\right)  =\tfrac{1}%
{2}\left(  A_{1},A_{2}\,|\,B_{1},\cdots,B_{k}\right)  +\tfrac{1}{2}\left(
A_{2},A_{1}\,|\,B_{1},\cdots,B_{k}\right)  $. \ However, we have not found a
compelling (nontrivial) physical example where this is the case.

\subsection{Illustrative quantum examples}

As in the classical situation, it is useful to consider explicit examples of
quantized dynamical systems described by quantum Nambu brackets, to gain
insight and develop intuition. \ However, for quantum systems, it is more than
useful; it is crucial to examine detailed cases to appreciate how quandaries
that have been hinted at in the past are actually resolved, especially since
the exact classical phase-space geometry in \S 2.1 is no longer applicable.
\ Similar studies have been attempted before, but have reached conclusions
sharply opposed to ours\footnote{``The quantization of Nambu structures turns
out to be a non-trivial problem, even (or especially) in the simplest
cases.''\ \cite{DitoFlato}}. \ Here, we demonstrate how the simplest Nambu
mechanical systems are quantized consistently and elegantly by conventional
operator methods.

\paragraph{\underline{SU(2) as a special case}}

The commutator algebra of the charges ($L_{0}\equiv L_{z}\;,\;L_{\pm}\equiv
L_{x}\pm iL_{y}$) is
\begin{equation}
\left[  L_{+},L_{-}\right]  =2\hbar L_{0}\;,\;\;\;\left[  L_{0},L_{-}\right]
=-\hbar L_{-}\;,\;\;\;\left[  L_{0},L_{+}\right]  =\hbar L_{+}\;,
\end{equation}
giving rise to $\left[  L_{-},L_{0}^{2}\right]  =\left\{  \left[  L_{-}%
,L_{0}\right]  ,L_{0}\right\}  =\hbar\left\{  L_{-},L_{0}\right\}  ,\;$etc.
\ The invariant quadratic Casimir is%
\begin{equation}
I=L_{+}L_{-}+L_{0}\left(  L_{0}-\hbar\right)  =L_{-}L_{+}+\left(  L_{0}%
+\hbar\right)  L_{0}\;.
\end{equation}
We use the algebra and the commutator resolution of the 4-bracket
\begin{equation}
\left[  A,B,C,D\right]  =\left\{  \left[  A,B\right]  ,\left[  C,D\right]
\right\}  -\left\{  \left[  A,C\right]  ,\left[  B,D\right]  \right\}
-\left\{  \left[  A,D\right]  ,\left[  C,B\right]  \right\}  \;,
\end{equation}
to obtain \cite{CurtrightZachos} the quantization of
(\ref{so(3)ClassicalBeauty}):
\begin{equation}
\left[  A,L_{0},L_{+},L_{-}\right]  =2\hbar\left[  A,I\right]  \;,
\label{su(2)QuantumBeauty}%
\end{equation}
and the more elaborate%
\begin{equation}
\left[  A,I,L_{+},L_{-}\right]  =2\hbar\left\{  \left[  A,I\right]
,L_{0}\right\}  =2\hbar\left[  \left\{  A,L_{0}\right\}  ,I\right]  \;.
\label{su(2)Entwined}%
\end{equation}
Since $I$ and $L_{0}$ commute, the nested commutator-anticommutator can also
be written using a 3-bracket: $\ \left[  \left\{  A,L_{0}\right\}  ,I\right]
=\left[  A,I,L_{0}\right]  -\left[  L_{0}I,A\right]  .$

So for $SU\left(  2\right)  $ invariant systems with $H=I/2$,
(\ref{su(2)QuantumBeauty}) gives rise to the complete analog of classical time
development as a derivation, (\ref{so(3)ClassicalBeauty}), namely
\begin{equation}
i\hbar^{2}\frac{dA}{dt}=\hbar\left[  A,H\right]  =\frac{1}{4}\left[
A,L_{0},L_{+},L_{-}\right]  \;, \label{4QNBDerivation}%
\end{equation}
where the QNB in question happens to be a derivation too
\cite{CurtrightZachos}, and thus satisfies an effective FI (see (\ref{EFI})).
\ By contrast, (\ref{su(2)Entwined}) gives rise to%
\begin{equation}
i\hbar^{2}\left\{  \frac{dA}{dt},L_{0}\right\}  =\hbar\left\{  \left[
A,H\right]  ,L_{0}\right\}  =\frac{1}{4}\left[  A,I,L_{+},L_{-}\right]  \;.
\label{4QNBEntwinedDerivation}%
\end{equation}
Since the latter of these is manifestly not a derivation, one should not
expect, as we have stressed, Leibniz rule and classical-like fundamental
identities to hold. \ Of course, since a derivation is entwined in the
structure, substitution $A\rightarrow A\mathcal{A}$ and application of
Leibniz's rule to just the time derivation alone will necessarily yield
correct but complicated expressions (but not particularly informative results,
given the consistency of the structure already established). However, in more
general contexts, and in fanciful situations where the operators playing the
role of $L_{0}$ were invertible, one might envision trying to solve for the
time derivative through formal resolvent methods (see the Appendix), e.g.%
\begin{equation}
4i\hbar^{2}\frac{dA}{dt}=\sum_{n=0}^{\infty}\left(  -L_{0}\right)
^{n}\,\left[  A,I,L_{+},L_{-}\right]  \,\left(  L_{0}\right)  ^{-n-1}\;,
\end{equation}
where the new bracket implicitly defined by the RHS would now be a derivation
(here, just the commutator with $I$). \ It is clear from our discussion,
however, that the QNBs and the entwinings they imply, are still the preferred
presentation of the quantized Nambu mechanics.

It might be useful to view solving for $dA/dt$ as a problem of implementing
scale transformations in the generalized Jordan algebra context. \ Evidently
there is little if any discussion of that mathematical problem in the
literature \cite{Ayupov}. \ 

The physics described by the first QNB, (\ref{4QNBDerivation}), is standard
time evolution, just encoded in an unusual way as a quantum 4-bracket.
\ However, the other QNB, in (\ref{4QNBEntwinedDerivation}), illustrates the
idea discussed in the Introduction. \ Physically, this Nambu bracket is an
entwined form of time evolution, where the Jordan algebraic eigenvalues
$\sigma$ of $L_{0}$, defined by $\left\{  A,L_{0}\right\}  =\sigma A$, set the
time scales for the various sectors of the theory: \ i.e. the formalism gives
dynamical time scales. \ To see this, resolve the identity, $1=\sum_{\lambda
}\mathbb{P}_{\lambda}$, in terms of $L_{0}$ projections, $\mathbb{P}_{\lambda
}^{2}=\mathbb{P}_{\lambda}$, and use this in turn to resolve any operator $A$
as a sum of left- and right-eigenoperators of $L_{0}$,
\begin{equation}
A=\sum_{\lambda,\rho}\mathbb{P}_{\lambda}A_{\lambda\rho}\mathbb{P}_{\rho
}\;,\;\;\;A_{\lambda\rho}=\mathbb{P}_{\lambda}A\mathbb{P}_{\rho}\;.
\label{TheProjects}%
\end{equation}
These eigenoperators obey%
\begin{equation}
L_{0}A_{\lambda\rho}=\lambda A_{\lambda\rho}\;,\;\;\;A_{\lambda\rho}L_{0}=\rho
A_{\lambda\rho}\;.
\end{equation}
Following such a decomposition, since $\left[  \mathbb{P}_{\lambda},H\right]
=0=\left[  \mathbb{P}_{\lambda},L_{0}\right]  $, (\ref{4QNBEntwinedDerivation}%
) can be written as a sum of terms%
\begin{equation}
i\hbar^{2}\left\{  \frac{dA_{\lambda\rho}}{dt},L_{0}\right\}  =\hbar\left\{
\left[  A_{\lambda\rho},H\right]  ,L_{0}\right\}  =\hbar\left[  \left\{
A_{\lambda\rho},L_{0}\right\}  ,H\right]  =i\hbar^{2}\left(  \lambda
+\rho\right)  \frac{dA_{\lambda\rho}}{dt}\;.
\end{equation}
That is to say, the sum of the left- and right-eigenvalues of the operator
$A_{\lambda\rho}$ gives a Jordan eigenvalue $\sigma=\lambda+\rho$,%
\begin{equation}
\left\{  A_{\lambda\rho},L_{0}\right\}  =\left(  \lambda+\rho\right)
A_{\lambda\rho}\;,
\end{equation}
and this Jordan eigenvalue sets the time scale for the instantaneous evolution
of the eigenoperator. \ Since a general operator is a sum of eigenoperators,
this construction will in general give a mixture of time scales. \ Stated
precisely:
\begin{equation}
i\hbar^{2}\left\{  \frac{dA}{dt},L_{0}\right\}  =i\hbar^{2}\sum_{\lambda,\rho
}\left(  \lambda+\rho\right)  \mathbb{P}_{\lambda}\frac{dA_{\lambda\rho}}%
{dt}\mathbb{P}_{\rho}=\tfrac{1}{4}\left[  A,I,L_{+},L_{-}\right]  \;.
\label{JordanTimeScales}%
\end{equation}
This highlights the differences between (\ref{4QNBEntwinedDerivation}) and
conventional time evolution for operators in the Heisenberg picture, as in
(\ref{4QNBDerivation}), since the time scales in (\ref{JordanTimeScales})
depend on the angular momentum eigenvalues. \ 

This example illustrates our introductory remarks about different time scales
for the different invariant sectors of a system. \ It also shows why the
action of the 4-bracket in question is not a derivation. \ The simple Leibniz
rule for generic $A$ and $\mathcal{A}$, that would equate $\left[
A\mathcal{A},I,L_{+},L_{-}\right]  $ with $A\left[  \mathcal{A},I,L_{+}%
,L_{-}\right]  $ $+$ $\left[  A,I,L_{+},L_{-}\right]  \mathcal{A}$, will fail
for (non-vanishing) products $A_{\lambda_{1}\mu}\mathcal{A}_{\mu\rho_{2}},$
unless $\lambda_{1}+\mu=\mu+\rho_{2}=\lambda_{1}+\rho_{2}$, i.e. unless
$\lambda_{1}=\mu=\rho_{2}$. \ If restricted and applied to a single angular
momentum sector, the 4-bracket under consideration does indeed give just the
time derivative of all diagonal (i.e. those with $\lambda=\rho$) angular
momentum eigenoperators on that sector. \ When projected onto any such sector,
the action of this 4-bracket is therefore a derivation, since the time scale
will be fixed. \ But without such a projection, when acting on the full
Hilbert space of the system, where more than one value of $\lambda$ is
encountered, this 4-bracket is \emph{not} a simple derivation, not even if it
acts on only diagonal $L$ eigenoperators, if two or more $L_{0}$ Jordan
eigenvalues are involved. \ At best, we may think of it as some sort of
dynamically scaled derivation, since it gives time derivatives scaled by
angular momentum eigenvalues.

The quantum bracket in (\ref{4QNBEntwinedDerivation}) has a kernel, just as
its classical limit does, but the quantum case evinces the linear
superpositions inherent in quantum mechanics. \ This is evident in
(\ref{JordanTimeScales}), where a given eigenoperator is left unchanged by the
bracket if $\lambda=-\rho$, rather than simply $\lambda=0=\rho$. \ This
quantum effect is linked to the fact Jordan algebras are not division rings,
as discussed in the Appendix. \ For this and other reasons, having to do with
the fact that the 4-bracket in (\ref{4QNBEntwinedDerivation}) is not a
derivation, it is not possible to simply divide the LHS of
(\ref{4QNBEntwinedDerivation}) by $L_{0}$ and then absorb the $1/L_{0}$ on the
RHS directly into the bracket, as we did in previous classical cases, such as
(\ref{TakingLogs}). \ While the result in (\ref{4QNBDerivation}) does indeed
have the expected form that such naive manipulations would produce (such
manipulations being valid for the classical limits of the expressions), that
result cannot be derived in this way. \ It is legitimately obtained only
through the commutator resolution, as above.

Other choices for the invariants in the 4-bracket lead to some even more
surprising results, and offer additional insight into the quantum tricks that
the QNBs are capable of playing. \ For example,%
\begin{align}
\left[  A,L_{0}^{2},L_{+},L_{-}\right]   &  =2\hbar\left\{  \left[
A,L_{0}^{2}\right]  ,L_{0}\right\}  +\hbar\left\{  \left[  A,L_{+}\right]
,\left\{  L_{-},L_{0}\right\}  \right\}  +\hbar\left\{  \left[  A,L_{-}%
\right]  ,\left\{  L_{+},L_{0}\right\}  \right\}  \;\nonumber\\
&  =2\hbar\left\{  \left[  A,I\right]  ,L_{0}\right\}  +2\hbar^{3}\left[
L_{0},A\right]  +\hbar\left[  L_{-},\left[  L_{+},\left[  A,L_{0}\right]
\right]  \right]  +\hbar\left[  L_{+},\left[  L_{-},\left[  A,L_{0}\right]
\right]  \right]  \;. \label{L0^2}%
\end{align}
This shows an interesting effect in addition to the dynamical time scales
evident in the first term of the last line, namely: \ \emph{quantum rotation
terms}, as given by the last three $O\left(  \hbar^{4}\right)  $ terms in that
last line. \ If $A$ is not invariant under rotations about the polar axis, so
that $\left[  L_{0},A\right]  \neq0$, the last three terms in (\ref{L0^2}%
)\ may generate changes in $A$, even though $A$ is time-invariant. \ The
effect is a purely quantum one: \ It disappears completely in the classical
limit. \ The QNB in question is algebraically covariant, but not algebraically
invariant (as opposed to \emph{time} invariant). \ Therefore, it may and does
lead to nontrivial tensor products when it acts on other algebraically
covariant $A$s. \ (As a general rule-of-thumb, if the QNB is allowed to do
something, then it will.)

Mathematically, it is sometimes useful to think of nested multi-commutators,
such as those in (\ref{L0^2}), as higher partial derivatives. \ This
manifestation provides another reason why general QNBs are not derivations
(i.e. first derivatives only) and do not obey the simple Leibniz rule.

Combining (\ref{L0^2}) with (\ref{su(2)Entwined}) also yields
\begin{equation}
\left[  A,\left(  L_{+}L_{-}\right)  ,L_{+},L_{-}\right]  =2\hbar^{2}\left[
A,I\right]  -2\hbar^{3}\left[  L_{0},A\right]  -\hbar\left[  L_{-},\left[
L_{+},\left[  A,L_{0}\right]  \right]  \right]  -\hbar\left[  L_{+},\left[
L_{-},\left[  A,L_{0}\right]  \right]  \right]  \;. \label{L+L-}%
\end{equation}
Since every commutator is inherently $O\left(  \hbar\right)  $, the first term
on the RHS of this last result is $O\left(  \hbar^{3}\right)  $, while the
last three are $O\left(  \hbar^{4}\right)  $. \ All vanish in the classical
limit,
\begin{equation}
\lim_{\hbar\rightarrow0}\left(  \frac{1}{i\hbar}\right)  ^{2}\left[  A,\left(
L_{+}L_{-}\right)  ,L_{+},L_{-}\right]  =0\;.
\end{equation}
This illustrates how nontrivial QNBs can collapse to nothing as CNBs. \ The
$O\left(  \hbar^{4}\right)  $ terms are quantum rotations, as in (\ref{L0^2}),
but with changed signs.

The two results (\ref{L0^2}) and (\ref{L+L-}) are simple illustrations of the
failure of QNBs to obey the elementary Leibniz rule. \ As derivators,
\begin{equation}
\left(  L_{0},L_{0}\,|\,L_{+},A,L_{-}\right)  =2\hbar^{3}\left[
L_{0},A\right]  +\hbar\left[  L_{-},\left[  L_{+},\left[  A,L_{0}\right]
\right]  \right]  +\hbar\left[  L_{+},\left[  L_{-},\left[  A,L_{0}\right]
\right]  \right]  \;,
\end{equation}
whose RHS is $O\left(  \hbar^{4}\right)  $ and thus vanishes in the classical
limit as $O\left(  \hbar^{2}\right)  $, and
\begin{equation}
\left(  L_{+},L_{-}\,|\,L_{+},A,L_{-}\right)  =2\hbar^{2}\left[  A,I\right]
-2\hbar^{3}\left[  L_{0},A\right]  -\hbar\left[  L_{-},\left[  L_{+},\left[
A,L_{0}\right]  \right]  \right]  -\hbar\left[  L_{+},\left[  L_{-},\left[
A,L_{0}\right]  \right]  \right]  \;,
\end{equation}
whose RHS is inherently $O\left(  \hbar^{3}\right)  $, due to the $\left[
A,I\right]  $ term, and also vanishes in the classical limit\footnote{If $f$
is a function in the $su\left(  2\right)  $ enveloping algebra, then $\left[
f,I\right]  =0,$ and $\left(  J_{+},J_{-};J_{+},f,J_{-}\right)  $ is again
inherently $O\left(  \hbar^{4}\right)  $.}. \ The LHS of either of these
expressions would vanish identically were the 4-brackets involved actually
derivations, but they are not. \ There are some special situations, such as
when $A$ is in the enveloping algebra and is invariant under rotations about
the polar axis, i.e. $A=A\left(  L_{0},\left(  L_{+}L_{-}\right)  \right)  $,
for which the derivators do vanish. \ However, for general $A$, including most
of those in the enveloping algebra which are not polar invariants, these
derivators do not vanish, and so in general the QNBs in (\ref{L0^2}) and
(\ref{L+L-})\ are not derivations.

Another simple example of a nontrivial QNB, with a trivial classical limit, is%
\begin{equation}
\left[  L_{0}L_{+},L_{0}^{2},L_{+},L_{-}\right]  =-4\hbar^{3}\left(
2L_{0}-\hbar\right)  L_{+}\;.
\end{equation}
This would again vanish were the generic 4-bracket a derivation, but as
already stressed, it is not. \ The corresponding $^{4}\mathbf{\Delta}$ here
gives
\begin{equation}
2\left(  L_{0},L_{+}\,|\,L_{0}^{2},L_{+},L_{-}\right)  =\left[  \left\{
L_{0},L_{+}\right\}  ,L_{0}^{2},L_{+},L_{-}\right]  =-8\hbar^{3}\left\{
L_{0},L_{+}\right\}  \;, \label{L0L+Derivator}%
\end{equation}
a purely quantum effect for 4-brackets. \ Its classical limit is
\begin{equation}
\lim_{\hbar\rightarrow0}\frac{1}{\hbar^{2}}\left[  \left\{  L_{0}%
,L_{+}\right\}  ,L_{0}^{2},L_{+},L_{-}\right]  =-8\lim_{\hbar\rightarrow
0}\hbar\left\{  L_{0},L_{+}\right\}  =0\;.
\end{equation}
The RHS has one too many powers of $\hbar$ to contribute classically. \ A
class of such results, evocative of those found in deformed Lie algebras, is
given by%
\begin{align}
\left(  g\left(  L_{0}\right)  ,L_{+}\,|\,L_{0}^{2},L_{+},L_{-}\right)   &
=2\hbar L_{+}\,\left(  I+\hbar L_{0}-L_{0}^{2}\right)  \,\left(  g\left(
L_{0}+2\hbar\right)  -2g\left(  L_{0}+\hbar\right)  +g\left(  L_{0}\right)
\right) \nonumber\\
&  -4\hbar^{2}L_{+}\,\left(  \hbar+2L_{0}\right)  \,\left(  g\left(
L_{0}+2\hbar\right)  -g\left(  L_{0}+\hbar\right)  \right)  \;.
\end{align}
The choice $g\left(  L_{0}\right)  =L_{0}$ reduces to the particular case in
(\ref{L0L+Derivator}).

\paragraph{\underline{4-bracket sums for any Lie algebra}}

How does the quantum 4-bracket method extend to other examples, perhaps even
to models that are not integrable? \ In complete parallel with the classical
example, any Lie algebra will allow a commutator with a quadratic Casimir to
be rewritten as a sum of 4-brackets. \ Suppose
\begin{equation}
\left[  Q_{a},Q_{b}\right]  =i\hbar f_{abc}Q_{c}\;,
\end{equation}
in a basis where $f_{abc}$ is totally antisymmetric. \ Then, through the use
of the commutator resolution (\ref{4QNBasCommutators}), for a
structure-constant-weighted sum of quantum 4-brackets, we find
\begin{equation}
f_{abc}\left[  A,Q_{a},Q_{b},Q_{c}\right]  =3f_{abc}\left\{  \left[
A,Q_{a}\right]  ,\left[  Q_{b},Q_{c}\right]  \right\}  =3i\hbar f_{abc}%
f_{bcd}\left\{  \left[  A,Q_{a}\right]  ,Q_{d}\right\}  \;.
\end{equation}
Again, for simple Lie algebras, use (\ref{CAdjoint}) to obtain a commutator
with the Casimir $Q_{a}Q_{a}$.%
\begin{equation}
3i\hbar\,c_{\text{adjoint}}\,\left\{  \left[  A,Q_{a}\right]  ,Q_{a}\right\}
=3i\hbar\,c_{\text{adjoint}}\,\left[  A,Q_{a}Q_{a}\right]  \;.
\end{equation}
Thus we obtain the quantization of the classical result in (\ref{AnyLie4CNB}%
),
\begin{equation}
f_{abc}\left[  A,Q_{a},Q_{b},Q_{c}\right]  =3i\hbar\,c_{\text{adjoint}%
}\,\left[  A,Q_{a}Q_{a}\right]  \;. \label{AnyLie4QNB}%
\end{equation}
This becomes (\ref{AnyLie4CNB})\ in the classical limit, with an appropriate
factor of $2$ included, as in (\ref{4QNBClassicalLimit}). \ A slightly
different route to this result is to use the left- and right-sided resolutions
of the 4-bracket into 3-brackets, (\ref{QNB4}), and then note that the
trilinear invariant reduces to the quadratic Casimir.
\begin{equation}
f_{abc}\left[  Q_{a},Q_{b},Q_{c}\right]  =3i\hbar\,c_{\text{adjoint}}%
\,Q_{a}Q_{a}\;.
\end{equation}

Thus, as in (\ref{4QNBDerivation}), this particular linear combination of
quantum 4-brackets acts as a derivation. \ As a corollary, we have the
4-bracket Effective Fundamental Identity (EFI) \cite{CurtrightZachos}, i.e.
one with three of the entries being related by a Lie algebra.%
\begin{align}
f_{abc}\left[  Q_{a},Q_{b},Q_{c},\left[  A,B,\cdots,D\right]  \right]   &
=f_{abc}\left[  \left[  Q_{a},Q_{b},Q_{c},A\right]  ,B,\cdots,D\right]
\nonumber\\
&  +f_{abc}\left[  A,\left[  Q_{a},Q_{b},Q_{c},B\right]  ,\cdots,D\right]
\nonumber\\
&  +\cdots+f_{abc}\left[  A,B,\cdots,\left[  Q_{a},Q_{b},Q_{c},D\right]
\right]  \;. \label{EFI}%
\end{align}
By using (\ref{AnyLie4QNB}), all models with dynamics based on simple Lie
algebras with $H=\frac{1}{2}Q_{a}Q_{a}$ can be quantized through the use of
summed quantum 4-brackets to describe their time evolution as a
derivation:\ \
\begin{equation}
i\hbar^{2}\frac{dA}{dt}=\hbar\left[  A,H\right]  =\frac{1}%
{6i\,c_{\text{adjoint}}}\;f_{abc}\left[  A,Q_{a},Q_{b},Q_{c}\right]  \;.
\label{dA/dtAnyLie}%
\end{equation}
This special combination of\ sums of 4-brackets leads to an exception to the
generic QNB feature of dynamically scaled time derivatives. \ It shows that
QNBs can be used to describe conventional time evolution for many systems, not
only those that are superintegrable or integrable.

\paragraph{\underline{U(n) and isotropic quantum oscillators}}

The previous results on Nambu-Jordan-Lie algebras can be applied to harmonic
oscillators. \ For the isotropic oscillator the NJL approach quickly leads to
a compact result. \ A set of operators can be chosen that produces\ only one
term in the sum of Jordan-Kurosh products.

Consider the $n$ dimensional oscillator using the standard raising/lowering
operator basis, but normalized in a way ($\sqrt{2}a\equiv x+ip,\;\sqrt
{2}b\equiv x-ip$) that makes the classical limit more transparent:
\begin{equation}
\left[  a_{i},b_{j}\right]  =\hbar\delta_{ij}\;,\;\;\;\left[  a_{i}%
,a_{j}\right]  =0=\left[  b_{i},b_{j}\right]  \;. \label{u(n)raise&lower}%
\end{equation}
Construct the usual bilinear charges that realize the $U\left(  n\right)  $
algebra:
\begin{equation}
N_{ij}\equiv b_{i}a_{j}\;,\;\;\;\left[  N_{ij},N_{kl}\right]  =\hbar\left(
N_{il}\delta_{jk}-N_{kj}\delta_{il}\right)  \;. \label{u(n)NumberAlgebra}%
\end{equation}
Then the isotropic Hamiltonian is
\begin{equation}
H=\omega\sum_{i=1}^{n}\left(  N_{i}+\tfrac{1}{2}\hbar\right)  \;,\;\;\;N_{i}%
\equiv N_{ii}\;,
\end{equation}
which gives the $n^{2}$ conservation laws
\begin{equation}
\left[  H,N_{ij}\right]  =0\;. \label{HConservation}%
\end{equation}
Consideration of the isotropic oscillator dynamics using QNBs yields the main
result for oscillator $2n$-brackets.

\paragraph{\underline{Isotropic oscillator quantum brackets}}

(the $U(n)$ quantum reductio ad dimidium) \ Let $N=N_{1}+N_{2}+\cdots+N_{n}$,
and intercalate the $n-1$ non-diagonal operators $N_{i\;i+1}$, for
$i=1,\cdots,n-1$, into a $2n$-bracket with the $n$ mutually commuting $N_{j}$,
for $j=1,\cdots,n$, along with an arbitrary $A$ to find\footnote{Analogously
to the classical case, the non-diagonal charges are not hermitian. \ But the
proof leading to (\ref{TLCresult}) also goes through if non-diagonal charges
have their subscripts transposed. \ This allows replacing $N_{i\;i+1}$ with
hermitian or anti-hermitian combinations $N_{i\;i+1}\pm N_{i+1\;i}$ in the LHS
$2n$-bracket, to obtain the alternative linear combinations $N_{i\;i+1}\mp
N_{i+1\;i}$ in the GJP on the RHS of (\ref{TLCresult}).}
\begin{align}
\left[  A,N_{1},N_{12},N_{2},N_{23},\cdots,N_{n-1},N_{n-1\;n},N_{n}\right]
&  =\hbar^{n-1}\left\{  \left[  A,N\right]  ,N_{12},N_{23},\cdots
,N_{n-1\;n}\right\} \nonumber\\
&  =\hbar^{n-1}\left[  \left\{  A,N_{12},N_{23},\cdots,N_{n-1\;n}\right\}
,N\right]  \;. \label{TLCresult}%
\end{align}
This result shows that the QNB on the LHS will indeed vanish, not just when
$A$ is one of the $2n-1$ charges listed along with $A$ in the bracket, but
also if $A$ is one of the remaining $U\left(  n\right)  $ charges, by virture
of the explicit commutator with $N=H/\omega$ appearing on the RHS of
(\ref{TLCresult}) and (\ref{HConservation}). \ The classical limit of
(\ref{TLCresult}) is (\ref{classicalTLCresult}), of course.

The non-diagonal operators $N_{i\;i+1}$ do not all commute among themselves,
nor with all the $N_{j}$, but their non-Abelian properties are encountered in
the above Jordan and Nambu products in a minimal way. \ Only adjacent entries
in the list $N_{12},N_{23},N_{34},\cdots,N_{n-1\;n}$ fail to commute. \ Also
in the list of $2n-1$ generators within the original QNB, each $N_{j}$ fails
to commute only with the adjacent $N_{j-1\;j}$ and $N_{j\;j+1}$. \ Such a list
of invariants constitutes a ``Hamiltonian path'' through the
algebra\footnote{There are other Hamiltonian paths through the algebra. \ As
previously mentioned in the case of CNBs, a different set of $2n-1$ invariants
which leads to an equivalent reductio ad dimidium can be obtained just by
taking an arbitrarily ordered list of the mutually commuting $N_{i}$, and then
intercalating non-diagonal generators to match adjacent indices on the $N_{i}%
$. \ That is, for any permutation of the indices $\left\{  \sigma_{1}%
,\cdots,\sigma_{n}\right\}  $, we have: \ $\left[  A,N_{\sigma_{1}}%
,N_{\sigma_{1}\sigma_{2}},N_{\sigma_{2}},N_{\sigma_{2}\sigma_{3}}%
,N_{\sigma_{3}},\cdots,N_{\sigma_{n-1}},N_{\sigma_{n-1}\;\sigma_{n}}%
,N_{\sigma_{n}}\right]  =\hbar^{n-1}\left\{  \left[  A,N\right]
,N_{\sigma_{1}\sigma_{2}},N_{\sigma_{2}\sigma_{3}},\cdots,N_{\sigma
_{n-1}\;\sigma_{n}}\right\}  =\hbar^{n-1}\left[  \left\{  f,N_{\sigma
_{1}\sigma_{2}},N_{\sigma_{2}\sigma_{3}},\cdots,N_{\sigma_{n-1}\;\sigma_{n}%
}\right\}  ,N\right]  \;$.}.

\noindent\textbf{Proof: \ }Linearity in each argument and total antisymmetry
of the Nambu bracket allow us to replace any one of the $N_{i}$ by the sum
$N$. \ Replace $N_{n}\rightarrow N$, hence obtain
\begin{equation}
\left[  A,N_{1},N_{12},N_{2},\cdots,N_{n-1},N_{n-1\;n},N_{n}\right]  =\left[
A,N_{1},N_{12},N_{2},\cdots,N_{n-1},N_{n-1\;n},N\right]  \;.
\end{equation}
Now, since $\left[  N,N_{ij}\right]  =0$, the commutator resolution of the
$2n$-bracket implies that $N$ must appear ``locked'' in a commutator with $A$,
and therefore $A$ cannot appear in any other commutator. \ But then $N_{1}$
commutes with all the remaining free $N_{ij}$ except $N_{12}$. \ So $N_{1}$
must be locked in $\left[  N_{1},N_{12}\right]  $. \ Continuing in this way,
$N_{2}$ must be locked in $\left[  N_{2},N_{23}\right]  $, etc., until finally
$N_{n-1}$ is locked in $\left[  N_{n-1},N_{n-1\;n}\right]  $. \ Thus all $2n$
entries have been paired and locked in the indicated $n$ commutators, i.e.
they are all ``zipped-up''. \ Moreover, these $n$ commutators can and will
appear as products ordered in all $n!$ possible ways with coefficients $+1$
since interchanging a pair of commutators requires interchanging two pairs of
the original entries in the bracket. \ We conclude that
\begin{equation}
\left[  A,N_{1},N_{12},N_{2},\cdots,N_{n-1},N_{n-1\;n},N_{n}\right]  =\left\{
\left[  A,N\right]  ,\left[  N_{1},N_{12},\right]  ,\cdots,\left[
N_{n-1},N_{n-1\;n}\right]  \right\}  \;.
\end{equation}
Now all the paired $N_{ij}$ commutators evaluate as $\left[  N_{i-1}%
,N_{i-1\;i}\right]  =\hbar N_{i-1\;i}$, so we have
\begin{equation}
\left[  A,N_{1},N_{12},N_{2},\cdots,N_{n-1},N_{n-1\;n},N_{n}\right]
=\hbar^{n-1}\left\{  \left[  A,N\right]  ,N_{12},\cdots,N_{n-1\;n}\right\}
\;.
\end{equation}
Finally the commutator with $N$ may be performed either before or after the
Jordan product of $A$ with all the $N_{i-1\;i}$, since again $\left[
N,N_{ij}\right]  =0$. \ Hence
\begin{equation}
\left\{  \left[  A,N\right]  ,N_{12},\cdots,N_{n-1\;n}\right\}  =\left[
\left\{  A,N_{12},\cdots,N_{n-1\;n}\right\}  ,N\right]
\;.\;\;\;\text{{\small QED}}%
\end{equation}
In analogy with the classical situation, the quantum invariants which are in
involution (i.e. the Cartan subalgebra of $U\left(  n\right)  $) are separated
out of the QNB into a single commutator involving their sum, the oscillator
Hamiltonian, while the invariants which do not commute ($n-1$ of them,
corresponding in number to the rank of $SU\left(  n\right)  $) are swept into
a generalized Jordan product. \ Thus, we have been led to a more complicated
Jordan-Kurosh eigenvalue problem for $U\left(  n\right)  $ invariant dynamics,
as the entwined effect of several mutually non-commuting $N_{ij}$s. \ The
individual $N_{ij}$\ may not be diagonalized simultaneously, so it may not be
obvious what the general formalism of projection operators will lead to in
this case, as compared to (\ref{TheProjects}) et seq., but in fact it can be
carried through by rearranging the terms in the Jordan product, as we shall explain.

Our QNB description of time evolution for isotropic quantum oscillators
therefore becomes
\begin{equation}
\omega\left[  A,N_{1},N_{12},N_{2},\cdots,N_{n-1},N_{n-1\;n},N_{n}\right]
=i\hbar^{n}\left\{  \frac{dA}{dt},N_{12},\cdots,N_{n-1\;n}\right\}  \;,
\label{OscillatorEntwinedTimeDerivative}%
\end{equation}
whose classical limit is precisely (\ref{u(n)CNBdA/dt}). \ This result encodes
both dynamical time scales and, in higher orders of $\hbar$, group rotation
terms, as a consequence of the generalized Jordan eigenvalue problem involving
noncommuting elements, $N_{12},\cdots,N_{n-1\;n}$. \ 

The specific oscillator realization of $U\left(  3\right)  $ explicates this
last point. \ The RHS of (\ref{OscillatorEntwinedTimeDerivative}) becomes%
\begin{equation}
\left\{  \frac{dA}{dt},N_{12},N_{23}\right\}  =3\left\{  \left(  N_{2}%
+\tfrac{1}{2}\hbar\right)  N_{13},\frac{dA}{dt}\right\}  +\tfrac{1}{2}%
\hbar\left[  N_{13},\frac{dA}{dt}\right]  +\left[  \left[  N_{23},\frac
{dA}{dt}\right]  ,N_{12}\right]  \;. \label{GJP=JP+GroupRotations}%
\end{equation}
We have rearranged terms so as to produce just a simple Jordan product
(anticommutator), not a generalized one, and rotations of the time derivative.
\ This leads to a Jordan spectral problem involving only a commutative
product, $N_{2}N_{13}=N_{13}N_{2}$, to set the dynamical time scales for the
various invariant sectors of the theory. \ The additional rotation terms in
(\ref{GJP=JP+GroupRotations}) are similar to those encountered previously in
the $SU\left(  2\right)  $ examples, (\ref{L0^2}) and (\ref{L+L-}), with a
notable difference: \ The rotation is performed on $dA/dt$, not $A$. \ As in
those previous $SU\left(  2\right)  $ cases, however, the rotations are higher
order in $\hbar$ than the anticommutator term, and so they drop out of the
classical limit. \ Decompositions similar to (\ref{GJP=JP+GroupRotations}%
)\ apply to all the other $U\left(  n\right)  $ cases described by
(\ref{OscillatorEntwinedTimeDerivative}), as can be seen from the list of
operators in the GJP of that equation, by noting that only adjacent
$N_{i\;i+1}$elements in the list fail to commute.

It should also be apparent from the form of (\ref{GJP=JP+GroupRotations}) that
one cannot simply divide the LHS by $N_{12}N_{23}$ and then naively sweep the
$\left(  N_{12}N_{23}\right)  ^{-1}$ factor on the RHS into a logarithm.
\ This is permitted in the classical limit, as in (\ref{TakingLogs}), but
operators are not as easily manipulated on Hilbert space. \ Perhaps it is
useful to think of this as a problem of implementing scale transformations in
the generalized Jordan-Kurosh algebra context, but here the rotation terms
complicate the problem. \ These terms also complicate the issue of the quantum
bracket's kernel, although that issue for just the first RHS term in
(\ref{GJP=JP+GroupRotations}) is the familiar one for Jordan algebras.

The result (\ref{OscillatorEntwinedTimeDerivative}) helps to clarify why the
Leibniz rules fail when time evolution is expressed using QNBs for the
isotropic oscillator, for here this failure has been linked to the
intervention of a Jordan product involving non-commuting invariants. \ The
Leibniz failure can be summarized in derivators. \ For the $U\left(  n\right)
$ case,%
\begin{gather}
\left(  A,\mathcal{A}\,|\,N_{1},N_{12},N_{2},N_{23},\cdots,N_{n-1}%
,N_{n-1\;n},N_{n}\right)  =\hbar^{n-1}\left[  \left\{  A\mathcal{A}%
,N_{12},N_{23},\cdots,N_{n-1\;n}\right\}  ,N\right] \\
-\hbar^{n-1}A\left[  \left\{  \mathcal{A},N_{12},N_{23},\cdots,N_{n-1\;n}%
\right\}  ,N\right]  -\hbar^{n-1}\left[  \left\{  A,N_{12},N_{23}%
,\cdots,N_{n-1\;n}\right\}  ,N\right]  \mathcal{A}\;.\nonumber
\end{gather}
For $U\left(  2\right)  ,$ this reduces to%
\begin{equation}
\left(  A,\mathcal{A}\,|\,N_{1},N_{12},N_{2}\right)  =\hbar\left[  N,A\right]
\left[  N_{12},\mathcal{A}\right]  -\hbar\left[  N_{12},A\right]  \left[
N,\mathcal{A}\right]  \;.
\end{equation}
In the classical limit, the derivator vanishes, as expected.

\paragraph{\underline{SO(n+1) and quantum particles on n-spheres}}

For a quantum particle moving freely on the surface of an n-sphere, it is a
delicate matter to express the quantum $so\left(  n+1\right)  $ charges in
terms of the canonical coordinates and momenta, but it can be done (cf. the
discussion in \cite{CurtrightZachos}, and references cited therein). \ The
quantum version of the classical PB algebra is then obtained without any
modifications.%
\begin{equation}
\left[  P_{a},P_{b}\right]  =i\hbar L_{ab}\;,\;\;\;\left[  L_{ab}%
,P_{c}\right]  =i\hbar\left(  \delta_{ac}P_{b}-\delta_{bc}P_{a}\right)
\;,\;\;\;\left[  L_{ab},L_{cd}\right]  =i\hbar\left(  L_{ac}\delta_{bd}%
-L_{ad}\delta_{bc}-L_{bc}\delta_{ad}+L_{bd}\delta_{ac}\right)  \;.
\end{equation}
Hence the symmetric Hamiltonian has the same quadratic form as the classical
expression, (\ref{HforNsphere}). \ So a natural choice for the QNB involves
the same set of $2n-1$ invariants as selected in the classical case:
\ $\left[  A,P_{1},L_{12},P_{2},L_{23},P_{3},\cdots,P_{n-1},L_{n-1\;n}%
,P_{n}\right]  $.

The QNB results in an entwined time derivative, with attendant dynamical time
scales, and quantum rotation terms.%
\begin{align}
\frac{1}{\left(  i\hbar\right)  ^{n}n}\left[  A,P_{1},L_{12},P_{2}%
,L_{23},P_{3},\cdots,P_{n-1},L_{n-1\;n},P_{n}\right]   &  =\left(  -1\right)
^{n-1}\left\{  P_{2},P_{3},\cdots,P_{n-1},\frac{dA}{dt}\right\} \nonumber\\
&  +\;\;\text{quantum rotation terms\ \ .} \label{QuantumSphere}%
\end{align}
where $i\hbar dA/dt=\left[  A,H\right]  $. \ As in the previous $SU\left(
2\right)  $ and $U\left(  n\right)  $ quantum examples, the operator
entwinement on the RHS is not trivially eliminated through simple logarithms,
as it is in the classical situation in going from (\ref{ClassicalSphere}) to
(\ref{ClassicalSphereLogs}), but leads to Jordan-Kurosh spectral problems on
the Hilbert space of the system. \ The kernel of the quantum bracket is
similarly impacted.

If one of the $P$s or $L$s in the $2n$-bracket of (\ref{QuantumSphere}) were
replaced by $H$, the occurrence of an entwined time derivative would be
manifest (see (\ref{3sphereI}) below for the 3-sphere case). \ Otherwise, with
the invariants as chosen, it is laborious to obtain $\left[  A,H\right]  $ by
direct calculation. \ Likewise, the explicit form of the quantum rotation
terms in (\ref{QuantumSphere}), for general $n$, are laboriously obtained by
direct calculation, and will be given elsewhere. \ They may be constructed
through an embedding of the orthogonal group into the unitary group treated
previously. Suffice it here to say that they are higher order in $\hbar$, as
expected, and to consider the case of the 3-sphere, for comparison to the
chiral charge methods given below.

For the 3-sphere, it is convenient to define the usual duals (sum repeated
indices),%
\begin{equation}
L_{i}=\tfrac{1}{2}\,\varepsilon_{ijk}L_{jk}\;.
\end{equation}
Then the algebra becomes the well-known (compare to (\ref{Axial&Isospin})\ and
(\ref{Axial&IsospinPBs}))%
\begin{equation}
\left[  L_{i},L_{j}\right]  =i\hbar\varepsilon_{ijk}L_{k}\;,\;\;\;\left[
L_{i},P_{j}\right]  =i\hbar\varepsilon_{ijk}P_{k}\;,\;\;\;\left[  P_{i}%
,P_{j}\right]  =i\hbar\varepsilon_{ijk}L_{k}\;.
\end{equation}
and the group invariant Hamiltonian is $H=\frac{1}{2}\left(  P_{i}P_{i}%
+L_{i}L_{i}\right)  $. \ By direct calculation, we then obtain (cf. third RHS
line in (\ref{GJP3}), and also recall for a particle on the 3-sphere,
$L_{i}P_{i}=0$)%
\begin{align}
\left[  A,L_{1},L_{2},P_{1},P_{2},P_{3}\right]   &  =3i\hbar^{3}\left\{
\frac{dA}{dt},P_{3}\right\}  -3\hbar^{2}\left\{  \left[  A,L_{i}P_{i}\right]
,L_{3}\right\} \label{Quantum4Sphere}\\
&  +\tfrac{1}{4}\hbar^{2}\sum_{i}\left(  \left[  \,\left[  \left[
A,L_{i}-P_{i}\right]  ,L_{i}-P_{i}\right]  \,,\,L_{3}+P_{3}\,\right]  -\left[
\,\left[  \left[  A,L_{i}+P_{i}\right]  ,L_{i}+P_{i}\right]  \,,\,L_{3}%
-P_{3}\,\right]  \right)  \;.\nonumber
\end{align}

Once again, the quantum rotation terms represent higher order corrections, in
powers of $\hbar$, corresponding to group rotations of $A$. \ For example, if
$A$ is the 3-sphere bilinear
\begin{equation}
A_{ab}\equiv\left(  L_{a}+P_{a}\right)  \left(  L_{b}-P_{b}\right)  \;,
\end{equation}
then $dA_{ab}/dt=0$ for a particle moving freely on the surface of the sphere,
but the corresponding quantum group rotation induced by the 3-sphere 6-bracket
is \emph{not} zero. \ The six bracket reduces entirely to those quantum
rotation terms. \ Explicitly, we find
\begin{equation}
\left[  A_{ab},P_{1},L_{3},P_{2},L_{1},P_{3}\right]  =4i\hbar^{5}\sum
_{c}\left(  \varepsilon_{b2c}A_{ac}-\varepsilon_{a2c}A_{cb}\right)  \;.
\end{equation}
As in all previous cases, the quantum rotations disappear in the classical
limit,
\begin{equation}
\lim_{\hbar\rightarrow0}\;\left[  A_{ab},P_{1},L_{3},P_{2},L_{1},P_{3}\right]
/\hbar^{3}=0\;.
\end{equation}

\paragraph{\underline{SO(4)=SU(2)$\times$SU(2) as another special case}}

We consider this particular example in more detail, as a bridge to general
chiral models, choosing bracket elements that exhibit dynamical time scales
both with and without group rotations. \ Use $L_{i}$ and $R_{i}$ for the
mutually commuting $su\left(  2\right)  $ charges. \
\begin{equation}
\left[  L_{i},L_{j}\right]  =i\hbar\varepsilon_{ijk}L_{k}\;,\;\;\;\left[
R_{i},R_{j}\right]  =i\hbar\varepsilon_{ijk}R_{k}\;,\;\;\;\left[  L_{i}%
,R_{j}\right]  =0\;.
\end{equation}
(Again, compare to (\ref{Axial&Isospin})\ and (\ref{Axial&IsospinPBs}) and
note the normalization here differs from that used earlier: $\mathcal{L}%
_{i}=-2L_{i},\;\mathcal{R}_{i}=-2R_{i}.$) \ Define the usual quadratic
Casimirs for the left and right algebras,
\begin{equation}
I_{L}=\sum_{i}L_{i}^{2}\;,\;\;\;I_{R}=\sum_{i}R_{i}^{2}\;.
\end{equation}
Then, for a Hamiltonian of the form $H\equiv F\left(  I_{L},I_{R}\right)  ,$
where $F$ is any function of the left and right Casimirs, we find%
\begin{equation}
\left[  A,H,R_{1},R_{2},L_{1},L_{2}\right]  =\left(  i\hbar\right)
^{2}\left\{  \left[  A,H\right]  ,L_{3},R_{3}\right\}  =\left(  i\hbar\right)
^{2}\left[  \left\{  A,L_{3},R_{3}\right\}  ,H\right]  \;, \label{3sphereI}%
\end{equation}
This is the quantum analogue of the classical result in (\ref{HinBracket}).
\ Aside from trivial normalization factors, the difference lies in the
particular ordering of operators in the quantum expression. \ Physically,
(\ref{3sphereI}) is simply time evolution generated by the Hamiltonian, with
the Jordan-Kurosh, simultaneous eigenvalues of $L_{3}$ and $R_{3}$ setting the
time scales for the various sectors of the theory. \ In particular this is
true for $H\propto I_{L}$ or $H\propto I_{R}$, as is relevant for a quantum
particle moving freely on the 3-sphere.

The dynamical time scale structure produced by this quantum 6-bracket is a
simple extension of the structure found previously for the $SU\left(
2\right)  $ example. \ The Jordan-Kurosh eigenvalues, $\sigma$, are now
defined by
\begin{equation}
\left\{  A_{\sigma},L_{3},R_{3}\right\}  =\sigma A_{\sigma}\;,
\end{equation}
A complete set of operators consists of all doubly-projected eigenoperators,
$A_{\lambda_{1}\rho_{1},\lambda_{2}\rho_{2}}$, where%
\begin{align}
L_{3}A_{\lambda_{1}\rho_{1},\lambda_{2}\rho_{2}}  &  =\lambda_{1}%
A_{\lambda_{1}\rho_{1},\lambda_{2}\rho_{2}}\;,\;\;\;A_{\lambda_{1}\rho
_{1},\lambda_{2}\rho_{2}}L_{3}=\lambda_{2}A_{\lambda_{1}\rho_{1},\lambda
_{2}\rho_{2}}\;,\nonumber\\
R_{3}A_{\lambda_{1}\rho_{1},\lambda_{2}\rho_{2}}  &  =\rho_{1}A_{\lambda
_{1}\rho_{1},\lambda_{2}\rho_{2}}\;,\;\;\;A_{\lambda_{1}\rho_{1},\lambda
_{2}\rho_{2}}R_{3}=\rho_{2}A_{\lambda_{1}\rho_{1},\lambda_{2}\rho_{2}}\;.
\end{align}
Hence $\left\{  A,L_{3},R_{3}\right\}  =\sigma_{12}A$, with%
\begin{equation}
\sigma_{12}=2\lambda_{1}\rho_{1}+\lambda_{1}\rho_{2}+\rho_{1}\lambda
_{2}+2\lambda_{2}\rho_{2}\;,
\end{equation}
since%
\begin{equation}
\left\{  A,L_{3},R_{3}\right\}  =\left\{  L_{3},R_{3}\right\}  A+L_{3}%
AR_{3}+R_{3}AL_{3}+A\left\{  L_{3},R_{3}\right\}  \;.
\end{equation}
So the time scales for the various sectors of the theory are set jointly by
the eigenvalues of $L_{3}$ and $R_{3}$ . \ 

The simple Leibniz rule for generic $A$ and $\mathcal{A}$, that would equate
$\left[  A\mathcal{A},H,R_{1},R_{2},L_{1},L_{2}\right]  $ with \newline%
$A\left[  \mathcal{A},H,R_{1},R_{2},L_{1},L_{2}\right]  $ $+$ $\left[
A,H,R_{1},R_{2},L_{1},L_{2}\right]  \mathcal{A}$, will fail for products
$A_{\lambda_{1}\rho_{1},\lambda_{2}\rho_{2}}\mathcal{A}_{\lambda_{2}\rho
_{2},\lambda_{3}\rho_{3}},$ unless%
\begin{equation}
\sigma_{12}=\sigma_{23}=\sigma_{13}\;.
\end{equation}
There are no higher-order quantum group rotation terms in this particular
case, due to our choice for the invariants in the bracket $\left[
A,H,R_{1},R_{2},L_{1},L_{2}\right]  $. \ The more general situation is
revealed by a different choice, as follows.

\paragraph{\underline{\textbf{3-sphere chiral 6-brackets}}}

We take all five of the fixed elements in the 6-bracket to be charges in the
$su\left(  2\right)  \times su\left(  2\right)  $ algebra, and not Casimirs,
to find%

\begin{equation}
\left[  A,L_{1},L_{2},L_{3},R_{1},R_{2}\right]  =\frac{3}{2}\left(
i\hbar\right)  ^{2}\left[  \left\{  A,R_{3}\right\}  ,I_{L}\right]  +\frac
{1}{2}\left(  i\hbar\right)  ^{2}\sum_{i}\left[  \,\left[  \left[
A,L_{i}\right]  ,L_{i}\right]  \,,\,R_{3}\,\right]  \;, \label{3sphereR}%
\end{equation}
or, equivalently,
\begin{equation}
\left[  A,R_{1},R_{2},R_{3},L_{1},L_{2}\right]  =\frac{3}{2}\left(
i\hbar\right)  ^{2}\left[  \left\{  A,L_{3}\right\}  ,I_{R}\right]  +\frac
{1}{2}\left(  i\hbar\right)  ^{2}\sum_{i}\left[  \,\left[  \left[
A,R_{i}\right]  ,R_{i}\right]  \,,\,L_{3}\,\right]  \;. \label{3sphereL}%
\end{equation}
The first terms (single commutators) on the RHSs of (\ref{3sphereR}) and
(\ref{3sphereL}) are inherently $O\left(  \hbar^{3}\right)  $, and give scaled
time derivatives, while the second terms (triple commutators) are $O\left(
\hbar^{5}\right)  $, and give additional group rotations. \ Perhaps these
results may be interpreted as group \emph{covariant} Hamiltonian flows, with
``quantum connections'' given as the triple commutator, higher-order effects
in $\hbar$. \ Note the LHS of (\ref{3sphereR}) manifestly vanishes when $A$ is
one of $L_{1},\;L_{2},\;L_{3},\;R_{1},$ or $R_{2},$ while the RHS manifestly
vanishes for the remaining choice $A=R_{3}$, as well as $R_{1}$ and $R_{2}$.
\ Similarly, the RHS of (\ref{3sphereL}) manifestly vanishes for $A=L_{j}$,
$j=1,2,3,$ including the one case excepted by the LHS of that equation.

We may add or subtract (\ref{3sphereR}) and (\ref{3sphereL}) to gain
$L\longleftrightarrow R$ symmetry between left- and right-hand sides, but the
resulting quantum expressions do not permit easy conversions into logarithms,
as in the classical case (cf. (\ref{3SphereDifference})). \ Nonetheless, for
the free particle on the 3-sphere, with $H=2I_{L}=2I_{R}$, we may write the
sum and difference as%
\begin{equation}
\left[  A,L_{1},L_{2},R_{3}\pm L_{3},R_{1},R_{2}\right]  =\frac{-3i\hbar^{3}%
}{4}\left\{  \frac{dA}{dt},L_{3}\pm R_{3}\right\}  +\frac{1}{2}\left(
i\hbar\right)  ^{2}\sum_{i}\left(  \left[  \,\left[  \left[  A,R_{i}\right]
,R_{i}\right]  \,,\,L_{3}\,\right]  \pm\left[  \,\left[  \left[
A,L_{i}\right]  ,L_{i}\right]  \,,\,R_{3}\,\right]  \right)  \;.
\label{R+LEntwinedTimeDer}%
\end{equation}
This is the by-now-familiar form, consisting of an entwined time derivative
and group rotations.

As a simple example to isolate and accentuate the group rotation effects, take
$A$ to be any bilinear $A_{ab}\equiv L_{a}R_{b}$ of specific left and right
charges. \ Since commutators are indeed derivations, all functions of the six
possible $L_{a}$ and $R_{b}$ charges commute with the Casimirs, so the first
terms on the RHS's of (\ref{3sphereR}) and (\ref{3sphereL}) vanish for
$A=A_{ab}$ (i.e. $A_{ab}$ for a particle moving freely on the surface of a
3-sphere has no time derivatives). \ The second terms on the RHS's of
(\ref{3sphereR}) and (\ref{3sphereL}) do \emph{not} vanish for $A=A_{ab}$ but
are just rotations of the $L_{a}$ and $R_{b}$ charges, respectively, about the
$z$ axis.
\begin{equation}
\sum_{i}\left[  \,\left[  \left[  A_{ab},L_{i}\right]  ,L_{i}\right]
\,,\,R_{3}\,\right]  =2i\hbar^{3}\sum_{c}\varepsilon_{b3c}A_{ac}%
\;,\;\;\;\sum_{i}\left[  \,\left[  \left[  A_{ab},R_{i}\right]  ,R_{i}\right]
\,,\,L_{3}\,\right]  =2i\hbar^{3}\sum_{c=1,2,3}\varepsilon_{a3c}A_{cb}\;.
\end{equation}
So, for this particular example,
\begin{equation}
\left[  A_{ab},L_{1},L_{2},L_{3},R_{1},R_{2}\right]  =-i\hbar^{5}\sum
_{c}\varepsilon_{b3c}A_{ac}\;,\;\;\;\left[  A_{ab},R_{1},R_{2},R_{3}%
,L_{1},L_{2}\right]  =-i\hbar^{5}\sum_{c}\varepsilon_{a3c}A_{cb}\;.
\end{equation}
Let us establish this result in detail by proceeding from the $SU\left(
2\right)  \times SU\left(  2\right)  $ chiral form of the reductio ad
dimidium, namely
\begin{equation}
\left[  A,L_{1},L_{2},L_{3},R_{1},R_{2}\right]  =\left(  i\hbar\right)
^{2}\sum_{j}\left\{  \left[  A,L_{j}\right]  ,L_{j},R_{3}\right\}  \;.
\end{equation}
This enables us to compute how the bilinear is transformed by the 6-bracket.
\ First \
\begin{equation}
\left[  L_{a}R_{b},L_{1},L_{2},L_{3},R_{1},R_{2}\right]  =\left(
i\hbar\right)  ^{2}\sum_{j}\left\{  \left[  L_{a}R_{b},L_{j}\right]
,L_{j},R_{3}\right\}  =\left(  i\hbar\right)  ^{3}\sum_{j,k}\varepsilon
_{ajk}\left\{  L_{k}R_{b},L_{j},R_{3}\right\}  \;.
\end{equation}
But then
\begin{align}
\left\{  L_{k}R_{b},L_{j},R_{3}\right\}   &  =L_{k}R_{b}\left\{  L_{j}%
,R_{3}\right\}  +L_{j}\left\{  L_{k}R_{b},R_{3}\right\}  +R_{3}\left\{
L_{k}R_{b},L_{j}\right\} \nonumber\\
&  =2L_{k}L_{j}R_{b}R_{3}+L_{j}L_{k}\left\{  R_{b},R_{3}\right\}  +\left\{
L_{k},L_{j}\right\}  R_{3}R_{b}\;,
\end{align}
so, summing repeated indices,%
\begin{equation}
\varepsilon_{ajk}\left\{  L_{k}R_{b},L_{j},R_{3}\right\}  =i\hbar
\varepsilon_{ajk}\varepsilon_{kjm}L_{m}\left(  R_{b}R_{3}-\frac{1}{2}\left\{
R_{b},R_{3}\right\}  \right)  =\frac{1}{2}\left(  i\hbar\right)
^{2}\varepsilon_{ajk}\varepsilon_{kjm}L_{m}\varepsilon_{b3c}R_{c}=\hbar
^{2}L_{a}R_{c}\varepsilon_{b3c}\;.
\end{equation}
This confirms by direct calculation that the chosen bracket does not just
produce entwined time derivatives, but more elaborately, the bracket combines
entwined $dA/dt$ with infinitesimal group rotations of $A$. \ Since group
rotations are symmetries of the system's dynamics, this is not an inconsistent
combination (cf. covariant derivatives in Yang-Mills theory).

\paragraph{\underline{Quantum G$\times$G chiral particles}}

Consider next models whose dynamics are invariant under chiral groups,
$G\times G$. \ For example, a particle moving freely on the group manifold is
of this type. Let $n$ be the dimension of the group $G$, and write the charge
algebra underlying the group $G\times G$ as
\begin{equation}
\left[  L_{i},L_{j}\right]  =i\hbar f_{ijk}L_{k}\;,\;\;\;\left[  R_{i}%
,R_{j}\right]  =i\hbar f_{ijk}R_{k}\;,\;\;\;\left[  L_{i},R_{j}\right]  =0\;.
\end{equation}
Then for odd $n\equiv1+2s$, with sums over repeated indices understood to run
from $1$ to $n$,%
\begin{gather}
\left[  A,L_{1},\cdots,L_{n},R_{1},\cdots,R_{n-1}\right]  =\frac{1}{n!\left(
n-1\right)  !}\,\varepsilon_{i_{1}\cdots i_{n}}\varepsilon_{j_{i}\cdots
j_{n-1}n}\left[  A,L_{i_{1}},\cdots,L_{i_{n}},R_{j_{1}},\cdots,R_{j_{n-1}%
}\right] \nonumber\\
=K_{n}\,\varepsilon_{i_{1}\cdots i_{n}}\varepsilon_{j_{i}\cdots j_{n-1}%
n}\left\{  \left[  A,L_{i_{1}}\right]  ,\left[  L_{i_{2}},L_{i_{3}}\right]
,\cdots,\left[  L_{i_{n-1}},L_{i_{n}}\right]  ,\left[  R_{j_{1}},R_{j_{2}%
}\right]  ,\cdots,\left[  R_{j_{n-2}},R_{j_{n-1}}\right]  \right\}  \;,
\label{nOddQuantumGxG}%
\end{gather}
where $K_{n}=\left[  4^{s}\left(  s!\right)  ^{2}\right]  ^{-1}$ (the same
numerical combinatoric factor introduced earlier in the classical example
(\ref{Kn})) incorporates the number of equivalent ways to obtain the list of
commutators in the generalized Jordan bracket as written\footnote{The number
of ways of choosing the $n$ commutators in (\ref{nOddQuantumGxG}) is $n\left(
n-2\right)  \left(  n-4\right)  \cdots\left(  1\right)  \times\left(
n-2\right)  \left(  n-4\right)  \cdots\left(  1\right)  $, so\newline%
$K_{n}=\frac{n\left(  n-2\right)  \left(  n-4\right)  \cdots\left(  1\right)
\times\left(  n-2\right)  \left(  n-4\right)  \cdots\left(  1\right)
}{n!\left(  n-1\right)  !}\;.$}. \ So
\begin{gather}
\left[  A,L_{1},\cdots,L_{n},R_{1},\cdots,R_{n-1}\right]  =K_{n}\,\left(
i\hbar\right)  ^{n-1}\varepsilon_{i_{1}\cdots i_{n}}\varepsilon_{j_{i}\cdots
j_{n-1}n}\times\\
\times\left(  f_{i_{2}i_{3}k_{1}}\cdots f_{i_{n-1}i_{n}k_{s}}\right)  \left(
f_{j_{1}j_{2}m_{1}}\cdots f_{j_{n-2}j_{n-1}m_{s}}\right)  \left\{  \left[
A,L_{i_{1}}\right]  ,L_{k_{1}},\cdots,L_{k_{s}},R_{m_{1}},\cdots,R_{m_{s}%
}\right\}  \;.\nonumber
\end{gather}
This leads to some mixed symmetry tensors that are familiar from classical
invariant theory for Lie groups,
\begin{equation}
\tau_{n\left\{  m_{1}\cdots m_{s}\right\}  }\equiv\varepsilon_{j_{1}\cdots
j_{n-1}n}\;f_{j_{1}j_{2}m_{1}}\cdots f_{j_{n-2}j_{n-1}m_{s}}\;.
\end{equation}
Need has not dictated obtaining elegant expressions for these tensors, except
in special cases, but\ undoubtedly they exist\footnote{S Meshkov has suggested
that similar tensors and invariants constructed from them appear in nuclear
shell theory.}.

In terms of these the reduction becomes
\begin{align}
&  \left[  A,L_{1},\cdots,L_{n},R_{1},\cdots,R_{n-1}\right] \nonumber\\
&  =K_{n}\,\left(  i\hbar\right)  ^{n-1}\tau_{i_{1}\left\{  k_{1}\cdots
k_{s}\right\}  }\tau_{n\left\{  m_{1}\cdots m_{s}\right\}  }\left\{  \left[
A,L_{i_{1}}\right]  ,L_{k_{1}},\cdots,L_{k_{s}},R_{m_{1}},\cdots,R_{m_{s}%
}\right\}  \;. \label{OddChiralTLCresult}%
\end{align}
Results for even $n$ are similar, only in that case the arbitrary $A$ must be
locked in a commutator with an $R$.

As in the classical case, (\ref{AnotherWay}), a somewhat simpler choice for
the invariants in the maximal bracket requires us to compute (note the range
of the sums)
\begin{align}
&  \left[  A,F\left(  I_{L},I_{R}\right)  ,L_{1},\cdots,L_{n-1},R_{1}%
,\cdots,R_{n-1}\right] \nonumber\\
&  =\frac{1}{\left(  n-1\right)  !\left(  n-1\right)  !}\,\sum_{\text{all
}i,j=1}^{n-1}\varepsilon_{i_{1}\cdots i_{n-1}}\varepsilon_{j_{i}\cdots
j_{n-1}n}\left[  A,F\left(  I_{L},I_{R}\right)  ,L_{i_{1}},\cdots,L_{i_{n-1}%
},R_{j_{1}},\cdots,R_{j_{n-1}}\right]  \;,
\end{align}
where $F\left(  I_{L},I_{R}\right)  $ is any function of the left and right
Casimirs. \ The RHS here vanishes for even $n$, so again we take odd $n$, say
$n=1+2s$. \ Then by the commutator resolution, since $\left[  F\left(
I_{L},I_{R}\right)  ,L_{i}\right]  =0=\left[  F\left(  I_{L},I_{R}\right)
,R_{i}\right]  $, we can write
\begin{align}
&  \left[  A,F\left(  I_{L},I_{R}\right)  ,L_{1},\cdots,L_{n-1},R_{1}%
,\cdots,R_{n-1}\right] \\
&  =K_{n}\,\sum_{\text{all }i,j=1}^{n-1}\varepsilon_{i_{1}\cdots i_{n-1}%
}\varepsilon_{j_{i}\cdots j_{n-1}n}\left\{  \left[  A,F\left(  I_{L}%
,I_{R}\right)  \right]  ,\left[  L_{i_{1}},L_{i_{2}}\right]  ,\cdots,\left[
L_{i_{n-2}},L_{i_{n-1}}\right]  ,\left[  R_{j_{1}},R_{j_{2}}\right]
,\cdots,\left[  R_{j_{n-2}},R_{j_{n-1}}\right]  \right\}  \;,\nonumber
\end{align}
where again $1/K_{n}=4^{s}\left(  s!\right)  ^{2}$. \ So, for a Hamiltonian of
the form $H=F\left(  I_{L},I_{R}\right)  ,\;$we have%
\begin{align}
&  \left[  A,H,L_{1},\cdots,L_{n-1},R_{1},\cdots,R_{n-1}\right] \nonumber\\
&  =\frac{1}{\left(  s!\right)  ^{2}}\,\left(  -4\hbar^{2}\right)  ^{s}%
\sum_{\text{all }k,m=1}^{n}\sigma_{\left\{  k_{1}\cdots k_{s}\right\}  }%
\sigma_{\left\{  m_{1}\cdots m_{s}\right\}  }\left\{  \left[  A,H\right]
,L_{k_{1}},\cdots,L_{k_{s}},R_{m_{1}},\cdots,R_{m_{s}}\right\}  \;,
\label{GxGI}%
\end{align}
with $2s\equiv n-1$ and the completely symmetric tensor $\sigma_{\left\{
k_{1}\cdots k_{s}\right\}  }$ defined as in the classical situation
(\ref{SigmaTensor}). \ Note the range of the sum in (\ref{SigmaTensor}) is
truncated from that in (\ref{GxGI}), although the sum may be trivially
extended just by adding a fixed extra index to the Levi-Civita symbol.
\begin{equation}
\sigma_{\left\{  k_{1}\cdots k_{s}\right\}  }=\sum_{\text{all }i=1}%
^{n}\varepsilon_{ni_{1}\cdots i_{n-1}}\;f_{i_{2}i_{3}k_{1}}\cdots
f_{i_{n-2}i_{n-1}k_{s}}\;. \label{SigmaTensorToo}%
\end{equation}

The commutator of $A$ with the function of Casimirs can be computed after the
generalized Jordan product (GJP), again since $\left[  H,L_{i}\right]
=0=\left[  H,R_{i}\right]  $. \ So, with the sums over repeated $k$s and $m$s
understood,
\begin{align}
&  \left[  A,H,L_{1},\cdots,L_{n-1},R_{1},\cdots,R_{n-1}\right] \nonumber\\
&  =K_{n}\left(  i\hbar\right)  ^{n-1}\,\sigma_{\left\{  k_{1}\cdots
k_{s}\right\}  }\sigma_{\left\{  m_{1}\cdots m_{s}\right\}  }\left[  \left\{
A,L_{k_{1}},\cdots,L_{k_{s}},R_{m_{1}},\cdots,R_{m_{s}}\right\}  ,H\right]
\;. \label{QuantumGxG}%
\end{align}
The GJP spectral equation,
\begin{align}
&  \lambda A_{\lambda}+\text{\ higher-order quantum rotation terms}\nonumber\\
&  =K_{n}\left(  i\hbar\right)  ^{n-1}\,\sigma_{\left\{  k_{1}\cdots
k_{s}\right\}  }\sigma_{\left\{  m_{1}\cdots m_{s}\right\}  }\left\{
A_{\lambda},L_{k_{1}},\cdots,L_{k_{s}},R_{m_{1}},\cdots,R_{m_{s}}\right\}  \;,
\end{align}
must now be solved to find the time scales $\lambda$ that govern the QNB
generated time evolution,
\begin{align}
&  i\hbar\lambda\frac{d}{dt}A_{\lambda}+\text{\ higher-order quantum rotation
terms}\nonumber\\
&  =\lambda\left[  A_{\lambda},H\right]  =\left[  A_{\lambda},H,L_{1}%
,\cdots,L_{n-1},R_{1},\cdots,R_{n-1}\right]  \;.
\end{align}

All this extends in a straightforward way to the algebras of symmetry groups
involving arbitrary numbers of factors. \ Rather than pursue that
generalization, however, we focus instead on unitary factors, where the $\tau$
and $\sigma$ tensors simplify. \ For a touch of variety, we take the left and
right group factors to be different unitary groups.

\paragraph{\underline{U(n)$\times$U(m) models}}

For systems with $U\left(  n\right)  \times U\left(  m\right)  $ group
invariant dynamics, with the proper choice of charge basis, the
structure-constant-weighted sums of the previous formulas can be made to
reduce to single terms as in the case of the previous $U\left(  n\right)  $
example. \ We take the oscillator basis for each of the algebras, so that the
charges obey the commutators%
\begin{equation}
\left[  N_{ij},N_{kl}\right]  =\hbar\left(  N_{il}\delta_{jk}-N_{kj}%
\delta_{il}\right)  \;,\;\;\;\left[  M_{ab},M_{cd}\right]  =\hbar\left(
M_{ad}\delta_{bc}-M_{cb}\delta_{ad}\right)  \;,\;\;\;\left[  N_{ij}%
,M_{ab}\right]  =0\;.
\end{equation}
for $i,j,k,l=1,\cdots,n,$ and $a,b,c,d=1,\cdots,m.$ \ As before, we denote the
mutually commuting diagonal charges as $N_{jj}=N_{j}$ and $M_{aa}=M_{a}$, with
(central charge) sums $N=\sum_{j=1}^{n}N_{jj}$, \ $M=\sum_{a=1}^{m}M_{aa}$.
\ Then, as for the single $U\left(  n\right)  $ results, we have either%
\begin{align}
&  \left[  A,N_{1},N_{12},N_{2},\cdots,N_{n-1},N_{n-1\;n},N_{n},M_{1}%
,M_{12},M_{2},\cdots,M_{m-1},M_{m-1\;m}\right] \nonumber\\
&  =\hbar^{n+m-2}\left\{  \left[  A,N\right]  ,N_{12},\cdots,N_{n-1\;n}%
,M_{12},\cdots,M_{m-1\;m}\right\} \nonumber\\
&  =\hbar^{n+m-2}\left[  \left\{  A,N_{12},\cdots,N_{n-1\;n},M_{12}%
,\cdots,M_{m-1\;m}\right\}  ,N\right]  \;,
\end{align}
or similarly with $M\longleftrightarrow N$, as well as other such relations
that follow from choosing different Hamiltonian paths through the algebras.

Replacement of one of the diagonal charges with an arbitrary function of the
left and right Casimirs, as well as the two central sums, leads to similar
results. \ These may now be used to discuss time development for systems whose
Hamiltonians are of the form
\begin{equation}
H=F\left(  N,M,I_{N},I_{M}\right)  \;,
\end{equation}
for which%
\begin{align}
&  \left[  A,H,N_{1},N_{12},N_{2},\cdots,N_{n-1},N_{n-1\;n},M_{1},M_{12}%
,M_{2},\cdots,M_{m-1},M_{m-1\;m}\right] \nonumber\\
&  =\hbar^{n+m-2}\left[  \left\{  A,N_{12},\cdots,N_{n-1\;n},M_{12}%
,\cdots,M_{m-1\;m}\right\}  ,H\right] \nonumber\\
&  =\hbar^{n+m-2}\left\{  \left[  A,H\right]  ,N_{12},\cdots,N_{n-1\;n}%
,M_{12},\cdots,M_{m-1\;m}\right\} \nonumber\\
&  =i\hbar^{n+m-1}\left\{  \frac{dA}{dt},N_{12},\cdots,N_{n-1\;n}%
,M_{12},\cdots,M_{m-1\;m}\right\}  \;. \label{QuantumU(n)xU(m)}%
\end{align}
The effect of the remaining, noncommuting charges in the generalized Jordan
product is once again to set time scales for the various invariant sectors of
the theory. \ So, if
\begin{equation}
\left\{  A_{\sigma},N_{12},\cdots,N_{n-1\;n},M_{12},\cdots,M_{m-1\;m}\right\}
=\sigma A_{\sigma}+\text{ \ \ higher-order quantum rotation terms}\;,
\end{equation}
then
\begin{align}
i\hbar^{n+m-1}\sigma\frac{dA_{\sigma}}{dt}  &  =\left[  A_{\sigma}%
,H,N_{1},N_{12},N_{2},\cdots,N_{n-1},N_{n-1\;n},M_{1},M_{12},M_{2}%
,\cdots,M_{m-1},M_{m-1\;m}\right] \nonumber\\
&  +\text{ \ \ higher-order quantum rotation terms}\;,
\end{align}
with quite elaborate sums of such terms describing the time evolution of
general operators. \ 

All this extends to the algebras of symmetry groups involving arbitrary
numbers of unitary group factors. \ 

\subsection{Summary Table}

For convenience, we summarize the results of the preceding sections as a Table
of key formulas.\bigskip

\begin{center}%
\begin{tabular}
[c]{|c|cc|}\hline\hline
\multicolumn{1}{||c|}{Model symmetry} & \multicolumn{1}{||c}{Classical
dynamics} & \multicolumn{1}{||c||}{Quantum dynamics}\\\hline\hline
$SO(3)$ & (\ref{so(3)ClassicalBeauty}) &
\multicolumn{1}{|c|}{(\ref{4QNBDerivation}) (\ref{4QNBEntwinedDerivation}%
)}\\\hline
Any Lie (4-bracket sum) & (\ref{AnyLie4CNB}) &
\multicolumn{1}{|c|}{(\ref{AnyLie4QNB})}\\\hline
$U(N)$ (oscillators) & (\ref{u(n)CNBdA/dt}) &
\multicolumn{1}{|c|}{(\ref{OscillatorEntwinedTimeDerivative})}\\\hline
$SO(N+1)$ & (\ref{ClassicalSphere}) &
\multicolumn{1}{|c|}{(\ref{QuantumSphere})}\\\hline
$SO(4)=SU(2)\otimes SU(2)$ & (\ref{HinBracket}) (\ref{3SphereR3Kernel})
(\ref{3SphereL3Kernel}) (\ref{3SphereDifference}) &
\multicolumn{1}{|c|}{(\ref{3sphereI}) (\ref{3sphereR}) (\ref{3sphereL})
(\ref{R+LEntwinedTimeDer})}\\\hline
$G\otimes G$ & \multicolumn{1}{|c|}{(\ref{GxGMaximalCNB})
(\ref{GxGClassicalTimeEvolution})} & (\ref{QuantumGxG})
(\ref{QuantumU(n)xU(m)})\\\hline
\end{tabular}
\bigskip
\end{center}

An empirical methodology suggested by the above examples argues for the
following checklist in quantizing a general classical system of the type
(\ref{ClassicalScaledTime}). \ If $V$ is trivial (i.e. numerical), the QNB
corresponding to the CNB involved is a prime candidate for an
\textquotedblleft exceptional\textquotedblright\ derivation quantization,
provided the derivation property checks (and thus the EFI). \ In the generic
case, if $V$ is a function of the invariants, manipulation of the classical
expression may be useful, to result in a simpler $V$, and in new CNB entries
which would still combine into the Hamiltonian in the PB resolution. The
corresponding QNB would then be expected to yield the entwined structures to
be studied as above, with the Hamiltonian appearing as an entwined commutator
(time derivative), and with the respective time scale eigenvalue problems to
be solved.

\section{Conclusions}

In this paper, we have demonstrated and illustrated through simple, explicit
examples, how Nambu brackets provide a consistent, elegant description, both
classically and quantum mechanically, especially of superintegrable systems
using even QNBs. \ This description can be equivalent to classical and quantum
Hamiltonian mechanics, but it is broader in its conceptualizations and may
have more possible uses. \ In particular, we have explained in detail how QNBs
are consistent, after all, given due consideration to multiple time scales set
by invariants entwining the time derivatives, and how reputed inconsistencies
have instead involved unsuitable and untenable conditions. \ We have also
emphasized additional complications that distinguish odd QNBs.

We believe the physical interpretations of entwined time derivatives, with
their dynamical time scales, and group rotations, in the general situation,
explain the perceived failure of the classical Leibniz rules and the classical
FI in a transparent way, and are the only ingredients required for a
successful non-Abelian quantum implementation of the most general Nambu
brackets as descriptions of dynamics. \ Perhaps this approach is equivalent to
the Abelian deformation approach \cite{DitoFST}, but that has not been shown.
\ However, ultimately it should not be necessary to have to argue, physically,
that if the Abelian deformation approach to quantization of Nambu brackets is
indeed logically complete and consistent, then it must give specific results
equivalent to the more traditional noncommutative operator approach given
here. \ There is, after all, not very much freedom in the quantization of free
particles and simple harmonic oscillators!

Moreover, Hanlon and Wachs \cite{Hanlon} announced the result that even QNB
algebras (designated by them as ``Lie $k$-algebras'') are ``Koszul'' (also see
\cite{Azcarraga96a}), and therefore have duals which are commutative and
totally associative. \ Is it possible that the Abelian deformation
quantization of Nambu brackets is precisely this dual, and is in that sense
equivalent mathematically to the non-Abelian structures we have discussed?

Other such mathematical issues and areas for further study are raised by our
analysis, such as: \ A complete mathematical classification of Jordan-Kurosh
eigenvalue problems; a corresponding treatment of quantum rotation terms; a
study of both classical and quantum topological effects in terms of Nambu
brackets; and the behavior of the brackets in the large N limit (as one way to
obtain a field theory).

There are also several open avenues for physical applications, the most
promising involving membranes and other extended objects. \ In that regard,
given the quantum dichotomy of even and odd brackets, it would appear that
extended objects with alternate-dimensional world-volumes are more amenable to
QNBs. \ While volume preserving diffeomorphisms are based on classical
geometrical concepts, perhaps relying too strongly on associativity, their
ultimate generalization to noncommutative geometries, and their uses in field,
string, and membrane theories, should be possible. \ We hope the developments
in this paper contribute towards completion of such enterprises.

\begin{acknowledgments}
We thank Y Nambu and Y Nutku for helpful discussions, and J de Azc\'{a}rraga
for pointing out related earlier work, as well as a critical reading of the
manuscript. \ We also thank D Fairlie for informative remarks, in particular
for reminding us of Laplace's theorem on general minor expansions. \ This
research was supported in part by the US Department of Energy, Division of
High Energy Physics, Contract W-31-109-ENG-38, and in part by NSF Award 0073390.
\end{acknowledgments}

\appendix

\section{Formal Division}

\ 

We are often interested in solving non-linear algebraic equations in both Lie
and special Jordan algebras. \ This is hampered by the fact that these are not
division rings. \ 

Nevertheless, there is a \emph{formal} series solution to construct inverses
in both special Jordan and Lie algebras as contained in an associative
embedding algebra $\mathfrak{A}$. \ For the former, consider\footnote{Jordan
would include a factor of $1/2$ in the definition of $\circ$.}
\begin{equation}
a\circ b=b\circ a\equiv ab+ba=c\;.
\end{equation}
Suppose $b$ and $c$ are given. \ We wish to solve for $a$. \ We assume the
inverse $b^{-1}$ exists in the enveloping algebra. \ So we seek to construct
either right $b_{R}^{-1}$ or left $b_{L}^{-1}$ inverses under Jordan
multiplication $\circ,$ so that $a=c\circ b_{R}^{-1}=b_{L}^{-1}\circ c$. \ A
formal series solution for $b_{R}^{-1}$\ is obtained from the inverse $b^{-1}$
in the enveloping algebra by writing $a=cb^{-1}-bab^{-1}$, and iterating.
\ Thus,%
\begin{equation}
a=cb^{-1}-bcb^{-2}+b^{2}ab^{-2}=\left(  \sum_{n=0}^{\infty}\left(  -1\right)
^{n}b^{n}cb^{-n}\right)  b^{-1}\equiv c\circ b_{R}^{-1}\;.
\end{equation}
Similarly, for the left inverse $a=b^{-1}c-b^{-1}ab$, so%
\begin{equation}
a=b^{-1}c-b^{-2}cb+b^{-2}ab^{2}=b^{-1}\left(  \sum_{n=0}^{\infty}\left(
-1\right)  ^{n}b^{-n}cb^{n}\right)  \equiv b_{L}^{-1}\circ c\;.
\end{equation}
Requiring formally that these two inverses give the same $a$ leads to an
expression that involves only Jordan products of elements from the enveloping
algebra,%
\begin{equation}
a=\frac{1}{2}\sum_{n=0}^{\infty}\left(  -1\right)  ^{n}\left(  b^{n}%
cb^{-n-1}+b^{-n-1}cb^{n}\right)  =\frac{1}{2}\sum_{n=0}^{\infty}\left(
-1\right)  ^{n}\left(  b^{n}cb^{n}\right)  \circ b^{-2n-1}\;.
\end{equation}
However, it involves an infinite number of such products. \ This raises
convergence issues, even when $b^{-1}$ exists in the enveloping algebra.

As the simplest possible illustration of the convergence issues, suppose $b$
and $c$ commute, $bc=cb$. \ Then either of the above series for left or right
inverses gives an ill-defined result, $a=b^{-1}c\,\sum_{n=0}^{\infty}\left(
-1\right)  ^{n}$. \ Evidently, the proper way to interpret the (Borel
summable) series in this case is $\sum_{n=0}^{\infty}\left(  -1\right)
^{n}=\frac{1}{1-\left(  -1\right)  }=\frac{1}{2}$, to produce the solution to
the original equation when $ab=ba$ as well as $bc=cb$. \ Namely, $2ab=c$ and
$a=\frac{1}{2}b^{-1}c$. \ Nevertheless, convergence is a problem. \ 

For the Lie case, the same formal approach may be considered. \ Let
\begin{equation}
a\diamond b=-b\diamond a\equiv ab-ba=c\;.
\end{equation}
Suppose $b$ and $c$ are given and solve for $a$. \ That is, construct either
right $b_{R}^{-1}$ or left $b_{L}^{-1}$ inverses under Lie multiplication,
$\diamond,$ so that $a=c\diamond b_{R}^{-1}=-b_{L}^{-1}\diamond c$. \ Again,
these are given by formal series solutions obtained by writing $a=cb^{-1}%
+bab^{-1}$ and iterating. \ Thus,%
\begin{equation}
a=cb^{-1}+bcb^{-2}+b^{2}ab^{-2}=\left(  \sum_{n=0}^{\infty}b^{n}%
cb^{-n}\right)  b^{-1}\equiv c\diamond b_{R}^{-1}\;.
\end{equation}
Similarly, for the left inverse $a=-b^{-1}c+b^{-1}ab$, so%
\begin{equation}
a=-b^{-1}c-b^{-2}cb+b^{-2}ab^{2}=-b^{-1}\left(  \sum_{n=0}^{\infty}%
b^{-n}cb^{n}\right)  \equiv-b_{L}^{-1}\diamond c\;.
\end{equation}
Requiring that these two inverses give the same $a$ leads to an expression
that involves only Lie products of elements from the enveloping algebra,%
\begin{equation}
a=\frac{1}{2}\sum_{n=0}^{\infty}\left(  b^{n}cb^{-n-1}-b^{-n-1}cb^{n}\right)
\equiv\frac{1}{2}\sum_{n=0}^{\infty}\left(  b^{n}cb^{n}\right)  \diamond
b^{-2n-1}\;.
\end{equation}
But, once more, it involves an infinite number of such products. \ Again this
raises convergence issues, even when $b^{-1}$ exists in the enveloping algebra.

As an illustration of convergence issues in this case, follow Wigner's counsel
and take $2\times2$ matrices, $b=\sigma_{y}=b^{-1}$ and $c=2i\sigma_{z}$.
\ Then, for even $n$, $\left(  b^{n}cb^{n}\right)  \diamond b^{-2n-1}=\left[
c,b\right]  =4\sigma_{x}$; while for odd $n$, $\left(  b^{n}cb^{n}\right)
\diamond b^{-2n-1}=\left[  bcb,b\right]  =-\left[  c,b\right]  =-4\sigma_{x}$.
\ Again, the series gives an ill-defined result,$\ a=\frac{1}{2}\;4\sigma
_{x}\;\sum_{n=0}^{\infty}\left(  -1\right)  ^{n}$. \ This shows clearly that
convergence is again a problem. \ As before, the proper way to interpret the
sum in this particular example is $\sum_{n=0}^{\infty}\left(  -1\right)
^{n}=\frac{1}{2}$, to produce the obvious solution to the original equation,
$a=\sigma_{x}.$ \ \ 

The failed convergence for these series is accompanied by a basic problem:
\ divisors of zero. \ Even when $b$ is invertible in the enveloping algebra,
so that the only solution of $ab=0$ is $a=0$, this does \emph{not} hold for
the Jordan or Lie products. \ The Lie case is most familiar and easily seen.
$\ a\diamond b=0$ always has an infinite number of nonvanishing solutions
$a\neq0$. \ Namely, $a=\kappa b$ for any parameter $\kappa\neq0$. \ Moreover,
there can\ and will be other independent solutions for higher dimensional
enveloping algebras. \ That is to say, Lie algebras are not division rings,
even when they only involve commutators of invertible elements from
$\mathfrak{A}$. \ The same is true for the Jordan case, in general. \ For
instance, using the $2\times2$ matrices as an example, again with
$b=\sigma_{y}$, we have $a\circ b=0\;$for$\;a=\kappa\sigma_{z}+\lambda
\sigma_{x}$ for any parameters $\kappa$ and $\lambda$. \ So, as stressed
already, special Jordan algebras are not division rings -- not even when they
involve only anticommutators of invertible elements from $\mathfrak{A}$. \ 

Perhaps one way to avoid these difficulties and place the formal series
constructions on a firmer footing would be through regularizing
\emph{deformations} of the algebras. \ This works for the specific Jordan or
Lie examples above, as illustrations of the method. \ Rather than the Jordan
or Lie products, take $ab+\lambda ba=c$. \ (This deformation was actually
analyzed in Jordan's original paper \cite{Jordan}, before he settled on the
$\lambda=1$ case.) \ This yields a convergent series for the right inverse
$b_{R}^{-1}$ when $\left|  \lambda\right|  <1$, and a convergent series for
the left inverse $b_{L}^{-1}$ when $\left|  \lambda\right|  >1$. \ For the
right inverse, write $a=cb^{-1}-\lambda bab^{-1},$ and iterate. \ Thus%
\begin{equation}
a=cb^{-1}-\lambda bcb^{-2}+\lambda^{2}b^{2}ab^{-2}=\left(  \sum_{n=0}^{\infty
}\left(  -\lambda\right)  ^{n}b^{n}cb^{-n}\right)  b^{-1}\;.
\end{equation}
For the simplest situation where $bc=cb$, this gives
\begin{equation}
a=b^{-1}c\;\left(  \sum_{n=0}^{\infty}\left(  -\lambda\right)  ^{n}\right)
=b^{-1}c\;\frac{1}{1+\lambda}\;.
\end{equation}
Now the correct result emerges in the limit $\lambda\rightarrow1$, but
strictly speaking this is \emph{not }within the radius of convergence of the
series. \ The series must first be summed to obtain a meromorphic function, by
analytic continuation, and the limit applied to that function.

The same method works for the simple Lie example given above. \ Again, suppose
$b=\sigma_{y}=b^{-1}$ and let $c=2i\sigma_{z}$. \ The series for the right
inverse now gives $a=\frac{1}{2}\,4\,\sigma_{x}\,\sum_{n=0}^{\infty}%
\lambda^{n}=\frac{2}{1-\lambda}\sigma_{x}$. \ The limit $\lambda\rightarrow-1$
converts both this solution and the original equation $ab+\lambda ba=c$ into
the Lie problem of interest $a\diamond b=c$.


\begin{thebibliography}{99}                                                                                               %


\bibitem {Aitken}A C Aitken, \textit{Determinants and Matrices}, Greenwood
Publishing Group, 1983 (reprint).

\bibitem {Awata}H Awata, M Li, D Minic, and T Yoneya, \textquotedblleft On the
Quantization of Nambu Brackets\textquotedblright\ JHEP \textbf{02} (2001) 013
[hep-th/9906248].


\bibitem {Ayupov}S Ayupov, A Rakhimov, and S Usmanov, \textit{Jordan, real,
and Lie structures in operator algebras}, Kluwer Academic Publishers, Boston, 1997.

\bibitem {Azcarraga96a}J A de Azc\'{a}rraga, A. M. Perelomov, and J C
P\'{e}rez Bueno, \textquotedblleft The Schouten-Nijenhuis bracket, cohomology
and generalized Poisson structures\textquotedblright\ J Phys \textbf{A29}
(1996) 7993-8010 [hep-th/9605067].

\bibitem {Azcarraga96b}J A de Azc\'{a}rraga, and J C P\'{e}rez Bueno,
\textquotedblleft Higher-order simple Lie algebras\textquotedblright\ Commun
Math Phys 184 (1997) 669-681 [hep-th/9605213].

\bibitem {Azcarraga97}J A de Azc\'{a}rraga, J M Izquierdo, and J C P\'{e}rez
Bueno, \textquotedblleft On the generalizations of Poisson
structures\textquotedblright\ J Phys \textbf{A30} (1997) L607-L616
[hep-th/9703019]
; \textquotedblleft An introduction to some novel applications of Lie algebra
cohomology in mathematics and physics\textquotedblright\ Rev R Acad Cien Serie
A Mat \textbf{95} (2001) 225-248 [physics/9803046].


\bibitem {Baker}L Baker and D B Fairlie, \textquotedblleft Hamilton-Jacobi
equations and Brane associated Lagrangians\textquotedblright\ Nucl Phys
\textbf{B596 }(2001) 348-364 [hep-th/0003048].


\bibitem {Baranovich}M Baranovich and M S Burgin, \textquotedblleft Linear
$\Omega$-Algebras\textquotedblright\ Russian Math Surveys \textbf{30} (1975) 65-113.

\bibitem {Bayen}F Bayen and M Flato, \textquotedblleft Remarks concerning
Nambu's generalized mechanics\textquotedblright\ Phys Rev \textbf{D11} (1975) 3049-3053.

\bibitem {BraatenCurtrightZachos}E Braaten, T Curtright, and C Zachos,
\textquotedblleft Torsion and Geometrostasis in Nonlinear Sigma
Models\textquotedblright\ Nucl Phys \textbf{B260 }(1985) 630-688.


\bibitem {Calogero}F Calogero, \textquotedblleft Partially Superintegrable
Systems and Nonlinear Harmonic Oscillators\textquotedblright\ talk at the
\textit{Workshop on Superintegrability in Classical and Quantum Systems},
Centre de recherches math\'{e}matiques, Universit\'{e} de Montr\'{e}al, 16 -
21 September 2002.

\bibitem {Chatterjee}R Chatterjee, \textquotedblleft Dynamical Symmetries and
Nambu Mechanics\textquotedblright\ Lett Math Phys \textbf{36} (1996) 117-126
[hep-th/9501141].


\bibitem {Curtright}T Curtright, \textquotedblleft Schr\"{o}dinger's
Cataplex\textquotedblright\ pp 121-131 \textit{Quantum gravity, generalized
theory of gravitation, and superstring theory-based unification}, B
Kursunoglu, S Mintz, and A Perlmutter (editors), Kluwer Academic, New York
2000\ (28th Orbis Scientiae Conference, 1999) [quant-ph/0011101].


\bibitem {CurtrightFairlie}T Curtright and D Fairlie, \textquotedblleft Extra
Dimensions and Nonlinear Equations\textquotedblright\ to appear in J Math Phys
[math-ph/0207008].


\bibitem {CurtrightUematsuZachos}T Curtright, T Uematsu, and C Zachos,
\textquotedblleft Geometry and duality in supersymmetric $\sigma
$-models\textquotedblright\ Nucl Phys \textbf{B469} (1996) 488-512
[hep-th/0011137].


\bibitem {CurtrightZachos}T L Curtright and C\ K Zachos, \textquotedblleft
Deformation Quantization of Superintegrable Systems and Nambu
Mechanics\textquotedblright\ N J Phys \textbf{4} (2002) 83 [hep-th/0205063];
\textquotedblleft Nambu Dynamics, Deformation Quantization, and
Superintegrability\textquotedblright\ talk at the \textit{Workshop on
Superintegrability in Classical and Quantum Systems}, Centre de recherches
math\'{e}matiques, Universit\'{e} de Montr\'{e}al, 16 - 21 September 2002
[math-ph/0211021];
\ \textquotedblleft Quantizing Dirac and Nambu Brackets\textquotedblright%
\ talk at the Coral Gables Conference, 11-14 December 2002 [hep-th/0303088].

\bibitem {DitoFlato}G Dito and M Flato, \textquotedblleft Generalized Abelian
Deformations: \ Application to Nambu Mechanics\textquotedblright\ Lett Math
Phys \textbf{39} (1997) 107-125 [hep-th/9609114].


\bibitem {DitoFST}G Dito, M Flato, D Sternheimer, and L Takhtajan,
\textquotedblleft Deformation Quantization and Nambu
Mechanics\textquotedblright\ Comm Math Phys \textbf{183} (1997) 1-22
[hep-th/9602016].


\bibitem {Estabrook}F B Estabrook, \textquotedblleft Comments on Generalized
Hamiltonian Dynamics\textquotedblright\ Phys Rev \textbf{D8} (1973)
2740-2743.\newpage

\bibitem {Filippov}V T Filippov, \textquotedblleft n-Lie
Algebras\textquotedblright\ Sib Mat Zh \textbf{26} (1985) 126-140 (English
translation: Sib Math Journal \textbf{26} (1986) 879-891).

\bibitem {Francoise}J-P Fran\c{c}oise, \textquotedblleft Perturbation theory
of superintegrable systems\textquotedblright\ talk at the \textit{Workshop on
Superintegrability in Classical and Quantum Systems}, Centre de recherches
math\'{e}matiques, Universit\'{e} de Montr\'{e}al, 16 - 21 September 2002.

\bibitem {Gautheron}P Gautheron, \textquotedblleft Some Remarks Concerning
Nambu Mechanics\textquotedblright\ Lett Math Phys \textbf{37} (1996) 103-116.

\bibitem {Gonera}C Gonera and Y Nutku, \textquotedblleft Super-integrable
Calogero-type systems admit maximal number of Poisson
structures\textquotedblright\ Phys Lett \textbf{A285} (2001) 301-306.

\bibitem {Hanlon}P Hanlon and M Wachs, \textquotedblleft On Lie
k-Algebras\textquotedblright\ Adv Math \textbf{113} (1995) 206-236.

\bibitem {Hietarinta}J Hietarinta, \textquotedblleft Nambu tensors and
commuting vector fields\textquotedblright\ J Phys \textbf{A30} (1997) L27-L33.

\bibitem {Higgins}P Higgins, \textquotedblleft Groups with multiple
operators\textquotedblright\ Proc London Math Soc \textbf{6} (1956) 366--416.

\bibitem {Hoppe}J Hoppe, \textquotedblleft On M Algebras, the Quantization of
Nambu Mechanics, and Volume Preserving Diffeomorphisms\textquotedblright\ Helv
Phys Acta \textbf{70} (1997) 302-317 [hep-th/9602020].


\bibitem {Jacobson}N Jacobson, \textquotedblleft Lie and Jordan triple
systems\textquotedblright\ Am J Math 71 (1949) 149-170;\newline%
\textquotedblleft General representation theory of Jordan
algebras\textquotedblright\ Trans Am Math Soc \textbf{70} (1950) 509--530.

\bibitem {Jordan}P Jordan, \textquotedblleft\"{U}ber die Multiplikation
quantenmechanischer Gr\"{o}\ss en\textquotedblright\ Z Phys \textbf{80} (1933)
285-291; \newline P Jordan, J von Neumann, and E Wigner, \textquotedblleft On
an algebraic generalization of the quantum mechanical
formalism\textquotedblright\ Ann Math \textbf{35} (1934) 29-64.


\bibitem {Kalnay}A J Kalnay and R Tascon, \textquotedblleft On Lagrange,
Hamilton-Dirac, and Nambu mechanics\textquotedblright\ Phys Rev \textbf{D17}
(1978) 1552-1562.

\bibitem {Kerner}R Kerner, \textquotedblleft Ternary Algebraic Structures and
their Applications in Physics\textquotedblright\ [math-ph/0011023].


\bibitem {Kurosh}A G Kurosh, \textquotedblleft Multioperator Rings and
Algebras\textquotedblright\ Russian Math Surveys \textbf{24} (1969) 1-13.

\bibitem {Lister}W Lister, \textquotedblleft A structure theory of Lie triple
systems\textquotedblright\ Trans Am Math Soc \textbf{72} (1952) 217--242.

\bibitem {Matsuo}Y Matsuo and Y Shibusa, \textquotedblleft Volume preserving
diffeomorphism and noncommutative branes\textquotedblright\ JHEP \textbf{02}
(2001) 006 [hep-th/0010040].


\bibitem {Minic}D Minic, \textquotedblleft M Theory and Deformation
Quantization\textquotedblright\ [hep-th/9909022].


\bibitem {MinicTze}D Minic and C-H Tze, \textquotedblleft Nambu Quantum
Mechanics: A Nonlinear Generalization of Geometric Quantum
Mechanics\textquotedblright\ Phys Lett \textbf{B536} (2002) 305-314
[hep-th/0202173].


\bibitem {Mukunda}N Mukunda and E Sudarshan, \textquotedblleft Relation
between Nambu and Hamiltonian mechanics\textquotedblright\ Phys Rev
\textbf{D13} (1976) 2846-2850.

\bibitem {Nambu}Y Nambu, \textquotedblleft Generalized Hamiltonian
Dynamics\textquotedblright\ Phys Rev \textbf{D7} (1973) 2405-2412.


\bibitem {Nutku}Y Nutku, \textquotedblleft Quantization of superintegrable
systems with Nambu-Poisson brackets\textquotedblright\ talk at the
\textit{Workshop on Superintegrability in Classical and Quantum Systems},
Centre de recherches math\'{e}matiques, Universit\'{e} de Montr\'{e}al, 16 -
21 September 2002; \ \textquotedblleft Quantization with Nambu-Poisson
brackets: \ The harmonic oscillator!\textquotedblright\ J Phys A, \emph{in
press}.

\bibitem {Pioline}B Pioline, \textquotedblleft Comments on the Topological
Open Membrane\textquotedblright\ Phys Rev \textbf{D66 }(2002) 025010
[hep-th/0201257].


\bibitem {Sahoo}D Sahoo and M C Valsakumar, \textquotedblleft Nambu mechanics
and its quantization\textquotedblright\ Phys Rev \textbf{A46} (1992)
4410-4412; \textquotedblleft Nonexistence of Nambu Quantum
Mechanics\textquotedblright\ Mod Phys Lett \textbf{A9} (1994) 2727-2732.

\bibitem {Sakakibara}M Sakakibara, \textquotedblleft Remarks on a Deformation
Quantization of the Canonical Nambu Bracket\textquotedblright\ Prog Theor Phys
\textbf{104} (2000) 1067-1071.


\bibitem {Schlesinger}M Schlesinger and J D Stasheff, \textquotedblleft The
Lie algebra structure of tangent cohomology and deformation
theory\textquotedblright\ J Pure Appl Algebra \textbf{38} (1985) 313-322.

\bibitem {Takhtajan}L Takhtajan, \textquotedblleft On foundation of the
generalized Nambu mechanics\textquotedblright\ Comm Math Phys \textbf{160}
(1994) 295--315 [hep-th/9301111].


\bibitem {Tempesta}P Tempesta, A V Turbiner, and P Winternitz,
\textquotedblleft Exact solvability of superintegrable
systems\textquotedblright\ J Math Phys \textbf{42} (2001) 4248-4257
[hep-th/0011209].


\bibitem {Vainerman}L Vainerman and R Kerner, \textquotedblleft On special
classes of n-algebras\textquotedblright\ J Math Physics \textbf{37} (1996) 2553-2565.

\bibitem {Xiong}Chuan-sheng Xiong, \textquotedblleft Remark on Quantum Nambu
Bracket\textquotedblright\ Phys Lett \textbf{B486} (2000) 228-231
[hep-th/0003292].


\bibitem {ZachosCurtright}C K Zachos and T L Curtright, \textquotedblleft
Deformation Quantization, Superintegrability, and Nambu
Mechanics\textquotedblright\ [hep-th/0210170];
\ \textquotedblleft Deformation Quantization of Nambu
Mechanics\textquotedblright\ talk at the Coral Gables Conference, 11-14
December 2002\ [quant-ph/0302106].
\end{thebibliography}
\end{document}